\documentclass[11pt]{article}

\usepackage{amsmath,graphicx}
\usepackage{epstopdf}
\usepackage{multirow}
\usepackage{amsfonts}
\usepackage{amssymb}
\usepackage{amscd}

\def\hybrid{\topmargin -20pt    \oddsidemargin 0pt
        \headheight 0pt \headsep 0pt
        \textwidth 6.25in       % A4 paper
        \textheight 9.5in       % A4 paper
        \marginparwidth .875in
        \parskip 5pt plus 1pt   \jot = 1.5ex}
\hybrid

\numberwithin{equation}{section}
\numberwithin{table}{section}
\renewenvironment{thebibliography}[1]
 { \small
   \begin{list}{[\arabic{enumi}]}
    {\usecounter{enumi} \setlength{\parsep}{0pt}
     \setlength{\itemsep}{5pt} \settowidth{\labelwidth}{#1.}
     \sloppy
    }}{\end{list}}
 
\newcommand{\refcite}{\cite}
\newcommand{\beq}{\begin{equation}}
\newcommand{\eeq}{\end{equation}}
\newcommand{\bea}{\begin{eqnarray}}
\newcommand{\eea}{\end{eqnarray}}
\newcommand{\ba}{\begin{array}}
\newcommand{\ea}{\end{array}}
\newcommand{\bt}{\begin{tabular}}
\newcommand{\et}{\end{tabular}}
\newcommand{\bc}{\begin{center}}
\newcommand{\ec}{\end{center}}

\def\nn{\nonumber} 
 
\def\cy {Calabi--Yau}

\def\clss {Clifford(6,6)} 
 
\def\0 {\nonumber} 
\def\del{\partial} 
\def\bi{\bar \imath}
\def\Ox{\Omega} 
\def\bj{\bar \jmath}
\def\minicent#1#2{
  \begin{minipage}{#1 cm}
    \begin{center}
     #2 
    \end{center}
  \end{minipage}
}
\newcommand{\sla}{\slash\!\!\!}

\newcommand{\ax}{\alpha}
\newcommand{\bx}{\beta}

\newcommand{\Oxb}{\bar{\Omega}}
\newcommand{\ep}{\epsilon}
\newcommand{\ka}{\kappa}

\newcommand{\cK}{\mathcal{K}}
\newcommand{\K}{\mathcal{K}}
\newcommand{\cN}{\mathcal{N}}

\newcommand{\KK}{\mathcal{K}}

\newcommand{\cM}{\mathcal M}
\newcommand{\M}{\mathcal{M}}

\newcommand{\I}{\text{Im}}
\newcommand{\R}{\text{Re}}
\newcommand{\dd}{d}
\newcommand{\Gtb}{\bar{G}_3\vphantom{\bar G}} 

\newcommand{\RR}{{\rm RR}}
\newcommand{\cof}{{\rm cof}}
\newcommand{\Ms}{M_{\rm{s\not usy}}}
\def\cy {Calabi--Yau}

\def\clss {Clifford$(6,6)$} 

\def\stt {SU(3)$\times$SU(3)}
\def\Phit{\tilde \Phi}
\def\tts{$T \oplus T^*$ }
\newcommand{\Dt}{\rm{D3}}
\newcommand{\Df}{\rm{D5}}
\newcommand{\Ds}{\rm{D6}}
\newcommand{\Ot}{\rm{O3}}
\newcommand{\Of}{\rm{O5}}
\newcommand{\Os}{\rm{O6}}

\newcommand{\sym}[2]%{\left<{#1},{#2}\right>}
   {\omega\!\left(#1,#2\right)}
\newcommand{\symr}[2]%{\left<{#1},{#2}\right>}
   {\omega_\rho\!\left(#1,#2\right)}
\newcommand{\symJ}[2]%{\left<{#1},{#2}\right>}
   {\omega_J\!\left(#1,#2\right)}
\newcommand{\mukai}[2]{\left<{#1},{#2}\right>}
\newcommand{\revmukai}[2]{\left<{#2},{#1}\right>}

\newcommand{\cref}{{\bf [check ref]}}

\newcommand{\F}{\mathcal{F}}

\newcommand{\N}{\mathcal{N}}

\begin{document}

%\markboth{Mariana Gra{\~n}a}
%{Fluxreview}

%%%%%%%%%%%%%%%%%%%%% Publisher's Area please ignore %%%%%%%%%%%%%%%
%
%\catchline{}{}{}{}{}
%
%%%%%%%%%%%%%%%%%%%%%%%%%%%%%%%%%%%%%%%%%%%%%%%%%%%%%%%%%%%%%%%%%%%%

%\begin{titlepage}

\rightline{\small LPTENS-05/26}
\rightline{\small CPHT-RR049.0805}
\begin{center}

\vskip 2cm
{\Large \bf Flux compactifications in string theory: }\\
\vskip 0.4cm 

{\Large \bf a comprehensive review}

\vskip 1.5cm

{\bf Mariana Gra{\~n}a}

\vskip 0.8cm
{\em Laboratoire de Physique Th\'eorique de l'Ecole Normale
Sup\'erieure\\
24 rue Lhomond, 75231 Paris Cedex, France}\\
\vskip 0.4cm

{\em Centre de Physique Th{\'e}orique, Ecole
Polytechnique \\
91128 Palaiseau Cedex, France}

\vskip 0.4cm
{\em Service de Physique Th\'eorique,                   
CEA/Saclay \\
91191 Gif-sur-Yvette Cedex, France}  

\vskip 0.3cm
{\tt mariana.grana@cea.fr}
\vskip2cm

\begin{abstract}

\vskip0.6cm

We present a pedagogical overview of flux compactifications in string
theory, from the basic ideas to the most recent developments.
We concentrate on closed string fluxes in type II theories.
We start by reviewing the supersymmetric flux configurations 
with maximally symmetric four-dimensional spaces.  We then discuss
the no-go theorems (and their evasion) for compactifications with fluxes. 
We analyze the resulting four-dimensional effective theories, as well
as some of its perturbative and non-perturbative corrections, focusing
on moduli stabilization. Finally, we briefly review statistical 
studies of flux backgrounds.

%\keywords{Keyword1; keyword2; keyword3.}
\end{abstract}

%\ccode{PACS numbers: 11.25.Hf, 123.1K}
%\end{titlepage}

\end{center}

\pagebreak

{\small
\tableofcontents }

\newpage

\section{Introduction} \label{sec:intro}

One of the central questions in string theory concerns 
the existence and viability of semi-realistic four-dimensional
vacua. The current paradigm of particle phenomenology
prefers an $\N=1$ matter sector with spontaneously broken
supersymmetry at low energies. A huge amount of effort in
string theory is devoted to finding such spontaneously broken $\N=1$ vacuum
with a Standard Model sector. 

As soon as the $E_8 \times E_8$ and $SO(32)$ heterotic theories
were constructed, vacuum configurations with four-dimensional $\N=1$
supersymmetry were found by compactifying the heterotic string
on Calabi-Yau manifolds \cite{CHSW}. 
Unbroken $\N=1$ supersymmetry at the compactification
scale in the heterotic theory is a very stringent
requirement.
If the vacuum is a product of four-dimensional
maximally symmetric space and some compact manifold, the 
former can only be Minkowski, and the latter
is  required to be Calabi-Yau. Furthermore, no vacuum
expectation value for the NS field strength is allowed. 
The situation improves when a warped factor multiplying
the space-time metric is taken into account \cite{Strominger}. 
The NS field can acquire a vacuum expectation value,
but the price to pay was too high at the time:
the internal manifold is no longer K\"ahler.
Not much was known about non K\"ahler manifolds,
and, as a consequence, 
the resulting four dimensional effective theory
was largely unknown.
The attraction was therefore concentrated on
flux-less Calabi-Yau or toroidal orbifold compactifications of the
heterotic theory (with, however, vevs for internal fluxes,
which break the gauge group to the
Standard Model or GUT groups) \cite{hetSM}.
Supersymmetry is spontaneously
broken in these models by four-dimensional 
non-perturbative effects, such as gaugino condensation \cite{hetgaugino}. 
Due to the lack of technologies to study non-perturbative
phenomena at string level, the structure of 
non-perturbative effects, as well as the possibility
to break supersymmetry spontaneously, are determined by field theoretic
considerations. In spite of the enormous 
progress achieved over the years, the mechanism
is not yet satisfactory, as it always leads to negative
cosmological constants, and  
suffers from other cosmological problems \cite{Quevedorev}.
However, heterotic or type I string internal fluxes can,
besides breaking the $SO(32)$ group to the Standard Model one,
trigger spontaneous supersymmetry breaking \cite{Bachas}. 
Small tadpoles for the metric
and dilaton are not canceled at the classical level
in this construction, 
but are hoped to be canceled by
higher loop or non perturbative corrections.
Nevertheless, one gets by this mechanism a satisfactory theoretical control
of supersymmetry breaking, and consequently 
a good description of the low energy physics.

The scene changed drastically after the discovery
of D-branes as non-perturbative BPS objects in
string theory \cite{PolD}. D-branes can serve as 
ingredients in constructing four-dimensional
standard-like models \cite{SM}. Additionally,
they constitute the previously missing sources for RR fluxes.
Very soon after their discovery, the possibility
of finding new supersymmetric vacua for type II string theories
with non-vanishing vacuum expectation values for RR field
strengths was envisaged \cite{PS,Michelson}. Solutions with background
fluxes became rapidly more interesting from the theoretical
and phenomenological point of view.

Non-vanishing vacuum expectation values for the field strengths
were shown to serve as a way
to partially break the $\N=2$ supersymmetry
of Calabi-Yau (non) compactifications down 
to $\N=1$ by mass deformation \cite{TaylorVafa}. 
In conformally flat six-dimensional spaces, fluxes can break
the $\N=4$ supersymmetry to $\N=3,2,1,0$ 
in a controlled and stable way \cite{GP1,Gubser,FP,Kachrububbles}. 
Fluxes became an even more attractive mechanism 
of partially breaking supersymmetry 
after AdS/CFT correspondence was conjectured \cite{AdsCFT}:
type IIB solutions with 3-form fluxes 
could realize string theory duals of confining
gauge theories \cite{PoSt,KS,MN}.
While partially breaking supersymmetry,
fluxes give vacuum expectation values
to some of the typically large number of massless fields (``moduli'') 
arising in string theory
compactifications \cite{TaylorVafa,DRS,GKP}. 
In some IIA scenarios discussed recently \cite{DKPZ,WGKT,IFC},
fluxes alone can stabilize all moduli classically in a regime where
the supergravity approximation can be 
trusted. Fluxes generate at the same time
warped metrics, which can realize large hierarchies \cite{GKP} 
as in Randall-Sundrum type models
\cite{RS,Ver}.

Fluxes cannot however be turned on at will
in compact spaces, as they give a positive contribution
to the energy momentum tensor \cite{MNnogo,GKP}.
As a consequence, negative tension sources (orientifold planes)
should be added 
for consistent compactifications of type II
theories. The number of units of fluxes 
allowed has therefore always an upper bound,
given by the geometry of the compactification manifold.
This still leaves nevertheless a huge amount
of freedom, making compactifications
in background fluxes one of the most
rich and attractive ingredients in the ultimate goal of
realizing string-based models of particle
physics and early universe cosmology.
Looking at the story from the opposite perspective, flux compactifications
are perhaps too rich.
Despite the flurry of activity in the field,
we still lack of an understanding of 
whether any of the large amount of available perturbative
vacua (the dense ``discreetuum'' \cite{BP}, or ``landscape'' \cite{landscape})
is in any sense preferred over the rest (either dynamical, cosmological or
antrhopically). In the absence of a vacuum selection
principle, a statistical study of the landscape
was advocated as possible guidance principle
for the search of the right vacuum \cite{AsDo,DeDo}.

The purpose of this review is to provide a pedagogical exploration
of the literature on flux compactifications, from the basic ideas to
the recent developments. 
We do not plan (and cannot be) exhaustive, as the 
field has evolved enormously, and the amount of literature
on the subject is huge. Although we give a large number of references, the citation list is clearly not exhaustive either.
We decided to concentrate on type II 
compactifications with $\N=1$ supersymmetric
flux vacua. 
We discuss the effective four dimensional theories, as well as some of its 
perturbative and non-perturbative corrections,
focusing on moduli stabilization.
We also give a brief overview of statistical studies of flux backgrounds.

Inevitably, many recent and not so recent developments in flux compactifications
will not be covered in this review. Among them, some of the    
main subjects not to be discussed (for practical reasons, not for lack of interest) are open string fluxes and open string 
moduli stabilization. Besides, not much will be said
about M-theory flux vacua, and their potential to
stabilize moduli. There 
has been a lot of very recent progress in 
understanding the open  
and M-theory landscapes \cite{mtl}, 
open moduli stabilization by open
and closed string fluxes \cite{LustMRS,BCMS,oms} and
moduli stabilization in M-theory \cite{DRS,mtm,AsKa},  
which is worth
a review by itself. Neither do we discuss
twisted moduli, and their stabilization mechanisms 
\cite{WGKT,CU,LMRS}.
In the final summary we mention other  topics not covered
in this review.

The paper is organized as follows.
In section \ref{sec:basic} we give the basic
definitions to be used all throughout the review. In section
\ref{sec:susy} we discuss type II $\N=1$ Minkowski backgrounds with
flux. In subsection \ref{sec:GCG} we review very briefly
generalized complex geometry, with the purpose of describing
the internal geometries of $\N=1$ vacua, which we do in
subsection \ref{sec:GCY}. The reader interested in the main theme of
compactifications on (conformally rescaled) Ricci-flat manifolds can skip
these two sections, which are not needed to understand most
of the rest of the review. In section \ref{sec:nogo} we discuss
the no-go theorems for compactifications with fluxes, and
the way string theory avoids them. In section \ref{sec:4D}
we review the four-dimensional effective theories in Calabi-Yau
and Calabi-Yau orientifold compactifications of type II 
theories. Flux generated
potentials and their superpotential origins are discussed
in subsections \ref{sec:fluxpot} and \ref{sec:fluxsup}. We end
up the section with a  
brief discussion of mirror symmetry in flux backgrounds, done
in subsection
\ref{sec:mirror}. In section \ref{sec:moduli} we review 
moduli stabilization by fluxes. We discuss the general mechanism
of flux stabilization in IIB and IIA Calabi-Yau orientifolds in subsections
\ref{sec:moduliconifold} and \ref{sec:modCYIIA}, and
illustrate with examples for the simpler cases of orientifolds
of tori in subsections \ref{sec:modtoriIIB} and \ref{sec:modtoriIIA}
for IIB and IIA respectively. In section \ref{sec:corrections} 
we discuss some corrections to the low energy effective action,
reviewing in subsections \ref{sec:KKLT} and \ref{sec:deSitter}
moduli stabilization including these corrections, and de Sitter vacua.
Finally, we give in section \ref{sec:stat} a very brief overview
of the distributions of flux vacua. We finish by a summary,
mentioning some topics not covered in the review.

\section{Basic definitions} \label{sec:basic}

In this section we give the basic definitions that will 
be used all along the review. The definitions of 
some parameters less frequently used,
as well as conventions, are left to the Appendix.

The massless bosonic fields of type II superstring theory are 
the dilaton $\phi$, the metric tensor $g_{MN}$ and the antisymmetric 2-tensor
$B_{MN}$ in the NS-NS sector.
The massless RR sector of type IIA contains a 1-form and 3-form potentials
$C_M$, $C_{MNP}$. That of type IIB 
comprises the axion $C_0$, the 2-form potential $C_{MN}$, and the
four-form field $C_{MNPQ}$ with self-dual five-form field strength.
In type IIB, the two scalars $C_0$ and $\phi$ can be combined 
into a complex field $\tau = C_0+
ie^{-\phi}$ which parameterizes an $SL(2,{\bf R})/ U\left(1\right)$ coset space.

The fermionic superpartners are two Majorana-Weyl
gravitinos $\psi^A_{M}$, $A=1,2$ of opposite chirality
in IIA ($\gamma_{11} \, \psi^1_{\rm{IIA} \, M}=\psi^1_{\rm{IIA}\, M} \ $; 
$ \gamma_{11} \, \psi^2_{\rm{IIA} \, M}=-\psi^2_{\rm{IIA}\, M}$)
and the same
chirality in 
type IIB ($\gamma_{11} \, \psi^A_{\rm{IIB} \, M}=\psi^A_{\rm{IIB}\, M}$);
 and two Majorana-Weyl
dilatinos $\lambda^A$ with opposite chirality than the gravitinos.

Type II theories have $D=10$,
$\mathcal
{N}$=2 supersymmetry with two Majorana-Weyl supersymmetry parameters $\epsilon^A$
of the same chirality as the corresponding gravitinos.

The field strength for the NS flux is defined
\beq \label{H}
H=dB \ .
\eeq
For the RR field strengths,
we will use the democratic formulation
of Ref.\refcite{Bergshoeff}, who actually considers 
all RR potentials ($C_1 ... C_9$ in IIA, and $C_0, C_2 ... C_{10}$ in IIB), 
imposing a self-duality constraint
on their field strengths to reduce the doubling of degrees of freedom. 
The RR field strengths are given by
\footnote{The notation $F^{(10)}$ is used
to distinguish them from the purely internal 
fluxes $F$ defined in (\ref{splitflux})
and  used all throughout the review. It should not be 
confused with the supraindices $(1),(3),(6),(8)$
in Tables \ref{ta:IIA} and \ref{ta:IIB} below, which denote 
a particular SU(3) representation. 
}
\beq \label{fl}
F^{(10)}=dC - H \wedge C + m \, e^B = \hat F- H \wedge C
\eeq
where $F^{(10)}$ is the formal sum of all even (odd) fluxes
in IIA (IIB), $\hat F=dC+m e^B$, and $m\equiv F^{(10)}_0=\hat F_0$ is the mass
parameter of IIA. These RR fluxes are 
constrained by the Hodge-duality relation
\beq \label{sd}
F^{(10)}_{n} = (-1)^{Int[n/2]} \star F^{(10)}_{10-n} \ ,
\eeq
where $\star$ is a ten-dimensional Hodge star.

The Bianchi identities for the NS flux and 
the democratic RR fluxes are
\beq \label{Bianchigen}
dH=0 \ , \quad dF^{(10)} - H \wedge F^{(10)} =0 \ . 
\eeq

When sources are present, there is 
no globally well-defined potential,
and the integral of the field strength
over a cycle does is not necessarily zero.  
When this is the case, there is
a non-zero {\it flux}. Charges are quantized
in string theory \footnote{From the pure
supergravity point of view, the charges are continuous parameters.
In the quantum theory, they are quantized, and the total number of quanta
will play a particularly important role in section \ref{sec:nogo}.}, and therefore the fluxes
have to obey Dirac quantization conditions.
Any flux with a standard Bianchi identity (NS or RR) should satisfy
\beq
\frac{1}{(2\pi \sqrt{\ax'})^{p-1}}\int_{\Sigma_p} \hat F_p \in \mathbb{Z}
\eeq
for any $p$-cycle $\Sigma_p$.

By Poincar\'e and Hodge duality, there are as many 
2- as 4-cycles in homology, while 3-cycles come in pairs $(A,B)$.
We therefore define electric and magnetic fluxes
for each field strength according to
\bea \label{quant}
\frac{1}{(2\pi)^2 \ax'} \int_{A_K} H_3 &=& m^K \ , \qquad 
\frac{1}{(2\pi)^2 \ax'} \int_{B^K} H_3= e_{K} \ ,  \qquad \quad K=1,..,\frac{h^3}{2}\nn \\
\frac{1}{(2\pi)^2 \ax'} \int_{A_K} \hat F_3 &=& m^K_{\rm{RR}} \ , \qquad
\frac{1}{(2\pi)^2 \ax'} \int_{B^k} \hat F_3= e_{\rm{RR}\,K} \ , \nn \\
\frac{1}{2\pi \sqrt \ax'} \int_{A_a} \hat F_2 &=& m^a_{\rm{RR}} \ , \qquad
\frac{1}{(2\pi \sqrt \ax')^3} \int_{B^a} \hat F_4= e_{\rm{RR}\,a} \ , \quad \quad
\ \, a=1,..,h^2.
\eea
We have not defined an integral flux for $F_1$ and $F_5$
because there are no non trivial 1 and 5-cycles 
in Calabi-Yau 3-folds (which will be the manifold
we will mostly deal with). The distinction between A
and B-cycles is conventional at this level. 
In non-compact Calabi-Yau's, the A-cycles are
compact, while the B-ones go off to infinity. 
 
We define the Poincare duals to the cycles as
$\ax_K=[B^K]\,$, $\bx^K=[A_K]$, $w_a=[B^a]$, $\tilde w^a=[A_a]$,
% this definition of duality has an opposite sign than that in
% Candela's lectures
or equivalently
\bea \label{norm3forms}
\int_{A_L} \alpha_K &=& \int \alpha_K \wedge \beta^L=\delta_K^L \ , \qquad 
\int_{B^K} \beta^L = \int \beta^L \wedge \alpha_K =- \delta_K^L \  \nn \\
 \int_{A_a} \omega_b &=& \int \omega_a \wedge \tilde \omega^b=\delta_a^b \ , \qquad 
\int_{B^a} \tilde \omega^b = \int \tilde \omega^b \wedge \omega_a =- \delta_a^b
\eea
where an integral without a subindex indicates an integral
over the whole six-dimensional manifold.

These relations imply that the field strengths can be expanded
in the following way 
\bea \label{HF}
\frac{1}{(2\pi)^2 \ax'} H_3 &=&  
 m^K \ax_K  -e_K \bx^K \ , \qquad
\frac{1}{(2\pi)^2 \ax'} \hat F_3 =   
  m^K_{\RR} \ax_K -e_{\RR \,K} \bx^K  \nn \\
\frac{1}{2\pi \sqrt{\ax'}} \hat F_2 &=&  m_{\rm{RR}}^a \omega_a  \ , \qquad \qquad
\frac{1}{(2\pi \sqrt{\ax'})^3} \hat F_4 = -  e_{\rm{RR}\,a} \, \tilde \omega^a  \ \nn \\
2\pi \sqrt{\ax'} \hat F_0 &=&   m_{\rm{RR}}^0 \ , \qquad \qquad \quad
\frac{1}{(2\pi \sqrt{\ax'})^5} \hat F_6 =   e_{\rm{RR}\,0} \rm{Vol_6}  
\eea
In most of the text, we will take  $(2\pi)^2 \alpha'=1$. 
Factors of  $(2\pi)^2 \alpha'$ are written explicitly
only in a few equations, when they are relevant.

There is a
a symplectic $Sp(2\, h^{(1,1)} + 2, \mathbb Z)$ and 
$Sp(2\, h^{(2,1)} + 2, \mathbb Z)$ invariance,
part of which correspond to electric-magnetic
duality. We can define the symplectic vectors
\beq \label{symplv}
N = (e_K,m^K) \ , \qquad
N^{\rm{IIB}}_{\RR}=(e_{\RR \, K}, m^K_{\RR}) \ , \qquad 
N^{\rm{IIA}}_{\RR}= (e_{\RR \, A}, m^A_{\RR})
\eeq
where $(e_{A \, \RR}, m^A_{\RR})=(e_{\RR \, 0}, e_{\RR \, a}, m^0_{\RR}, m^A_{\RR})$.

\section{Type II Supersymmetric Backgrounds with Flux} \label{sec:susy}

In this section we review what are the possible configurations
of fluxes and internal geometry that ${\cal N}=1$ supersymmetry allows. 
By analyzing integrability conditions, it was proved \cite{GaPa,London1,Tsimpis} 
that in the context of type II supergravity, a background
that is supersymmetric and whose fluxes satisfy Bianchi 
identities and equations of motion is a solution to the full equations
of motion (whenever there are no mixed external-internal components
of the Einstein tensor, which will be our case). In this section
we concentrate on supersymmetry conditions, while Bianchi identities
and the equations of motion for flux are discussed in section \ref{sec:nogo}.

The analysis of supersymmetry conditions in (unwarped)
compactifications of the heterotic string in the presence
of NS flux
has been carried out in the celebrated
paper \refcite{CHSW} 
The absence of warp factor (in the Einstein frame) 
enforces the
flux to vanish. Warped backgrounds with NS flux have been found to
be consistent with supersymmetry for
the heterotic theory in Ref.~\refcite{Strominger}
(see also \refcite{dWD}), and have been taken up
for type II theories 
using the language of G-structures (to be reviewed in section 
\ref{sec:Gstructures}) 
in Refs.\cite{London1,CCDLMZ,London2}.

Supersymmetric M-theory compactifications on four-folds
to three dimensions were first  
analyzed in Ref.~\refcite{BB}.
 M (and F-theory) compactifications with fluxes
to three and four dimensions
were discussed in Ref.~\refcite{DRS,GoCu}, and
analyzed using the language of
G structures in Refs.\refcite{BeCvM,KaMiTo,MaSpa}
(see Ref.\refcite{consthesis} for a review
and more references). Type IIA (and M-theory)
supersymmetric backgrounds on manifolds
of $G_2$ and SU(3) structure
were first studied in Refs.\refcite{KKbundles,holomonopole,DaPr}
using $G_2$ and SU(3) structure techniques.

Supersymmetric type IIB and F-theory
backgrounds preserving a particular type of 
${\cal N} =1$ supersymmetry with both NSNS and RR fluxes 
(such that the complex flux $G_3$ is imaginary self dual)
where studied in
Ref.~\refcite{GP1,Gubser} 
(for a review
and more references see \refcite{Freythesis}).
$\N=1$ type IIB flux backgrounds preserving
more general supersymmetries were studied in
Ref.\refcite{interpolating}, while 
the most general $\N=1$ supersymmetric ansatz in manifolds with   
 SU(3) structure (and some with SU(2) structure) 
has been studied in Refs.~\refcite{Frey,Dallagata,GMPT} 
(see also Ref.\refcite{BeCvIIB}).

Twenty years after the 
seminal paper by Candelas, Horowitz, Strominger, Witten,
and thanks to the work of many people,
the most general type II
backgrounds compatible with ${\cal N}=1$ supersymmetry  
on manifolds of SU(3) structure are now
known \cite{GMPT,Frey,Tsimpis,BeCvIIA,BeCvIIB},
and a lot is known about flux backgrounds
on SU(2) structure manifolds \cite{Dallagata,GCY2,LT2}. 
In this section we will review type II supersymmetric solutions
in the absence of flux, and in the following
sections we review their flux counterparts,
following mostly Refs.\refcite{GMPT,GCY2}.
Explicit examples of supersymmetric solutions
will be given later, mostly in section \ref{sec:moduli}.
Before starting to review the technical details, 
let us note that 
a classification of 
supersymmetric solutions from the Killing spinors
(and G-structures) has been carried out for example
in Refs.\refcite{GaPa}. Besides, some supersymmetric
 solutions with holographic duals were constructed
using the symmetries of the construction to make
an ansatz for the Killing spinors and the bosonic fields \cite{flow}.
This method is often referred to as ``algebraic Killing spinor technique''
\cite{killspi}. \\

In this review, we will discuss 
vacua whose four-dimensional space 
admits maximal 
space-time symmetry, i.e. Minkowski, anti-de Sitter
space ($AdS_4$) or de Sitter ($dS_4$). 
These have respectively Poincare, $SO(1,4)$ and
$SO(2,3)$ invariance. The most general ten-dimensional metric consistent with
 four-dimensional maximal symmetry is
\beq \label{metric}
ds^2= e^{2A(y)}\tilde{g}_{\mu \nu} dx^\mu dx^\nu + g_{mn} dy^m dy^n \, , \quad \mu=0,1,2,3 \, \quad m=1,...,6
\eeq 
where $A$ is a function of the internal coordinates called
{\it warp factor},  
$\tilde{g}_{\mu \nu}$
is a Minkowski, $dS_4$ or $AdS_4$ metric, and $g_{mn}$ is any six dimensional metric.

%{\bf may be not here?}
%Fluxes allow for supersymmetric compactifications to Minkowski and anti-de Sitter four-dimensional space Ref.~\refcite{Candelas}-Ref.~\refcite{...}, 
%while to get de Sitter space nonperturbative supersymmetry breaking
%ingredients like anti D-branes and gaugino 
%condensates are needed \cite{KKLT}.       
%This is easily understood by looking at the integrability conditions coming
%from supersymmetry equations, as we will see shortly.  
%{\bf end}

Demanding maximal symmetry requires the vacuum expectation value of the 
fermionic fields to vanish. The background should therefore be purely bosonic.
As far as the fluxes, we are allowed to turn on only those who have 
either no leg or four legs along space-time. Therefore the NSNS flux
$H_3$ can only be internal, while from the RR fluxes, only
$F_4$ in IIA and $F_5$ in IIB are allowed to have external components.
  
A supersymmetric vacuum where only bosonic fields have non-vanishing
 vacuum expectation
values should obey $<Q_{\ep} \chi>=<\delta_\ep \chi>=0$, where 
$Q$ is the supersymmetry generator, $\ep$ is the supersymmetry 
parameter and $\chi$ is any fermionic field. In type II theories,
the fermionic fields are two gravitinos $\psi^A_M$, $A=1,2$
and two dilatinos $\lambda^A$. In the
supergravity approximation, the
bosonic parts of their
supersymmetry
transformations in the string frame \footnote{Throughout the paper we use
mostly
 string frame. Whenever Einstein frame is used, it
will
be indicated explicitly}
\begin{equation} 
\label{eq:susyg} 
\delta \psi_M = \nabla_M \epsilon + \frac{1}{4} \sla{H_M} {\cal P} \epsilon + \frac{1}{16} e^{\phi}  
\sum_n  \sla \! {F^{(10)}_{n}} \, \Gamma_{M} {\cal P}_n \, \epsilon  \, ,
\end{equation} 
\begin{equation}  
\label{eq:susyd} 
\delta \lambda = \left(\sla{\partial} \phi + \frac{1}{2} \sla \! H {\cal P}\right) \epsilon 
+ \frac{1}{8} e^{\phi}  
\sum_n (-1)^{n} (5-n)  \sla \! {F^{(10)}_{n}} \,  
{\cal P}_n  \epsilon  \, .
\end{equation} 
In these equations $M=0,...,10$, $\psi_M$ stands for the column
vector $\psi_M={\psi_M^1 \choose \psi_M^2 }$ containing the
two Majorana-Weyl spinors of the same chirality in
type IIB, and of opposite chirality in IIA,  and similarly for
$\lambda$ and $\ep$. 
The $2 \times 2$ matrices ${\cal P}$ and ${\cal P}_n$ 
are different in IIA and  
IIB: for IIA ${\cal P} = \Gamma_{11}$ and 
${\cal P}_n = \Gamma_{11}^{(n/2)} \sigma^1$,
while for 
IIB  ${\cal P} = -\sigma^3$, ${\cal P}_n = \sigma^1$ for $\tfrac{n+1}{2}$ 
even and 
${\cal P}_n = i \sigma^2$ for $\tfrac{n+1}{2}$ odd.
A slash means a contraction with gamma matrices in the form
$\sla \! {F_{n}} = \frac{1}{n!} F_{P_1...P_N} \Gamma^{P_1...P_N}$, and 
 $H_M \equiv \frac{1}{2}H_{MNP} \Gamma^{NP}$. The NS and RR field strengths
are defined in (\ref{H}, \ref{fl}).
We are
using the democratic formulation of Ref. \cite{Bergshoeff}
for the RR fields, as explained in section \ref{sec:basic}.
%The part $G_n \cdot \gamma_{(n)}$ appears also in susy variations  
%for other fields.  
%$F^{(10)}_{n}=dC_{n-1} - H_3 \wedge C_{n-3}$ 
%are the modified RR field strengths  with
%non standard Bianchi identities, $dF^{(10)}_{n}=H_3 \wedge F_{n-2}$. More
%details about Bianchi identities are given in section \ref{sec:Bianchi}.

We want to study flux backgrounds that preserve maximal four 
dimensional symmetry. We therefore require
\beq \label{splitflux}
F^{(10)}_n = F_n + \rm{Vol}_4 \wedge \tilde F_{n-4} \ .
\eeq 
Using
the duality relation (\ref{sd}), the internal and external
components are related by \cite{GMPT}
\beq \label{dualhats}
\tilde F_{n-4} = (-1)^{Int[n/2]} *  F_{10-n} \ .
\eeq
where $*$ is a six-dimensional star. This allows to write the
supersymmetry transformation in terms of internal fluxes only 
$F_n$, $n=0,\dots,6$. For instance a non-zero $F^{(10)}_4$ 
with only $\mu$-type
indices is traded for a ``internal'' $F_6$  with $m$-type
indices. In~\eqref{eq:susyg}, \eqref{eq:susyd} this gives  twice the contribution
for each flux but now $n=0, \dots ,6$ only.

\subsection{Supersymmetric solutions in the absence of flux}

When no fluxes are present, demanding zero VEV for the gravitino variation
(\ref{eq:susyg})
requires the existence of a covariantly constant spinor on the ten-dimensional 
manifold,
i.e. $\nabla_M \ep =0$. The space-time component of this equation
reads
\beq
\tilde \nabla_\mu \epsilon + \frac{1}{2} (\tilde \gamma_\mu \gamma_5 \otimes
\sla \! \nabla A) \ep =0
\eeq
where we have used the standard decomposition 
of the
ten-dimensional gamma matrices (see Appendix \ref{ap:conventions})]
and $\tilde \nabla$ and $\tilde \gamma_{\mu}$ mean a covariant derivative 
and gamma matrix with respect to
$\tilde g_{\mu\nu}$.

This 
yields the following integrability condition
\beq \label{integra}
[\tilde \nabla_\mu, \tilde \nabla_\nu] \ep = -\frac{1}{2}
(\nabla_m A) (\nabla^m A)  \,
\gamma_{\mu\nu} \, \epsilon
\eeq
On the other hand, 
\beq
[\tilde \nabla_\mu, \tilde \nabla_\nu] \ep=\frac{1}{4} 
\tilde R_{\mu\nu\lambda\rho} \gamma^{\lambda \rho} \ep = \frac{k}{2}
\gamma_{\mu \nu}\, \epsilon
\eeq
where we have used that for a maximally symmetric space
$R_{\mu\nu\lambda\rho}=k(g_{\mu\lambda}g_{\nu\rho}-
g_{\mu \rho} g_{\nu \lambda})$, with $k$ negative
for AdS, zero for Minkowski and positive for dS.
Since $\gamma_{\mu \nu}$ is invertible, the integrability
condition reads
\beq
k + \nabla_m A \nabla^m A =0
\eeq
The only possible constant value of  $(\nabla A)^2$
on a compact manifold is zero, which implies 
that the the warp factor is constant 
and the four-dimensional manifold can only be Minkowski space.

%for non compact, AdS allowed

To analyze the internal component of the supersymmetry variation, 
we need to split the supersymmetry spinors into  four-dimensional
and  six-dimensional spinors. For reasons that will become
clear shortly, we will use only one internal Weyl spinor
(and its complex conjugate) to do the decomposition, which reads
\begin{equation}\begin{aligned}
\label{decompepsilon}
\epsilon^1_{\text{IIA}} &= \xi_+^1 \otimes \eta_+
   + \xi_-^1 \otimes \eta_- \ , \\
\epsilon^2_{\text{IIA}} &= \xi_+^2 \otimes \eta_- 
   + \xi_-^2 \otimes \eta_+ \ ,
\end{aligned}\end{equation}
for type IIA, where $\gamma_{11} \ep^{1}_{\text{IIA}}= \ep^{1}_{\text{IIA}}$
and $ \gamma_{11} \ep^{2}_{\text{IIA}}= -\ep^{2}_{\text{IIA}}$, and
the four and six-dimensional spinors obey
$\xi_-^{1,2}=(\xi_+^{1,2})^*$,
  and
$\eta_-=(\eta_+)^*$. (By a slight abuse of notation  we use plus
and minus to indicate both four-dimensional and six-dimensional
chiralities.) For type IIB both spinors have the same chirality, which we take to be positive,
resulting in the decomposition
\beq
\label{decompepIIB}
 \epsilon^A_{\text{IIB}} = \xi_+^A \otimes \eta_+ 
   + \xi_-^A \otimes \eta_- \ , \qquad A=1,2 \ .
\eeq
Inserting these decompositions in 
the internal component of the gravitino variation, Eq.(\ref{eq:susyg}), 
we get the following condition
\beq \label{covconst}
 \nabla_m \eta_{\pm}=0 \,.
\eeq
 The internal manifold should
therefore have a covariantly constant spinor. This is a very strong 
requirement from the topological and differential geometrical point of
view. It forces the manifold to have reduced holonomy. In the
following section we will explain this in more detail (a more detailed
pedagogical discussion of special holonomy 
relevant to the present context can be found for example in 
Ref. \refcite{Gubsertasi}). 
For the time being, we just state that for six-dimensional manifolds
the holonomy group should be SU(3), or a
subgroup of it. A six dimensional manifold with SU(3) holonomy  
is a Calabi-Yau manifold \cite{CY,CYst}. Such manifolds admit one
covariantly constant spinor. 
To have more than one covariantly 
constant spinor the holonomy group of the manifold should be smaller
than SU(3), and this results in a larger number of supersymmetries 
preserved. For most of this review, we shall consider the case of
manifolds having only one covariantly constant spinor (when
turning on fluxes, the covariant constancy condition will be relaxed,
but we will still work mostly with manifolds admitting only one non-vanishing
spinor)
This explains the use of only one internal spinor to
decompose the ten dimensional ones in Eqs. \eqref{decompepsilon} and 
\eqref{decompepIIB}. 

When there is one covariantly constant internal spinor, 
the internal gravitino equation tells us 
that there are two four-dimensional supersymmetry parameters,
$\xi^1$ and $\xi^2$. This compactification preserves therefore
eight supercharges, i.e. $\N =2$ in four dimensions.  
From the world--sheet point of view, a Calabi-Yau compactification
yields a super conformal field theory with (2,2) supersymmetry \cite{CYst}.

In summary, supersymmetric compactifications without
fluxes are only possible to unwarped Minkowski four-dimensional space,
with a Calabi-Yau manifold as internal space. These compactifications
preserve $\N =2$ in four dimensions.
Fluxes can break the $\N=2$ supersymmetry spontaneously 
down to $\N=1$ or even completely in a stable way. 
In the following sections we will review in detail how this works.
As we mentioned before, in order to decompose the ten-dimensional 
supersymmetry parameters,
the internal manifold should admit at least one nowhere vanishing 
internal spinor.  This 
restricts the class of allowed manifolds to those having reduced
structure. We will first review the concept of G-structures, which
is central for the development of flux compactifications, and then   
look at flux backgrounds preserving $\N=1$ supersymmetry.

\subsection{Supersymmetric backgrounds with fluxes} \label{sec:Gstructures}

In this section we review compactifications preserving the minimal amount
of supersymmetry, i.e. $\N=1$ in four dimensions. In order to have
some supercharges preserved, or even in the case when all of them are
completely broken spontaneously by the fluxes, we need to have globally
well defined supercurrents. This requires to have globally well
defined spinors on the internal manifold, which is only 
possible when its structure group is reduced. Let us start by
briefly reviewing the main facts about G-structures. For detailed
explanations, we refer the reader to the mathematical references
Ref.~\refcite{Salomon,Joyce,CS,FI}. For a review of G-structures in 
the context of compactifications with fluxes, see Ref.~\refcite{London2}. \\

In the absence of fluxes, supersymmetry requires a covariantly constant spinor 
on the internal manifold. This condition actually splits into two parts, first 
the existence of such a spinor (i.e., the existence of a non-vanishing globally well defined spinor), and second the condition that it is covariantly constant.
A generic spinor such as the supercurrent can be decomposed in the same way as the 
supersymmetry parameters, Eqs \eqref{decompepsilon} and
\eqref{decompepIIB}. The first condition implies then 
the existence of two four-dimensional supersymmetry parameters 
and thus an effective $\N =2$ four-dimensional action, while the second
implies that this action has an $\cN=2$ Minkowski vacuum.  
As far as the internal manifold is concerned, the fist condition is a 
topological requirement on the manifold, while the
second one is a differential condition on the metric, or rather, on its connection.
Let us first review the impactions of the first condition.

A  globally well defined non-vanishing
spinor exists only on manifolds that
have {\it reduced structure} \cite{Salomon,Joyce}. The structure group 
of a manifold is the 
group of transformations required to patch the orthonormal frame bundle. 
A Riemannian manifold of dimension $d$ has automatically structure group 
$SO(d)$. All vector, tensor and spinor representations can therefore be decomposed in 
representations of $SO(d)$. If the manifold has reduced structure group $G$, then
every representation can be further decomposed in representations of 
$G$. 

Let us concentrate on six dimensions, which is the case we are interested in,
and the group $G$ being $SU(3)$. On a manifold with $SU(3)$ structure, 
the spinor representation in six dimensions, 
in the ${\bf 4}$ of $SO(6)$, can be further decomposed in representations of
$SU(3)$ as $ {\bf 4} \rightarrow {\bf 3} + {\bf 1}$. There is  
therefore an $SU(3)$ singlet 
in the decomposition, which means that there is a spinor that depends trivially on
the tangent bundle of the manifold and is therefore well defined and non-vanishing.
The converse is also true: 
a six dimensional manifold that 
has a globally well defined non-vanishing spinor has structure group $SU(3)$.

We can now go ahead and decompose other $SO(6)$ representations, such as the 
vector ${\bf 6}$, 2-form ${\bf 15}$ and 3-form ${\bf 20}$ 
in representations of $SU(3)$. This
yields ${\bf 6} \rightarrow {\bf 3}+ {\bf \bar 3}$, 
${\bf 15} \rightarrow {\bf 8}+ {\bf 3}+ {\bf \bar 3} + {\bf 1}$, 
${\bf 20} \rightarrow {\bf 6}+ {\bf \bar 6} + {\bf 3}+ {\bf \bar 3}+ 
{\bf 1}+ {\bf 1} $. We can see that 
there are also singlets in the decomposition of 2-forms and 3-forms.
This means that there is also a non-vanishing globally well defined
real 2-form, and complex 3-form. These are called respectively $J$ and
$\Ox$. We can also see that there are no invariant vectors 
(or equivalently five-forms),
which means in particular that $J \wedge \Ox=0$. A six-form is on the
contrary a singlet (and there is only one of them, up
to a constant), which means that $J \wedge J \wedge J$ is
proportional to $\Ox \wedge \bar \Ox$. We use the convention
$J \wedge J \wedge J = \tfrac{3i}{4} \Ox \wedge \bar \Ox $.
$J$ and $\Omega$ determine a metric \footnote{$\Omega$
says what are the holomorphic and antiholomorphic coordinates,
and in expressed in these coordinates $g_{i \bj}=-i J_{i \bj}$}.

 Raising one of the indices
of $J$ we get an almost complex structure, which is a map
that squares to minus the identity, i.e. $J_m \,^p J_p \,^n = -\delta_m \,^n$.
A real matrix that squares to minus the identity has eigenvalues 
$\pm i$, coming in pairs. The existence of an almost complex structure
allows to introduce local holomorphic and antiholomorphic 
vectors $\partial_{z^i}$, $\partial_{z^{\bar i}}$, $i=1,2,3$, which
are the local eigenvectors with eigenvalues $+i$ and $-i$. 
If their dual one-forms $dz^i$ are integrable, i.e., there
exist local functions $f$ such that $dz=df$, and
if the transition
functions between the different patches are holomorphic, 
then the definition of complex coordinates
is globally consistent. In that case, the complex structure is said 
to be integrable, 
or equivalently the manifold is complex. 

The condition for integrability of the almost complex structure can be recast
in the vanishing of a tensor called Nijenhuis, defined as
\beq \label{Nijenhuis}
N_{mn}\,^p=2 \left(J_m\,^q \nabla_{[q} J_{n]}\,^p -
J_n\,^q \nabla_{[q} J_{m]}\,^p \right) \, .
\eeq 
A complex structure is integrable if its associated Nijenhuis tensor vanishes.
Due to the antisymmetrizations, the covariant derivatives in (\ref{Nijenhuis}) 
can actually be replaced by an ordinary derivatives.

The SU(3) structure is determined equivalently by the SU(3) invariant
spinor $\eta$, or by $J$ and $\Ox$. The latter can actually 
be obtained from the spinor by
\beq \label{bilinears}
J_{mn} = \mp \, 2i\, \eta^\dagger_{\pm} \gamma_{mn}  \eta_{\pm} \, \qquad
\Ox_{mnp}=-2i\, \eta^\dagger_- \gamma_{mnp} \eta_+
\eeq
 $J_{mn}$ is a 
(1,1)-form with respect to the almost complex structure $J_m\,^p$,, while $\Ox$ is a (3,0)-form \footnote{This can be seen from 
\eqref{bilinears} by the fact 
that $\eta_+$ is a Clifford vacuum annihilated by gamma
matrices with holomorphic indices, i.e. $\gamma^i \eta_+=0$.}.

We argued that supersymmetry imposes a topological plus a
differential condition on the manifold. So far we have reviewed the
topological condition, which amounts to the requirement that the
manifold has SU(3)
structure. Let us now see what the differential condition is.

In the case of Calabi-Yau 3-folds (this means a Calabi-Yau manifold
with three
complex dimensions), which are manifolds of SU(3) structure, the SU(3)
invariant spinor is also covariantly constant. The metric (or rather the 
Levi-Civita connection) is said to have SU(3) holonomy. The holonomy
group of a connection is the subgroup of O(n) that includes all
possible changes of direction that a vector suffers when being
parallely
transported around
a closed loop. In the case of a manifold with SU(3) structure,
one can show that there is always a metric compatible connection
(i.e., a connection satisfying $\nabla_m' g_{np}=0$), possibly
with torsion\footnote{The torsion is defined by 
$[\nabla_m', \nabla_n'] V_p= -R_{mnp}\,^q V_q -2 T_{mn}\,^q\nabla'_q
V_p$. },
which is 
also compatible with the structure and such that 
$\nabla_m' \eta=0$ \cite{Joyce}. This means that on a manifold with SU(3) structure
there is always a connection with or without torsion
that has SU(3) holonomy. In the case where this connection is
torsionless, the manifold is a Calabi-Yau. 

The torsion tensor
\beq
T_{mn}\,^p \in \Lambda^1 \otimes 
(su(3)\oplus su(3)^\perp)
\eeq 
where $\Lambda ^1$ is the space of 1-forms, and comes from the upper
index $p$, while the lower indices $mn$ span the space of two forms,
which
is isomorphic to $so(6)$, the Lie algebra of $SO(6)$. We have
also used $so(6)=su(3) \oplus su(3)^\perp$. Acting on SU(3) invariant
forms,
the $su(3)$ piece drops. The corresponding torsion is called the
intrinsic torsion, and contains the following representations
\bea \label{torsion}
T^0_{mn}\,^p \in \Lambda^1 \otimes su(3)^\perp &=& ({\bf 3} \oplus {\bf \bar
3}) \otimes  ({\bf 1} \oplus {\bf 3} \oplus {\bf \bar 3}) \nn \\
&=& ({\bf 1} \oplus {\bf 
1}) \oplus  ({\bf 8} \oplus {\bf 8}) \oplus ({\bf 6} \oplus {\bf \bar
6}) \oplus  2 \, ({\bf 3} \oplus {\bf \bar 3} ) \nn\\
&& \quad W_1 \quad  \quad\quad   W_2 \quad  \quad\quad W_3 \quad
\quad
 W_4, W_5  
\eea
$W_1, ... , W_5$ are the five torsion classes that appear in the 
covariant derivatives of the spinor, of $J$ and of $\Ox$. 
$W_1$ is a complex scalar, $W_2$ is  
a complex primitive (1,1) form (primitivity means 
$(W_2)_{mn} J^{mn}=0$), $W_3$ is a real primitive $(2,1) + (1,2)$
form and $W_4$ and $W_5$ are real vectors ($W_5$ is
actually a complex (1,0)-form, which has the same degrees of freedom).
Antisymetrizing the covariant derivative of $J$ and $\Ox$ and
decomposing into SU(3) representations, we can see that $dJ$
should contain $W_1,W_3$ and $W_4$, while $W_1, W_2$ and $W_5$ appear
in $d\Ox$ (see for example Ref.~\refcite{CS} for details). We can therefore
write
\bea
\label{dJdOmega}
   dJ &=& \tfrac{3}{2} Im\left(
      \bar W_1\Omega \right) + W_4\wedge J + W_3\ , \nn \\
   d\Omega &=& W_1 J^2 +W_2\wedge J +\bar W_5\wedge \Omega\ .
\eea
We give in Eq.(\ref{alvesre}) in the Appendix the inverse relations, namely
$W_i$ in terms of $dJ$, $d\Ox$, $J$ and $\Omega$.

A manifold of SU(3) structure is complex if $W_1=0=W_2$. We can understand
that this condition is necessary by noting that the pieces containing
$W_1$ and $W_2$ in $d\Ox$ are (2,2)-forms, while $\Ox$ itself is a 
(3,0)-form. In a complex manifold, the exterior derivative of
a (p, q)-form should only have (p+1, q) and (p, q+1) pieces, which
means that if the manifold was complex, $d\Omega$ could only be a
(3,1)-form. Therefore  for the manifold to
be complex, $W_1$ and $W_2$ must vanish. It can be shown that this condition is also sufficient,
and
is equivalent to requiring the Nijenhuis tensor defined in
Eq.(\ref{Nijenhuis}) to be zero. 
In a symplectic manifold, the fundamental 2-form, $J$, is closed. 
A symplectic manifold of SU(3) structure has therefore vanishing
$W_1$, $W_3$ and $W_4$. 
A K{\"a}hler manifold is complex and symplectic, which means that
the only possible nonzero torsion is $W_5$. In that case, the Levi Civita
connection has U(3) holonomy. Finally, a Calabi-Yau has SU(3)
holonomy, and all torsion classes vanishing. We summarize this 
and also give the vanishing torsion classes in other special manifolds
in Table \ref{ta:manifolds}.

\begin{table}
\begin{center}
\begin{tabular}{|c|c|}
\hline
{\bf Manifold} & {\bf Vanishing torsion class}\\
\hline
Complex & $W_1=W_2=0$ \\ 
\hline
Symplectic & $W_1=W_3=W_4=0$ \\ 
\hline
Half-flat & $\I W_1=\I W_2=W_4=W_5=0$ \\
\hline
Special Hermitean & $W_1=W_2=W_4=W_5=0$ \\ 
\hline
Nearly K\"ahler & $W_2=W_3=W_4=W_5=0$ \\
\hline
Almost K\"ahler & $W_1=W_3=W_4=W_5=0$ \\
\hline
K\"ahler & $ W_1=W_2=W_3=W_4=0$ \\
\hline
Calabi-Yau & $ W_1=W_2=W_3=W_4=W_5=0$ \\
\hline
``Conformal'' Calabi-Yau & $ W_1=W_2=W_3= 3 W_4-2 W_5=0$ \\
\hline
\end{tabular}
\caption{ \label{ta:manifolds}
\text{Vanishing torsion classes in special SU(3) structure manifolds.}}
\end{center}
\end{table}

\subsection{$\N =1$ Minkowski vacua (and beyond)} \label{sec:N=1}

We have discussed the topological condition 
on the internal manifold required in order to have some supersymmetry
preserved. In this section, we will see what is the differential 
condition that $\N=1$ supersymmetry imposes. This differential condition
will link the allowed torsion classes to the fluxes, i.e. given
an SU(3) structure manifold with certain non-vanishing torsion classes,
the allowed fluxes are completely determined by the torsion.

We know that imposing SU(3) structure on the manifold allows us to 
decompose the two ten-dimensional spinors as in
Eqs. (\ref{decompepsilon}), (\ref{decompepIIB}). 
We have to insert now these decompositions   in the gravitino and
dilatino variations, Eqs.(\ref{eq:susyg})-(\ref{eq:susyg}). But before
doing that, we should notice that if we leave the four dimensional
spinors $\xi ^{1,2}$ generic, then the supersymmetry preserved 
would be $\N=2$ instead of $\N=1$. We need therefore a relation
between $\xi ^1$ and $\xi ^2$. Demanding maximal four dimensional
symmetry only allows a trivial relation between $\xi^1$  and $\xi^2$, 
namely they should be proportional. The (complex) constant of proportionality 
can actually be a function of internal space, which can be 
included in the definition of six dimensional spinors. We will therefore
use
\begin{equation}\begin{aligned}
\label{epsilon}
\epsilon^1_{\text{IIA}} &= \xi_+ \otimes \, (a \, \eta_+)
   + \xi_- \otimes (\bar a \, \eta_-) \qquad \quad  
\epsilon^1_{\text{IIB}} = \xi_+ \otimes (a \, \eta_+) 
   + \xi_- \otimes (\bar a \, \eta_-)   \\
\epsilon^2_{\text{IIA}} &= \xi_+ \otimes \, (\bar b \,\eta_-) 
   + \xi_- \otimes (b \, \eta_+) \qquad \quad \,  
\epsilon^2_{\text{IIB}} = \xi_+ \otimes (b \, \eta_+) 
   + \xi_- \otimes (\bar b \, \eta_-) 
\end{aligned}\end{equation}
where $a$ and $b$ are complex functions. $\N=1$ supersymmetry
links $a$ and $b$, and how they are related tells us how the $\N=1$
vacuum sits in the underlying $\N=2$ effective four dimensional theory.  

When inserting these spinors in the supersymmetry variations, 
Eqs.(\ref{eq:susyg}, \ref{eq:susyd}), the four-dimensional pieces can be factored out, and we get 
equations involving the six-dimensional parts of the spinors.
It is useful to decompose the resulting spinors in a basis, given by
$\eta_+, \gamma^m \eta_-, \gamma^m \eta_+, \eta_-$, where
the first (last) two have positive (negative) chirality. We can write schematically the resulting equations for the positive chirality spinor as follows
\bea
\delta \Psi_{\mu}:&&P\, \eta_+ + P_m \gamma^m \eta_- = 0 \, ,\nn\\
\delta \Psi_{m}: && Q_m  \eta_+ + Q_{mn} \gamma^{n}
\eta_-  = 0 \, ,\nn\\
\delta \lambda: &&R \,\eta_+ +  R_{m} \gamma^{m} \eta_- =0 \, .
\label{eq:susyRQ}
\eea 
%In these expressions, $P$ contains a term involving the cosmological constant
%and another one coming from the singlet in RR fluxes; $P_m$ 
%includes a warp factor
%contribution and RR fluxes in ${\bf 3} + {\bf \bar 3}$ representations; 
%$Q_m$ contains torsion in  ${\bf 3} + {\bf \bar 3}$ representations, 
%i.e. $W_4$ and $W_5$; NSNS flux and RR flux also in  ${\bf 3} + {\bf \bar 3}$ representations, and a derivative of the function $a$;
%$Q_{mn}$
$P,Q$ and $R$ contain contributions coming from the torsion, the NS and RR
fluxes, warp factor $e^{2A}$, cosmological constant $\Lambda$ and 
derivatives of the functions $a$ and $b$ used in the decomposition
(\ref{epsilon}).  In Table \ref{rq} we indicate how the different
representations contribute to $P.Q$ and $R$. We are using that
$\eta_{+}$ is a Clifford(6) vacuum, annihilated by $\gamma^i \eta_+=0$. 
\begin{table}
\begin{center}
%\tbl{Decomposition of supersymmetry equations representations.}
\begin{tabular}{|c|c|c|c|c|c|c|}
\hline
& Torsion& NSNS flux & RR flux & $\Lambda$ & $\partial_m A$ & $\partial_m a$
\\ 
\hline
{\bf 1} & $Q_{\bi j}$ & $Q_{\bi j}, R$ & $P$ & $P$ & -& - \\
{\bf 3} & $Q_{i}$ & $Q_{i}, R_i$ & $P_i,Q_i,R_i$ &- & $P_i$ & $Q_i$ \\
{\bf 6} & $Q_{ij}$ & $Q_{ij}$ & $Q_{ij}$ &- &-  & - \\
{\bf 8} & $Q_{\bi j}$ & $Q_{\bi j}$ & $Q_{\bi j}$ &- & - & - \\ 
\hline 
\end{tabular}
\caption{\label{rq} \text{Decomposition of supersymmetry equations representations.}}
\end{center}
\end{table}

The explicit expressions (for the case
$\Lambda=0$) for these tensors are given in Ref.~\refcite{GMPT}.
Eqs. (\ref{eq:susyRQ}) give a relation between the torsion, fluxes, 
warp factor and cosmological constant in each representation. 
We will skip the details of the derivation done in \refcite{GMPT}
(see also Refs.\refcite{Dallagata,Frey,BeCvIIA,BeCvIIB,Tsimpis})
and quote the results. Tables \ref{ta:IIA} and \ref{ta:IIB},
taken from Ref.~\refcite{GMPT}, give all the possible 
$\N =1$ Minkowski vacua with SU(3) structure for type IIA and type IIB theories
\footnote{$+$ ($-$) in the first columns of Tables \ref{ta:IIA}, \ref{ta:IIB}
 correspond to $a=0$ ($b=0$),  $W_2^{\pm}$ are the real and imaginary 
parts of $W_2$ and all fluxes not written in the Tables are zero.}
(for $AdS_4$ vacua, see for example Refs. \refcite{Tsimpis,BeCvIIA,BeCvIIB}).
\begin{table}
%\tbl{Possible $\N=1$ vacua in IIA.}
\begin{center}
\begin{tabular}{|c||c|c|c||}
\hline 
{\bf IIA}& $a=0$ or $b=0$ (A)& \multicolumn{2}{|c||}{$a=b\, e^{i\beta}
$  (BC)} \\\hline\hline
{\bf 1}&\multicolumn{3}{|c||}{$W_1=H_3^{(1)}=0$}\\ \cline{2-4}
&\minicent3{\vspace{.2cm}$F_0^{(1)} = \mp F_2^{(1)} =F_4^{(1)} = \mp 
F_6^{(1)} $\vspace{.2cm}} &
\multicolumn{2}{|c||}{$F_{2n}^{(1)}=0$}\\\hline
{\bf 8}&  &generic $\beta$ & $\beta=0$\\\cline{3-4}
&$W_2= F_2^{(8)}= F_4^{(8)}=0$&\minicent3{\vspace{.2cm}$W_2^+=e^{\phi}
 F_2^{(8)}$ \\ $W_2^-=0
$\vspace{.2cm}}& 
\minicent5{\vspace{.2cm}$W_2^+=e^{\phi}
 F_2^{(8)}+e^{\phi}F_4^{(8)}$ \\ $W_2^-=
0 \ \ \ \ \ \ \ \ \ $}
% \minicent3{additional\\ $W_2^+=F_4^{(8)}$}
\\\hline
{\bf 6}&$W_3=\mp *_6 H_3^{(6)}$ & \multicolumn{2}{|c||}{$W_3= H_3^{(6)}=0$ }\\\hline
{\bf 3}&\minicent3{\vspace{.2cm} $\bar W_5=2W_4=\mp 2iH_3^{(\bar 3)}=\bar\del\phi$\\
$\bar\del A=\bar\del a=0$\vspace{.2cm}}&\multicolumn{2}{|c||}{
\minicent6{\vspace{.2cm}$F_2^{(\bar 3)}=2i \bar W_5=-2i \bar\partial A =
\frac23 i\bar\partial \phi$, $W_4=0$\vspace{.2cm}}}\\\hline\hline 
\end{tabular} 
\caption{\label{ta:IIA} \text{Possible $\N=1$ vacua in IIA.}}
\end{center}
\end{table}

\begin{table}
\begin{center}
%\tbl{Possible $\N=1$ vacua in IIB.}
\begin{tabular}{|c||c|c|c|c||c||}\hline 
{\bf IIB}& $a=0$ or $b=0$  (A) & \multicolumn{2}{|c|}{$a=\pm ib$  (B)}& 
$a=\pm b $ (C) & (ABC) \\\hline\hline
{\bf 1}&\multicolumn{5}{|c||}{$W_1=F^{(1)}_3=H_3^{(1)}=0$}\\\hline
{\bf 8}&\multicolumn{5}{|c||}{$W_2=0$}\\\hline
{\bf 6}&\minicent{2}{\vspace{.2cm}$F_3^{(6)}=0$\\ $W_3=\pm * H_3^{(6)}$\vspace{.2cm}}&
\multicolumn{2}{|c|}{\minicent{2.5}{$W_3=0$\\ 
$e^{\phi}F_3^{(6)}=\mp * H_3^{(6)}$}}&
\minicent{2.5}{$H_3^{(6)}=0$\\ $W_3= \pm e^{\phi} * F_3^{(6)}$}&
(\ref{eq:6int})\\\hline
{\bf 3}& {\minicent{2.3}{$\bar W_5=2W_4=\mp 2iH_3^{(\bar 3)}= 2 \bar\del\phi$\\
$\bar\del A=\bar\del a=0 $}}
 & 
\multicolumn{2}{|c|}{\minicent{4}{\vspace{.2cm}$ e^\phi F_5^{(\bar 3)}= 
\frac23 i \bar W_5 = i W_4=-2i\bar\partial A =-4i\bar\partial \log a$\\
$\bar \partial \phi =0$ \vspace{.2cm}}} &
{\minicent{3.4}{$\pm e^\phi F_3^{(\bar 3)}=2i \bar W_5=$\\
$- 2i\bar\partial A =- 4i\bar\partial \log a=$\\
$-i \bar \partial \phi$}}&
{(\ref{eq:intvec})}\\\cline{3-4}
&& 
F & \minicent{4}{\vspace{.2cm}$e^\phi F_1^{(\bar 3)}=2 e^\phi F_5^{(\bar 3)}=$\\ 
$i \bar W_5 = i W_4=i\bar\partial  \phi$\vspace{.2cm}}
&&\\\hline\hline
\end{tabular} 
\caption{\label{ta:IIB} \text{Possible $\N=1$ vacua in IIB.}}
\end{center}
\end{table}

The last column in Table \ref{ta:IIB} corresponds to 
intermediate (``ABC'') solutions, satisfying
\begin{eqnarray}
2 ab\, W_3 &=& e^{\phi} \, (a^2+b^2) \, *_6 F^{(6)}_{3} \nn \\
 (a^2 - b^2)\, W_3 &=& - (a^2+b^2) \,*_6H_3^{(6)}
\label{eq:6int}
 \\
 2 ab\, H_3^{(6)} &=& - e^{\phi} \,(a^2 - b^2) \,
F^{(6)}_{3}\nn
\end{eqnarray}

\begin{equation}
  \label{eq:intvec}
\begin{array}{l}
e^{\phi} F^{(\bar 3)}_3 = 
\frac{-4i \, ab (a^2+b^2) }{a^4-2ia^3b+2iab^3+b^4}  \, \bar \del  a
 \, , \\
e^{\phi} F^{(\bar 3)}_5 = 
\frac{-4 \, ab (a^2-b^2) }{a^4-2ia^3b+2iab^3+b^4}  \, \bar \del  a
 \, , \\
H_3^{(\bar 3)} = 
 \frac{-2i  (a^2 + b^2) (a^2-b^2)}{a^4-2ia^3b+2iab^3+b^4}  \, \bar \del a \, ,
\end{array} \quad
\begin{array}{l}
W_4 =  \frac{ 2 (a^2 - b^2)^2}{a^4-2ia^3b+2iab^3+b^4 } \bar \del a \, ,  \\
\bar W_5 =  \frac{2( a^4-4 a^2 b^2 +b^4)}{a^4-2ia^3b+2iab^3+b^4} \bar \del
a \, , \\
\end{array}
\quad
\begin{array}{l}
\bar \partial  A = - \frac{4 (ab)^2}{a^4-2ia^3b+2iab^3+b^4}
\bar \del a \, ,  \\
\bar \del \phi =   \frac{2 (a^2 + b^2)^2 }{a^4-2ia^3b+2iab^3+b^4}
 \bar \del a  \, . 
\end{array}
\end{equation}

$a$ and $b$ are two complex functions, satisfying $|a|^2+|b|^2=e^A$. 
There is also a gauge freedom in their phases: 
rescaling $\eta_+ \rightarrow e^{i \psi} \eta_+$
(or equivalently  $(a,b) \rightarrow  e^{i \psi} \, 
(a,b)$), then $\Ox \rightarrow e^{2i\psi} \Ox$, leaving $J$ invariant.
As a consequence, $(W_1, W_2) \rightarrow e^{2i \psi} (W_1,W_2)$ and 
$W_5 \rightarrow W_5 + 2 i d\psi$. From the tables we can 
see that $W_1$ is always
zero in vacua, while only one of $W_2^{\pm}$ is nonzero in 
some IIA solutions.
Furthermore, the transformation of $a,b$ and 
that of $W_5$ cancel out in the supersymmetry transformations.
This means that the overall phase of $ab$
can be fixed by rotating $W_2$ (table \ref{ta:IIA} is given in a
fixed gauge).  
As a consequence, from the four real parameters in $a,b$,
one is fixed by the normalization condition
and another one 
by the gauge choice and consequently  only two are physical. 
All $\N=1$ vacua with SU(3) structure can therefore be parameterized by 
two angles, as argued in Ref.~\refcite{Frey,GMPT}, in the form
\bea
a&=&e^{A/2}\, cos\,\alpha \, e^{i\tfrac\beta2} \nn \\
b&=&e^{A/2}\, sin\alpha \, e^{-i \tfrac\beta2}
\eea
 These two angles
parameterize a $U(1)_R$ subspace in the $SU(2)_R$ symmetry 
of the $\N=2$ underlying effective theory \cite{GLW}.

Note that in {\bf IIA} there are no intermediate solutions (the solutions 
on the second column of Table \ref{ta:IIA}, for which
the susy parameters are of ``interpolating'' type BC, do not depend on 
the interpolating parameter $\beta$). 
%The angle $\alpha$ in this case is fixed to either $0$, $\pi/2$
%or $\pi/4$. 
\underline{Type A} corresponds to a solution with NS flux
only (plus, in IIA, possible additional RR flux in singlet representations)
which is common to IIA, IIB and the heterotic theory,
found in Ref.~\refcite{Strominger} (see Ref. \refcite{Beck} for 
an extensive analysis). It involves a complex non K\"ahler
manifold ($W_1=W_2=0$, but $W_3 \neq 0$). In the second column,
\underline{Type BC},
the solution has RR flux only, and corresponds to the dimensional reduction
of an M-theory solution on a seven-dimensional manifold 
with $G_2$ holonomy \cite{holomonopole}. The fact that there are no 
intermediate solutions was explained in Ref.~\refcite{GMPT} by looking at 
the eleven dimensional origin of the solutions:  M-theory compactifications 
on seven manifolds with $G_2$ structure group where shown in
Ref.~\refcite{KMT} to forbid fluxes, thus leading to compactifications 
on manifolds of $G_2$ holonomy. Their dimensional reduction
gives the second 
column of Table \ref{ta:IIA}. In order to allow non-trivial 
fluxes, the structure group on
the seven dimensional manifold should be further reduced to $SU(3)$ or subgroups 
thereof. An $SU(3)$ structure in seven dimensions involves a vector
in addition to the fundamental 2-form and holomorphic 3-form
of its six dimensional counterpart
%, and the four-form
%flux has a leg along this direction
. 
If the reduction to 
six dimensions involves a second vector, then the resulting 
structure group of the six-dimensional manifold is
$SU(2)$ rather than $SU(3)$. In order to get $SU(3)$ structure
in six dimensions, the two vectors should coincide. In this case, the 
M-theory four-form flux reduces purely to NS three-form flux
(plus possibly some additional RR flux in singlet representations,
corresponding to M-theory flux along space-time)
giving the first column in Table \ref{ta:IIA}.

In {\bf IIB}, on the contrary, there are solutions with intermediate
values of $\alpha$ and $\beta$.
Types A, B and C are special because these angles are constant.
%,   
%equal to $(0,0)$ (or $(\pi/2,0)$ for the case $a=0$) for type A, 
%$(\pi/4,\pi/4)$ (or $(\pi/4,3\pi/4)$ for $a=-ib$) for type B 
%and $(\pi/4, 0)$ (or $(\pi/4,\pi/2)$ for $a=ib$) for type
%C. 
%In the intermediate cases the angles are not constant. 
\underline{Type A} solution in the first column is the same as
the first column in IIA (setting the RR singlets in the latter 
case to zero), and corresponds to the solution with
NS flux only \cite{Strominger}.  \underline{Type C}, 
 S-dual to type A,  has RR flux only, and 
the same non-vanishing torsion classes as type A. 
\underline{Type B}, on the
other hand, have, besides RR 5-form flux,
RR as well as NSNS 3-form fluxes. They are related by a 
Hodge duality \cite{GP1,Gubser}, usually
expressed in terms of the complex 3-form flux
\beq \label{defG}
G_3= F_3 - i e^{-\phi} H_3 = \hat F_3 - \tau H_3 \
\eeq
($\tau=C_0+i e^{-\phi}$ the complex combination
of axion-dilaton). In Type B solutions, $G_3$ 
is imaginary self dual and has no singlet (0,3) representation
(no flux gets a vev in a singlet representation in IIB, as
 Table \ref{ta:IIB} shows) 
\beq \label{ISD}
* \, G_3 = i \, G_3 \ \quad \text{and} \quad G_{(0,3)}=0 \ .
\eeq
The complex 3-form flux $G_3$, being imaginary self dual
and having no singlet or vector representation 
is therefore (2,1) and primitive with respect to
the complex structure defined by $\Omega$. 
For the solution on the first row in 
the {\bf 3} representation (the one not labeled with an ``F''), 
the six dimensional manifold is a conformal Calabi-Yau
(a manifold whose metric is related to that of a Calabi-Yau
by a conformal factor). This can be seen from the 
torsion classes $W_1=W_2=W_3=0$, $2 W_5= 3 W_4=-6 \, \bar \partial A$
(cf. Table \ref{ta:manifolds}),
which means that the six dimensional metric is
\beq \label{confCY}
 ds^2_6=e^{-2A} \widehat{ds}^2_6 \, \, (\rm{CY}) \ .
\eeq
The conformal factor is therefore the inverse of the warp factor,
and it is related to the RR 5-form flux. This 
class of $\N=1$ solutions is the most ``popular'' one, as
it involves a Calabi-Yau manifold, whose mathematical
properties are very well known. For most of this review, we will
concentrate on this very well explored class of solutions. 
They are dual to M-theory solutions on Calabi-Yau 4-folds found
by Becker and Becker \cite{BB}. 
Finally, there is an F-theory-like type B solution (labeled
by an ``F''), that involves 
imaginary self-dual 3-form flux, and additionally a 
non constant holomorphic axion-dilaton $\tau=\tau(z)$.
The six-dimensional manifold is still complex, but no longer 
conformal Calabi-Yau, as the torsions $W_4$ and $W_5$
are equal, rather than having a ration of $2/3$. 

Note that for the three types, A,B and C,
there is always a complex flux-torsion combination that is
(2,1) and primitive: 
% *H^{(6)}=-iH^{(6)} (6 is (1,2))
\bea
&\rm{Type A}&  : \ \, \ \ \ \ \ \  \ dJ \pm i\, H_3 \qquad \,  \text{is (2,1) and primitive}
% +\rightarrow a=0, - \rightarrow b=0 
\nn \\
&\rm{Type B}&  : \ \ \ \ \ \ \ \ \, F_3 \mp i e^{-\phi} \, H_3 \ \  \text{is (2,1) and primitive}
%, \pm \rightarrow a=\pm ib 
\nn \\
&\rm{Type C}&  :  \ d(e^{-\phi} J) \pm i\, F_3 \qquad  \ \, \text{is (2,1) and primitive}
% \pm \rightarrow a=\pm b
\eea
where the $\pm$ correspond to the two possible relations between $a$ and 
$b$ in Table \ref{ta:IIB}. \\

Dp-branes and O-planes preserve supersymmetries
such that $\epsilon^1=\gamma^{\perp} \epsilon^2$,
where ``$\perp$'' stands for directions perpendicular
to the D-brane or O-plane (see for example \refcite{JoeBB}). 
For D3-branes
and O3-planes, $\gamma^{\perp}$ is the product
of six gammas in Euclidean space, which has eigenvalues $\pm i$.
This means that D3-branes and O3-planes preserve supersymmetries
of the type $a=\pm i b$, which is of type B (the plus (minus)
corresponds to D3 (anti-D3)).
(Note also that when $a=\pm ib$, the complex spinor 
$\epsilon^1+ i \epsilon^2$ has a definite 
positive (negative) four-dimensional chirality.)
The antisymmetric product of two gamma matrices 
has also $\pm i$ eigenvalues, and
consequently D7-branes and O7-planes preserve type B supersymmetries.
The product of four gamma matrices have eigenvalues $\pm 1$, and therefore
the supersymmetries preserved by 
D5/D9-branes and O5/O9-planes is of type C.
branes wrapped on collapsed two cycles have the supersymmetries
of the lower dimensional brane, i.e. for example
D5-branes wrapped on the collapsed
2-cycles of the conifold has type B supersymmetries \cite{GKP}.
D3/D5 or D3/NS5 bound states have instead intermediate 
BC or AB supersymmetries \cite{interpolating}. 
In type IIA, O6 planes wrapped on Special
Lagrangian cycles preserve
type BC supersymmetries, where the
phase $\beta$ corresponds to the overall phase of
$\Omega$, which is a gauge choice, as argued below
Eq.(\ref{eq:intvec}).

The explicit IIB solutions worked out so far are mostly in 
the classes A,B or C. Starting from non compact cases,
the prominent ones that have a holographic dual interpretation
are Maldacena-Nu\~nez (MN) \cite{MN}, Klebanov-Strassler (KS) \cite{KS}
and Polchinski-Strassler (PS) solutions \cite{PoSt}.
MN, corresponding to NS5-branes wrapped on 2-cycles, 
is a non compact regular type A
$\N=1$ background. It's S-dual version,
constructed also in Ref.~\refcite{MN}, is a type C solution. 
KS is a regular non compact type B solution,
where the Calabi-Yau
in question is the conifold. This solution can be
``compactified''  by adding orientifold 3-planes \cite{GKP}, 
in the sense that it can 
be used as a local IR throat geometry of a 
compact Calabi-Yau,
as we will review in sections \ref{sec:nogo}, \ref{sec:moduliconifold}.
PS solution, corresponding to D5-branes or NS5-branes wrapped on finite
2-cycles with induced D3-charge does not have any of the above supersymmetries,
neither an interpolating type. Despite the exact solution is not
known yet (PS is constructed perturbatively), it is expected
not to have SU(3) structure, but a more reduced one. So do the
flow solutions of Ref. \refcite{flow}, which have SU(2) structure,
and where obtained using the algebraic killing spinor technique \cite{killspi}.

In the following, we give the reference to some of the type II
solutions discussed in the literature.

Type B backgrounds of IIB involving Calabi-Yau
hypersurfaces in weighted projective spaces  
were constructed for example in 
Refs.~\refcite{GKTT,GKT,DWRec}.
 There are many compact type B solutions involving
manifolds with a smaller structure group 
than $SU(3)$,  but
which still have supersymmetries obeying 
the type B condition, $a=ib$. In these cases,
 there is more than one well defined spinor, out
of which a subset can 
be preserved, leading to $\N=1$ up to $\N=4$ supersymmetries
in four dimensions. The latter, $\N=4$, is a solution with
5-form but no 3-form flux, like 
$AdS_5 \times S^5$. 
\footnote{From the $M_4 \times_w M_6$ point of view, 
$AdS_5 \times S^5$ solution corresponds to a conformally flat six dimensional manifold, 
with conformal factor $e^{-2A}$ and warp factor $e^{2A}= R^2/r^2$, 
where $r$ is a radial coordinate
in the six-dimensional space, and $R$ is proportional to the number
of units of 5-form flux.} 
Turning on 3-form fluxes, type B supersymmetries are
preserved if the complex 3-form flux is (2,1) and primitive with respect
to any of the complex structures defined
by the preserved spinors. A solution with $\N=3$ on a six-torus
(which has trivial structure group) 
was constructed in this fashion in Ref.~\refcite{FP}. 
$\N=2$ (and $\N=1$) solutions 
on orientifolds of $K^3 \times T^2$ 
(with structure group $SU(2)$), were studied
for example in Refs.~\refcite{DRS,TriTri}.
One of the first
constructions of $\N=1$ solutions on orientifolds of tori,
which will be discussed in detail in section \ref{sec:modtoriIIB}, 
is Ref.~\refcite{KST}.
Solutions with various $\N$ on tori, and in particular the 
possibility of connecting them by spherical domain walls
is discussed in Ref.~\refcite{Kachrububbles}.  
$\N=1$ type B flux solutions relevant to particle phenomenology, i.e.
containing chiral matter arising at the intersections of 
magnetized D-branes in representations close to
the Standard Model were constructed for example
in Ref.~\refcite{CUtorus,CvLi}
using tori (see also Ref.\refcite{Tdual1} for
their construction from type I), and in Ref.~\refcite{CU} involving conifolds.

Most of the remaining known compact solutions
(type C in IIB, with O5-planes, and type BC
in IIA, involving O6-planes) were
 obtained by 
T-dualizing type B solutions of IIB, and involve manifolds
of trivial structure. 
Refs.\refcite{KSTT,Schulz} for example, constructed 
IIB type C solutions on twisted tori 
(see \refcite{HullEdwards}), 
starting from  type B solutions on orientifolded (by O3's)
tori. The twisted tori in question are complex, non
K\"ahler (they have $W_3=e^{\phi} *F_3 \neq 0$).
IIA type BC solutions (supersymmetric
and non supersymmetric) on twisted tori
were also constructed in Refs. \refcite{DKPZ,IFC}
(the latter having some relevance to particle phenomenology) 
by minimizing the flux induced potential. 
Ref. \refcite{IFC} showed that the supersymmetric Minkowski vacua 
have $W_2=e^{\phi}F_2$
(implying that the internal manifold is not complex),
while AdS vacua have additionally $W_1 \neq 0$.
More references for compact solutions will be given 
in section \ref{sec:moduli}, when
discussing moduli stabilization

As for (non compact) solutions with intermediate supersymmetries, 
those corresponding to bound states of D3/D5 branes 
in flat space were 
obtained in Ref.~\refcite{BMM,CP} by T-dualizing D4-brane solutions of IIA. 
Ref.~\refcite{BGMPZ} found a one parameter family of regular
IIB solutions interpolating between Maldacena-Nu\~nez
(type C) and Klebanov-Strassler (type B), using the interpolating ansatz
for the metric and fluxes of Ref. \refcite{PaTs}.

Finally, let us comment that 
some of these flux solutions can be related by duality to 
flux-less solutions, as
nicely shown in Refs.~\refcite{Schulz2,Aspin}, and also 
to non geometric backgrounds \cite{nongeo}, as discussed in 
Ref. \refcite{nongeodual}.

\subsection{Internal manifold and generalized complex geometry}
\label{sec:GCG}

In this section we discuss a unifying mathematical
description of all internal manifolds arising in supersymmetric flux
backgrounds. This description involves generalized complex 
geometry \cite{GCY,Gualtieri}. Readers interested in 
flux compactifications on Ricci-flat  
manifolds like Calabi-Yau or tori can 
skip the following two subsections,
and go directly to section \ref{sec:nogo}. 
Sections \ref{sec:nogo} and on involve mostly
Ricci-flat manifolds, except some remarks
made in section \ref{sec:4D}, more precisely
at the end of  sections \ref{sec:CYspectrum}, \ref{sec:fluxpot},
\ref{sec:fluxsup} and in most of section \ref{sec:mirror}. \\

Looking back at Table \ref{ta:IIB} we can see that for all IIB vacua
with SU(3) structure, the internal manifold is complex 
(see Table \ref{ta:manifolds}).
For IIA, on the other hand, there is solution involving
a complex manifold, the one on the 
first column (type A), while all other solutions are 
symplectic. A single 
differential geometric description 
of the allowed internal manifolds 
once the back-reaction to the fluxes 
is taken to account should therefore 
unify complex and symplectic geometry.  
Amazingly enough, there is such a description: 
both are generalized complex manifolds
in generalized complex geometry. Generalized complex geometry,
proposed by Hitchin \cite{GCY} and developed in detail by his 
student Gualtieri \cite{Gualtieri}, was born out of the idea 
of adding the B-field to differential geometry. One of
the first outcomes is that complex and symplectic
manifolds are special cases of generalized complex manifolds,
which means that generalized complex geometry not only 
contains complex and symplectic geometry, but it also extends
it. In this section we review very briefly the basic ideas
that are useful in the context of flux compactifications.
For more detail, we refer the reader to the original references
Ref.~\refcite{GCY,Gualtieri}. Generalized complex geometry
has been used in the context of flux compactifications
from the space-time point of view in 
Refs.~\refcite{GMPT,GCYproceedings,Witt,JW,GCY2,GLW}; 
from the world-sheet perspective
in Refs.~\refcite{LMTZ}; in topological strings, 
D-branes and mirror symmetry 
\cite{KapustinLi,FMT,BB1};
most of these papers contain some introduction to generalized complex geometry.\\

Usual complex geometry deals with the tangent bundle of a manifold $T$,
whose sections are vectors $X$, and separately, with the cotangent bundle
$T^*$, whose sections are 1-forms $\zeta$.  
In generalized complex geometry the tangent
and cotangent bundle are joined as a single
bundle, $T \oplus T^*$.
Its sections
are the sum of a vector field plus a one-form $X + \zeta$. 
The  standard machinery of complex geometry can be generalized to
this bundle.
On this even-dimensional bundle, one can construct a 
generalized almost complex structure 
${\mathcal J}$, which is a map of $T \oplus  T^*$ to itself that 
squares to $-\Bbb I_{2d}$ ($d$ is real the dimension of
the manifold). This is analogous to an almost complex structure\footnote{In this 
subsection we denote by $I$ the almost complex structure
on $T$, to avoid confusion with the fundamental form $J$. In the rest of 
the paper we use $J$ for both, but when referring to the almost complex structure,
like for example in Eq.(\ref{Nijenhuis}), we write the indices explicitly.}
 $I_m\,^n$
which is a bundle map from $T$
to itself that squares to $-\Bbb I_d$.
As for an almost complex structure, ${\mathcal J}$ must also satisfy the 
hermiticity condition 
${\mathcal J}^{t} {\mathcal G} {\mathcal J} = {\mathcal G}$, 
with the respect to the natural metric on $T \oplus  T^*$, 
${\mathcal G}={{0 \ \ 1} \choose {1 \ \ 0}}$. 

Usual complex structures $I$ are
naturally embedded into generalized ones ${\mathcal J}$: 
take ${\mathcal J}$ to be
\beq \label{gcscom}
{\mathcal J}_1 \equiv 
\left(
\begin{array}{cc}
I & 0 \\ 0 & -I^t 
\end{array}
\right) \, ,
\eeq
with $I_m\,^{n}$ a regular almost complex structure (i.e. $I^2=-\Bbb I_d$).
This  ${\mathcal J}$ satisfies the desired properties, namely 
 ${\mathcal J}^2=-\Bbb I_{2d}$, 
${\mathcal J}^{t} {\mathcal G} {\mathcal J} = {\mathcal G}$.
Another example of generalized almost complex structure can be built
using a non degenerate 
two--form $J_{mn}$, 
\beq \label{gcssym}
{\mathcal J}_2\equiv
\left(
\begin{array}{cc}
0 & -J^{-1} \\ J & 0 
\end{array}
\right) \, .
\eeq

Given an almost complex structure $I_m\,^n$,
one can build holomorphic and antiholomorphic 
projectors $\pi_{\pm}=\frac12 (\Bbb I_{d}\pm i I)$. 
Correspondingly, projectors can be  
build out of a generalized almost complex structure, 
$\Pi_{\pm}=\frac12 (\Bbb I_{2d}\pm i {\mathcal J})$.
There is an
integrability condition for generalized almost complex
structures, analogous to the
integrability condition for usual almost complex structures.
For the usual complex structures, 
integrability, namely the vanishing of the 
Nijenhuis tensor, 
can be written as the condition $\pi_{\mp} [\pi_{\pm} X,\pi_{\pm}Y]=0$, i.e. the
Lie bracket of two holomorphic vectors should again be holomorphic.  
For generalized almost complex structures, integrability condition reads
exactly the same, with $\pi$ and $X$ replaced respectively
by $\Pi$ and $X+\zeta$, and the Lie bracket
replaced by  
the Courant bracket\footnote{The Courant
bracket is defined as follows: $[X+ \zeta , Y + \eta]_C=[X,Y] + {\mathcal L}_X \eta 
- {\mathcal L}_Y \zeta - \frac12 d(\iota_X \eta - \iota_Y \zeta)$.} 
on $T \oplus T^*$. 
The Courant bracket does not satisfy Jacobi identity in
general, but it does on the $i$--eigenspaces of ${\mathcal J}$.
In case
these conditions are fulfilled, we can drop the ``almost'' and speak of
generalized complex structures. 

For the two examples of generalized almost complex structure
given above, ${\mathcal J}_1$ and ${\mathcal J}_2$, integrability condition
turns into a condition on their building blocks, $I_m\,^n$ and $J_{mn}$. 
Integrability of ${\mathcal J}_1$
enforces $I$ to be an integrable almost complex structure
on $T$, and hence $I$ is a complex structure, or equivalently
the manifold is complex.
For ${\mathcal J}_2$, which was built from a two-form $J_{mn}$, integrability
imposes $dJ=0$, thus making $J$ into a symplectic form, and the manifold
a symplectic one.

These two examples are not exhaustive, and the most general
generalized complex structure is partially complex, partially
symplectic. Explicitly, a generalized 
complex manifold is locally equivalent to the product
$\Bbb C^k \times (\Bbb R^{d-2k},J)$, 
where $J=dx^{2k+1} \wedge dx^{2k+2} + ... + dx^{d-1} \wedge dx^d$ is
the standard symplectic structure and $k\le d/2$ is called rank, which 
can be constant or even vary over the manifold (jump by two at certain
special points or planes).  

There is an algebraic one to one
correspondence between generalized almost complex
structures and  pure spinors of Clifford(6,6). In string theory,
the picture of generalized almost complex structures emerges naturally from
the world--sheet point of view, while that of pure
spinors arises on the space-time side. 

Spinors on $T$ transform under Clifford(6), whose algebra is $\{\gamma^m,\gamma^n\}=2 g^{mn}$.
There is a representation of this algebra in terms of forms. 
We can take $\gamma^m=dx^m \wedge + g^{mn} \iota_n$.
\footnote{\label{iota} $\iota_{n}$: $\Lambda^p T^* \rightarrow
\Lambda^{p-1} T^*$, $\iota_{n} dx^{i_1} \wedge ... \wedge dx^{i_p} = p \delta^{[i_1}_n dx^{i_2} \wedge ... \wedge dx^{i_p]}$.} These satisfy 
the Clifford(d) algebra.
The algebra
of  Clifford(d,d) is instead
$$
\{ \Gamma^m, \Gamma^n\} =0\ , \qquad 
\{ \Gamma^m, \Gamma_n\} = \delta^m_n \ , \qquad
\{ \Gamma_m, \Gamma_n \}=0\ .
$$
$\Gamma^m$ and $\Gamma_m$ are independent, they cannot be obtained from 
one another by raising or lowering
indices with the metric.  
There is also a representation of this algebra in terms of forms, 
namely
\beq \label{66gammas}
\Gamma^m= dx^m \wedge \ , \qquad 
\Gamma_n = \iota_{n} \ . 
\eeq
The holomorphic 3-form $\Omega$ is a good vacuum
of Clifford(6,6), as it is annihilated by the
holomorphic $\Gamma^i$ and the antiholomorphic $\Gamma_{\bi}$. These are
half of the total gamma matrices, which implies that $\Ox$ is  
a {\it pure} \clss\ spinor. 
Acting with the other half,  $\Gamma^{\bi}$ and $\Gamma_{i}$
we get forms of all possible degrees.
Clifford(6,6) spinors are therefore equivalent to (p,q)-forms. 

Using the Clifford map, a \clss\ spinor can 
also be mapped to a 
bispinor \cite{GMPT,Witt}:
\begin{equation}
  \label{eq:clmap}
C\equiv\sum_k \frac{1}{k!}C^{(k)}_{i_1\ldots i_k} dx^{i_i}\wedge\ldots\wedge dx^{i_k}\qquad
\longleftrightarrow\qquad
\sla C \equiv
\sum_k \frac{1}{k!}C^{(k)}_{i_1\ldots i_k} \gamma^{i_i\ldots i_k}_{\alpha\beta} \ .
\end{equation}

On a space of SU(3) structure, there is a nowhere
vanishing SU(3) invariant 
Clifford(6) spinor $\eta$.
Out of it, we can construct
two nowhere vanishing SU(3,3) invariant 
bispinors by tensoring  $\eta$ with its dagger, namely \cite{GMPT,JW}
\beq
  \label{eq:genpure}
 \Phi_+ = \eta_+\otimes \eta_{+}^{\dagger}\ , \qquad 
 \Phi_- =  \eta_+\otimes \eta_{-}^{ \dagger}\ .
\end{equation}
(and its complex conjugates).
Using Fierz identities, this tensor product 
can be written in terms of the bilinears in Eq.(\ref{bilinears}) by
\beq
\eta_{+} \otimes \eta^{\dagger}_{\pm} = 
\frac{1}{4} \sum_{k=0}^6 \frac{1}{k!}  
\eta^{\dagger}_{\pm} \gamma_{i_1 ... i_k} \eta_{+} 
\gamma^{i_k ... i_1} 
\label{eq:fz}
\eeq  
Using the Clifford map (\ref{eq:clmap}) backwards, 
the tensor products in (\ref{eq:genpure}) are 
identified with regular forms. From now on, we
will use $\Phi_{\pm}$ to denote just the forms.

The subindices
plus and minus in $\Phi_{\pm}$ denote the 
Spin(6,6) chirality: positive corresponds
to an even form, and negative to an odd form.
Irreducible Spin(6,6) representations  are 
actually ``Majorana-Weyl'', namely they
are of definite parity --''Weyl''--
and real --''Majorana''--.  

The explicit expression
for the \clss spinors in (\ref{eq:genpure}) in
terms of the defining forms for the SU(3) structure
is
\beq \label{Fierz} 
\Phi_+=\eta_{+} \otimes \eta^{\dagger}_{+} = \frac{1}{8} e^{- i J}
\ , \qquad
\Phi_-=\eta_{+} \otimes \eta^{\dagger}_{-} = -\frac{i}{8} {\Omega} \ .
\eeq 

The forms in (\ref{eq:genpure}, \ref{Fierz}) are pure. This can be seen from
writing the usual gamma matrices acting on the left
of $\Phi$ (denoted as $ \buildrel\to\over{\gamma^m}$)  and on the right 
(denoted as $ \buildrel\leftarrow\over{\gamma^m}$) in terms
of the \clss\ gamma matrices (\ref{66gammas})
\begin{equation}
  \label{eq:gammamap}
 \buildrel\to\over{\gamma^m} = \frac12(dx^m \wedge +g^{mn}\iota_n)\ , \qquad
  \buildrel\leftarrow\over{\gamma^m} = \frac12(dx^m \wedge \pm g^{mn}\iota_n)\ \, ,
\end{equation}
where the $\pm$ sign depends on the parity of the spinor on which 
$ \buildrel\leftarrow\over{\gamma^m}$ acts.
We can check now that the forms (\ref{eq:genpure}) are indeed pure: 
the six gamma matrices that annihilate them are
\begin{equation}
  \label{eq:genann}
(\delta +iI)_m{}^n \gamma_n \,  \eta_+\otimes \eta_{\pm}^{\dagger}=0 \ ,
\qquad  
   \eta_+\otimes \eta_{\pm}^{\dagger}\,\gamma_n 
(\delta \mp iI)_m{}^n =0 \ .
\end{equation}
where $I$ is the almost complex structure on the tangent
bundle.

On a space of SU(3) structure on $T$, there are therefore
two SU(3,3) invariant pure forms (and their complex conjugates), 
$e^{i J}$ and $\Omega$. This implies that 
the structure group on \tts, which is generically SO(d,d), 
is reduced in this case to \stt\ \cite{Gualtieri,Witt} 
\footnote{\label{foot:compatibility}Two SU(3,3) invariant pure
spinors reduce the structure on \tts\ to \stt\ if they are compatible, namely
if they have three annihilators in common. Spinors of the form (\ref{eq:genpure})
are always compatible, as they have in common the 
three annihilators on the left of (\ref{eq:genann}).}.  

There is a one to one correspondence between a 
a pure spinor $\Phi$ and a generalized almost 
complex structure ${\mathcal J}$. It maps the $+i$ eigenspace of
${\mathcal J}$ to the annihilator space of the spinor $\Phi$. 
Under this correspondence
\bea \label{1-1}
\Phi_-=-\frac{i}{8}\, \Ox  \ & \leftrightarrow & \mathcal J_1 \nn \\
\Phi_+=\frac{1}{8} \, e^{-iJ} & \leftrightarrow & \mathcal J_2
\eea
where ${\mathcal J}_1$ and ${\mathcal J}_2$ are defined in 
(\ref{gcscom}, \ref{gcssym}).

Integrability condition for the generalized complex structure corresponds
on the pure spinor side to the condition 
\begin{center}
${\mathcal J}$ is integrable $\Leftrightarrow $ $\exists $ vector $v$ and 1-form $\zeta$ such that $d\Phi = (v \llcorner + \zeta \wedge) \Phi$ 
\end{center} 
A generalized Calabi-Yau \cite{GCY} 
is a manifold on which a closed pure spinor exists:
\begin{center}
Generalized Calabi-Yau $\Leftrightarrow $ $\exists $  $\Phi$ pure 
such that $d\Phi = 0$ 
\end{center}
From the previous property, a generalized Calabi-Yau has obviously an 
integrable complex structure. Examples of Generalized Calabi-Yau
manifolds are symplectic manifolds and complex 
manifolds with trivial torsion class $W_5$ (i.e., if
$W_1=W_2=0$, and $\bar W_5=\bar \partial f$ -cf.(\ref{dJdOmega})-
then $\Phi= e^{-f} \Omega$ is closed).

There is also the possibility of twisting by a closed three--form $H$.
Using a three--form, the Courant bracket can be modified\footnote{$[X+\zeta, Y+\eta]_H=[X+\zeta,Y+\eta]_C + \iota_X \iota_Y H$.}, and with it 
the integrability
condition. In terms of ``integrability'' of the pure spinors $\Phi$,
adding $H$ amounts to twisting the differential condititions 
for integrability and for Generalized Calabi-Yau. More precisely,
\begin{center} 
``Twisted'' Generalized Calabi-Yau $\Leftrightarrow $ $\exists $  $\Phi$ pure,
and $H$ closed  
such that $(d - H \wedge) \Phi = 0$ 
\end{center}
Decomposing $\Phi$ in forms, $\Phi=\sum_k \varphi_{k}$, the
twisted Generalized Calabi-Yau condition implies 
$d \varphi_{k} - H_3 \wedge \varphi_{k-2}= 0$
for every $k$. Note that this twisted exterior derivative 
appeared already in the definition of the modified
RR fields, Eq.(\ref{fl}) and in their Bianchi identities
$(d-H\wedge) F^{(10)}=0$.

Before relating this discussion to $\N=1$ flux vacua, let
us say that for very little price, one can
describe manifolds with SU(2) structure using the
same formalism. SU(2) structures in six dimensions
are defined by two nowhere vanishing
spinors $\eta^{1}, \eta^{2}$ that are never parallel.
A bilinear  of the form (\ref{bilinears}) with one gamma matrix
made out of them,
defines a complex vector, namely
\beq
\eta^{1\, \dagger}_+ \gamma^m \eta^2_-=v^m -i v'^m \ .
\eeq
Therefore, differently from
the SU(3) case, on manifolds with SU(2)
structure there is a nowhere vanishing vector 
\footnote{This is possible only in manifolds of vanishing Euler caracteristic, 
$\chi=0$.}.
It is possble to describe SU(3) and SU(2) structures
on the same footing. For that, we define
\cite{JW}
\beq
\eta^2_+= c \, \eta^1_+ + (v+iv')_m \gamma^m \eta^1_- \ ,
\eeq 
where $c$ is a function of the internal manifold, and
we let the norm of the vector $v+iv'$ vary
over the manifold. The SU(3) structure case
corresponds to $|v(p)+iv'(p)|=0, \forall p \in M$
(and therefore $c=1$ to keep $\eta^2$ normalized), while
in the SU(2) case $|v(p)+iv'(p)| \neq 0,  \forall p \in M$.
The intermediate cases, where the norm
vanishes at some points on the manifold, is better described 
as an \stt\ structure on \tts\ \cite{JW}, as we will explain shortly.

Using $\eta^1$ and $\eta^2$, we can build the pure
spinors (\ref{eq:genpure}), where the one daggered
is, say, $\eta^2$. Their explcit form is \cite{JW,GCY2}
\bea
  \label{eq:genpureforms}
\Phi_+&=&\eta^1_+ \otimes \eta^{2\, \dagger}_{+}
=\frac18(\bar c \, e^{-i\,j}-i  \omega)\wedge
e^{-iv\wedge v'}\ , \nn \\
\Phi_-&=&\eta^1_+ \otimes \eta^{2\, \dagger}_{-}=- \frac18( e^{-i\,j}+i c \, \omega)\wedge
(v+iv')\ .
\eea
These are given in terms
of the local SU(2) structure defined by $(\eta^1,\eta^2)$:
$j$ and $\omega$  are the (1,1) and (2,0)-forms 
on the local four dimensional space orthogonal to
$v$ and $v'$. The SU(3) structure defined
by $\eta^1$ is given by $J=j + v \wedge v'$,
$\Omega=w \wedge (v + i v')$. 

Very much like in (\ref{eq:genann}) one can show that $\Phi_{\pm}$
of (\ref{eq:genpureforms}) are pure: just replace $I$
by $I_1$ ($I_2$) in the eq. on the left (right) of (\ref{eq:genann}),
where $I_{1}$ ($I_2$) is the almost complex structures defined by
$\eta_{1}$ ($\eta_2$). One can similarly show that they
are compatible (see footnote \ref{foot:compatibility}).
Therefore, the existence of $\Phi_{\pm}$ implies that the
structure group on \tts\ is \stt. SU(3) and SU(2)
structures on $T$ are just particular cases of
\stt\ structures on \tts.

\subsection{$\N=1$ flux vacua as Generalized Calabi-Yau manifolds}
\label{sec:GCY}

Following Refs.\refcite{GCY,GCY2}, we show in this section that 
the internal manifold for
all the $\N=1$ vacua shown in Tables \ref{ta:IIA} and \ref{ta:IIB} 
(and also vacua with SU(2) structure, or more generally with 
\stt\ structure on \tts) are Generalized Calabi-Yau's. \\

As we reviewed in the previous sections, 
 as a result of
demanding $\delta \Psi_m=\delta \lambda =0$,
supersymmetry imposes
differential conditions on the internal spinor $\eta$. 
These differential conditions
turn into differential  conditions for the pure \clss\ spinors
$\Phi_{\pm}$, defined in (\ref{Fierz},\ref{eq:genpureforms}). 
We quote the results of  Ref. \refcite{GCY2}, skipping the 
technical details of the derivation.
$\N=1$ supersymmetry requires
\bea
\label{a}
e^{-2A+\phi}(d+H\wedge) (e^{2A-\phi} \tilde \Phi_+)&=&
0 \, ,
\\
e^{-2A+\phi}(d+H\wedge) (e^{2A-\phi}\Phit_-)&=&
 dA \wedge \bar\Phit_- 
-\frac 1{16}e^{\phi}\Big[ ( |a|^2 - |b|^2) F_{\mathrm{IIA}\,-}
 -i (|a|^2 + |b|^2)*F_{\mathrm{IIA}\,+}\Big]    \nn
\eea  
for type IIA, and
\bea
\label{b}
e^{-2A+\phi}(d-H\wedge) (e^{2A-\phi}\Phit_+)&=&
dA\wedge\bar\Phit_+ 
+\frac1{16} e^\phi
\Big[ (|a|^2 - |b|^2) F_{\mathrm{IIB}\,+} -i (|a|^2 + |b|^2)*F_{\mathrm{IIB}\,-}
\Big]
 \nonumber \\
e^{-2A+\phi}(d-H\wedge) (e^{2A-\phi}\Phit_-)&=&
0  \, ,
\eea 
for type IIB. 

In these equations 
\begin{equation}
  \label{eq:F}
  F_{\mathrm{IIA}\,\pm}=F_0\pm F_2+F_4\pm F_6\ , \qquad 
F_{\mathrm{IIB}\, \pm}=F_1 \pm F_3+F_5\ ,
\end{equation}
and $\tilde \Phi_{\pm}$ are unnormalized
\clss\ pure spinors. They
are constructed  as in (\ref{eq:genpureforms}), but out
of unnormalized spinors $\tilde \eta^{1,2}$ defined by
\beq
\tilde \eta^1_+ = a \, \eta^1_+ \ , \qquad  \tilde \eta^2_+ = b\, \eta^2_+ \ .
\eeq
These are the internal spinors that build the 
$\N=1$ supersymmetry parameter (cf. Eq. (\ref{epsilon}) and Tables
\ref{ta:IIA}, \ref{ta:IIB}, which are specialized to the
case $\eta^1=\eta^2$). 
$\tilde \Phi_{\pm}$ are therefore related to 
$\Phi_{\pm}$ in (\ref{Fierz}) or (\ref{eq:genpureforms})
by
\beq \label{Phit}
\Phit_+ = a \bar b \,\Phi_+ \ , \qquad \Phit_- = a  b \, \Phi_-
\eeq
$\N=1$ supersymmetry imposes the following relation between these norms
\begin{equation}
  \label{eq:norm}
  d |a|^2 = |b|^2 d A\ , \qquad d|b|^2 = |a|^2 dA\ , 
\end{equation}
for both IIA and IIB.
%inserting \ref{eq:intvec} we get $sin \alpha=\frac{K-e^{-2A}}{2k}$
%where k is some integration constant...?
According to the definitions given in the previous section, 
Eqs. (\ref{a}) and (\ref{b}) tell us that all $\N=1$ vacua
on manifolds with \stt\ structure on \tts 
(which includes the case of SU(3) and SU(2) structures on $T$) are 
{\it twisted Generalized Calabi-Yau's}. We can also see
from (\ref{a}), (\ref{b}) that
RR fluxes act  
an obstruction for integrability of the second pure spinor.

Specializing to the pure SU(3) structure case, i.e. for $\Phi_{\pm}$ given 
in Eq.(\ref{Fierz}), and looking at (\ref{1-1}), we see that the 
Generalized Calabi-Yau manifold
is complex\footnote{$H$ in Eq.(\ref{b}) does not ``twist''
the (usual) complex structure, as $(d-H \wedge) \Omega =0$ implies
in particular $d\Omega=0$.} in IIB and (twisted) symplectic in IIA.
For the general \stt\ case, $\N=1$ vacua can be realized
in hybrid complex--symplectic manifolds, i.e. manifolds
with $k$ complex dimensions and $6-2k$ (real) symplectic ones.
In particular, given the criralities of the preserved
\clss\ spinors, the rank  $k$ must be even in IIA 
 and odd in IIB (equal respectively to 0 and 3 in
the pure SU(3) case).  

One very important comment is in order: in the IIA and IIB type A solutions 
of Tables \ref{ta:IIA}-\ref{ta:IIB}, either $a$ or $b$ vanishes.
This implies, via Eq.(\ref{Phit}), that $\Phit_{\pm}=0$.
For this case, Eqs.(\ref{a},\ref{b}) just impose $F=0$, which is indeed
the case in type A solutions, but they do not say anything about
the integrability properties of the associated generalized almost
complex structures. We know that for type A the internal manifold
is complex \cite{Strominger}. 
However, the relation $W_3=*H_3$, satisfied
in type A solutions, say that the manifold {\it is not} twisted
symplectic (which needs $W_3=H_3$). Therefore, we should be  
careful when saying that IIA solutions in manifolds of SU(3)
structure are twisted symplectic: this is not true for
type A, which is a valid IIA solution. The more general statement
that all $\N=1$ vacua are twisted Generalized \cy's is nevertheless true,
as type A involves a complex manifold, which is a particular case 
of Generalized  
\cy.

A final comment before we move on to discuss the problems related to
compactification, is that $\N=2$ vacua with NS fluxes only,
where shown to satisfy Eqs. \ref{a}, \ref{b} 
for $\Phit_\pm=\Phi_\pm$, $F_{\rm{IIA}}=F_{\rm{IIB}}=A=0$  \cite{JW}.

\section{No-go theorems for compact solutions with fluxes} \label{sec:nogo} 

The integrated equations of motion (or integrability conditions, of the type
(\ref{integra})) yield restrictive no-go theorems that under quite
general conditions rule out warped compactifications to Minkowski
or de Sitter spaces in the
presence of fluxes \cite{MNnogo,IvPa,GKP,London1,London2}.
We will review first the general argument given in Ref.~\refcite{MNnogo},
and then discuss the no-go's from Bianchi identities, and the need to
introduce localized sources. 

\subsection{Four dimensional Einstein equation and no-go }

Ref.~\refcite{MNnogo}  
showed that for any solution, supersymmetric or not,
the flux contribution to the 
energy momentum tensor is always positive, ruling out compact internal spaces
when these are turned on.
This can be seen from the trace reversed Einstein's equation in ten dimensions
\beq \label{Einstein}
R_{MN}=T_{MN} - \frac{1}{8}g_{MN} T_L^L \, .
\eeq
For the metric (\ref{metric}), the four-dimensional components of Einstein's
equation imply
\beq
R_{\mu \nu}=\tilde R_{\mu \nu} - \tilde g_{\mu\nu} 
\left(\nabla^2 A +2 (\nabla A)^2 \right)=T_{\mu \nu} - \frac{1}{8} e^{2A} 
\tilde{g}_{\mu \nu} T_L^{\,L} \, .
\eeq
Contracting with $\tilde g^{\mu\nu}$ on both sides we find
\beq \label{nogo}
\tilde R + e^{2A} (-T_{\mu}^{\,{\mu}} + \frac{1}{2} T_L^{\,L} )=
4 \left(\nabla^2 A +2 (\nabla A)^2 \right)=2 e^{-2A} \nabla^2 e^{2A}
\eeq
where the contractions on the stress energy tensor are
done with the full ten-dimensional metric. For Minkowski and de Sitter
compactifications, $\tilde R \ge 0$. Defining
\beq
\hat T= -T_{\mu}^{\,{\mu}} + \frac{1}{2} T_L^{\,L}=\frac{1}{2} 
(-T_{\mu}^{\,{\mu}} + T_m^{\,m}) \,
\eeq   
%\cite{MN} show that if there is a (non flux) potential under the assumption
%that it is non-positive, i.e. $V \le 0$ (which is true for type II supergravity
% potential,
%except in massive IIA supergravity {\bf yes?}, for which \cite{MNnogo} derive
%an equivalent no-go by similar arguments), $\hat T$ is also positive.
%To see this, note that up to positive numerical constants
%\beq
%T_{MN} \sim -V g_{MN} \, , \qquad \hat T \sim -V \ge 0 
%\eeq
%where in the last equality the assumption $V \le 0$ was used. 
and using the expression for the energy momentum tensor for an n-form flux
\footnote{The energy momentum tensor for some of the fluxes 
have powers of $e^{\phi}$, which
do not affect the following argument.} 
\beq
T_{MN}=F_{MP_1...P_{n-1}} F_{N}^{P_1...P_{n-1}} -\frac{1}{2n}g_{MN} F^2 \, ,
\eeq
we arrive at
\beq \label{hatT}
\hat T = -F_{\mu P_1...P_{n-1}} F^{\mu P_1...P_{n-1}}+ 
\frac{n-1}{2n} F^2  \, .
\eeq
In this equation, fluxes along space-time
and internal fluxes make separate contributions
\footnote{We are referring here to the ten-dimensional fluxes $F^{(10)}$
(cf. Eq.(\ref{splitflux})), but we suppressed the label $(10)$ to 
lighten notation.}, which means that we can consider them independently. 
Let us first consider internal components of the flux, for which
the contribution to (\ref{hatT}) is 
\beq \label{Tintpos}
\hat T_{\rm{int}} = \frac{n-1}{2n}F^2 \ge 0 \ .
\eeq 
All internal fluxes give therefore a strictly positive contribution
to the trace of the energy momentum tensor\footnote{Eq.(\ref{Tintpos}) does not apply to $F_0$, the mass parameter of massive IIA supergravity. This flux
has nevertheless been shown independently to lead to similar no-go theorems \cite{MNnogo}).}, except for 
the one form flux, whose contribution vanishes.
As for fluxes along space-time, the first term in \ref{hatT} gives
\beq
F_{\mu P_1...P_{n-1}} F^{\mu P_1...P_{n-1}} = \frac{4}{n} F^2 \, .
\eeq  
We find therefore
\beq
\hat T_{\rm{ext}}=-\frac{(9-n)}{2n} F^2 \ge 0
\eeq
where we have used $F^2\le0$ since we are considering temporal components
of $F$. All fluxes with external components give also strictly positive
contributions, expect for $F_9$, which gives a vanishing contribution 
(the same as its dual, purely internal, $F_1$).

We have therefore shown that all fluxes
give a strictly positive contribution to the second term in 
\ref{nogo}, except for $F_1$, whose contribution vanishes, and $F_0$,
whose consideration was shown in Ref.~\refcite{MNnogo} to lead to similar no-go theorems.
Multiplying (\ref{nogo}) by $e^{2A}$ and integrating (\ref{nogo}) over the internal manifold, we get the (in)famous no-go theorems:
the right hand side vanishes, while the left hand side is non-negative
for de Sitter or Minkowski spaces. We see therefore that without taking localized sources
or higher derivative corrections to the equations of motion
into account, a compactification to de Sitter space, for which
the first term in the left hand side of \ref{nogo} is also positive, 
is completely ruled out. Compactifications to Minkowski space are allowed  
in the presence of one-form flux only, while in compactifications
to anti de-Sitter spaces, the cosmological constant is related to
the square of the fluxes. 
% In all these cases, the left hand side 
% of (\ref{nogo}) is
% zero pointwise, which means that the warp factor should additionally 
% be constant. ??? 

Let us now see, following Ref.~\refcite{GKP}, how the inclusion of localized sources
modifies the argument. Localized sources give an extra contribution
to (\ref{Einstein}) 
\beq
R_{MN}= T_{MN} - \frac{1}{8}g_{MN} T_L^{\,L} + 
T^{\,\rm{loc}}_{MN} - \frac{1}{8}g_{MN} T_L^{\,L\,\rm{loc}} 
\eeq 
Eq. (\ref{nogo}) gets accordingly an additional term
\beq \label{nogoloc}
\tilde R + e^{2A} \frac{1}{2}\left(\hat T^{\rm{flux}}
+ \hat T^{\rm{loc}}   \right)=
2 e^{-2A} \nabla^2 e^{2A} \, .
\eeq
In order to avoid the no-go theorem, the sources should
give a negative contribution to $\hat T$, canceling that of the fluxes.
For compactifications
to Minkowski space, localized sources should obey the following identity 
\beq \label{cancelT}
\sum_i \int \left(\hat T_i^{\rm{flux}} + \hat T_i^{\rm{loc}} \right) =0 \, .
\eeq
Ref.~\refcite{GKP}
showed that a p-brane extended along space-time
and wrapped over a (p-3)-cycle $\Sigma$ has
\beq
\hat T^{\rm{loc}}= \frac{7-p}{2} T_p \delta(\Sigma)
\eeq 
where 
\beq \label{branetension}
T_p=(2 \pi \sqrt \ax')^{-(p+1)} e^{\frac{p-3}{4} \phi} 
\eeq
is the
positive Einstein-frame brane tension. This implies that for $p < 7$
the branes also give a positive contribution to Einstein's equation.
In order to compactify we need to include negative
tension objects. String theory does have such
negative tension objects: orientifold planes, and can therefore
evade no-go theorems. 
Note that 
constructions involving only D7-branes and one form
flux (F-theory) avoid no-go theorems, as none of these
gives a contribution to the stress tensor.
But F-theory does have its "no-go", or rather, an upper
bound for the number D7-branes, which arises due to their
induced D3-charge. We will come back to this in
section \ref{sec:Bianchi}.

One final remark before moving on: the no-go theorems
discussed in this section apply to {\it any} solution,
regardless of its supersymmetry properties.

\subsection{Bianchi identities and equations of motion for flux: tadpole cancellation conditions}
\label{sec:Bianchi}

We saw that the integrated Einstein's equation 
gives a no-go theorem for compactifications with 
fluxes. In the case of supersymmetric flux backgrounds, on which 
we concentrate in this review, Einstein's equation is automatically
satisfied if in addition to supersymmetry, we demand
Bianchi identity and the the equation of motion for the fluxes.
A proof of this for IIA is given for example in Ref.\refcite{Tsimpis}, and 
assumes that there is no crossed time-spatial component
of the Einstein tensor, which is always the case in the
compactifications we are interested in.
The no-go theorems in supersymmetric solutions should therefore
arise from integrated Bianchi identities or integrated equations
of motion for the fluxes. This can be understood by
the BPS nature of supersymmetric solutions, in which
charges are equal to tensions. Einstein's equation
gives no-go theorems based on the effective tension of the fluxes, 
while Bianchi identities and the 
equations of motion for the fluxes restrict
the magnetic and electric charges of the solution. 
In the case of IIB
type B solutions
Bianchi identity (or the equation of motion) for the self-dual
flux $F_5$ is exactly the same as the trace of Einstein's
equation, Eq.(\ref{nogoloc}). For other solutions like
type A or C, Bianchi identities impose stronger conditions
than (\ref{nogoloc}), as the the latter is just a single piece
in the former. Besides imposing how much negative charge 
we need in order to allow a compact space, they tell us
how this charge should be localized, or in other words,
what supersymmetric cycles should  the orientifolds wrap.

The Bianchi identities for the NS flux and the "democratic" RR fluxes are
given in (\ref{Bianchigen}).
Due to the self-duality relation (\ref{sd}), Bianchi identity contains
also the equation  of motion for the fluxes, which is
\beq
d(\star F^{(10)}_n) \pm (-1)^{Int[n/2]} H \wedge F^{(10)}_{8-n} =0 \ ,
 \quad +(-):\rm{IIA (IIB)} \ ,
\eeq
or equivalently
\beq
d(\star F^{(10)}_n) + H \wedge \star F^{(10)}_{n+2} =0
\eeq
for both IIA and IIB. 

Inserting the decomposition (\ref{splitflux}), using (\ref{dualhats})
and the warped metric (\ref{metric})
yields the following Bianchi identities and equation of motion
for the internal RR fluxes 
\bea  \label{Bianchiint} 
&&(d  - H \wedge) \ \  F  =0 \nn \\
&&(d -  H \wedge)  \, (e^{4A} *F) =0
\eea
Integrating these over the appropriate cycles
leads to no-go type conditions: the integral of $dF$ over a compact cycle
is zero, while supersymmetry equations 
enforce relations of the type $F \sim *H$, 
which yields a positive number
after integrating over the same cycle. 
We will make this
more precise in the next section.
This is in general stronger than the no-go from Einstein's equation 
applied to supersymmetric solutions.
% In Einstein's equation, fluxes have a positive effective tension,
% which has to be cancelled with negative tension sources
% in order to allow for compact solution. In Bianchi
% identitities, supersymmetric 
% fluxes have effective positive charges, which
% in compact solutions has to be offset by negative charged sources.
Bianchi identities' no-go's can be again avoided
by including orientifolds, which are BPS sources of negative charge 
proportional to the tension. 
% In the folliwng,
% we consider Bianchi identities in the presence of localized sources (D-branes and orientifold planes).
The cancellation conditions are often referred to as 
cancellation of NS or RR tadpoles:
the net NS charge or RR charge of the solution
has to be zero, where the charges correspond to localized, smeared 
or effective sources (fluxes) extending along space-time.

Adding to (\ref{Bianchiint}) the contribution from the localized sources
we get
\beq \label{Bianchiloc}
dF_{n} = H_3 \wedge F_{n-2} + (2 \pi \sqrt{\alpha'})^{n-1}  
\, \rho^{\rm{loc}}_{8-n}
\eeq
% according to GKP there should be a $1/2\pi$, according to Schulz $dF_3$ it  shouldn't. from now on I'll ignore it
where $\rho^{\rm{loc}}_{8-n}$ is the dimensionless
charge density of the $8-n$-dimensional
(in space only) magnetic source for $F_{n}$, 
which contains
a $\delta^{n+1} (\vec x - \vec x_i)$.

%Let us see the tadpole cancelation conditions resulting from
%Bianchi identities (\ref{Bianchiloc}) in Calabi-Yau
%manifolds.

In type IIA, tadpole cancellation conditions come from 
D4, D6 and D8-brane sources extended along space-time. 
However, D4 and D8-branes would wrap 1 and 5-cycles respectively
in the
internal manifold. In Calabi-Yau,
which is the case we will deal mostly in this review,
there are no non-trivial 1 and 5-cycles, and
therefore such tadpole cancellation conditions do not arise. 
The only tadpole cancellation condition in Calabi-Yau
compactifications of IIA arise from D6-branes,
which are electric sources for $F_8^{(10)}$, and magnetic sources 
for it's dual field, $F_2^{(10)}=F_2$ (cf. Eq.(\ref{splitflux})). 
For localized sources consisting of D6-branes
and O6-planes extended along space-time and 
wrapped on a 3-cycle $\tilde \Sigma_3$, 
Bianchi identity (\ref{Bianchiloc}) integrated over the dual
cycle $\Sigma_3$ (i.e. intersecting
$\tilde \Sigma_3$ once)
yields the IIA tadpole cancellation condition\footnote{The charge of 
an Op-plane is $-2^{p-5}$ times the 
charge of a Dp-brane.}
\beq \label{tadpole6}
N_{\Ds} (\tilde \Sigma_3) -2\, N_{\Os}(\tilde \Sigma_3) 
+ \frac{F_0}{2 \pi \sqrt{\alpha'}} \, \int_{\Sigma_3} H_3 =0 \ ,
\eeq
where $F_0$ is the mass parameter of IIA, and $N_{D6}$,
$N_{O6}$ are
the number of D6-branes and O6-planes wrapped on the
cycle $\tilde \Sigma_3$, dual to $\Sigma_3$. Explicitly, 
a D6-brane wrapping  $n_{K}$ times the cycle $A_K$ and $n'^{K}$ times
the cycle $B^K$ (these are defined in  Eq.(\ref{quant})), 
or in other words a D6-brane wrapping $\tilde \Sigma_3=n_K A_K + n'^K B^K$, 
would contribute $-n_K$ units to (\ref{tadpole6}) when $\Sigma_3=B^K$, and 
$n'^K$ for $\Sigma_3=A_K$.

In type IIB, there are tadpole cancellation conditions
coming from D3, D5 and D7-branes. D7-branes, as we saw,
do not contribute to the energy momentum tensor, and neither does
the flux $F^{(10)}_1$ for which they are magnetic sources. 
 They do contribute
however to a tadpole for D3, as a wrapped D7-brane
has induced D3-charge is we take into account the first $\ax'$
correction to its action. This is best seen in the language of F-theory,
as we will review shortly. 

D5-branes extended along space-time and wrapped 
on an internal 2-cycle $\tilde \Sigma_2$ 
are electric sources for $F^{(10)}_7$ and
magnetic sources for $F^{(10)}_3=F_3$.
Bianchi identity (\ref{Bianchiloc}) integrated 
over the dual 4-cycle $\Sigma_4$ reads
\beq
N_{\Df} (\tilde \Sigma_2) -\, N_{\Of} (\tilde \Sigma_2) 
+ \frac{1}{(2 \pi)^2 \alpha'} \, 
\int_{\Sigma_4} H_3 \wedge F_1 =0 \ .
\eeq

D3-branes extended along space-time 
are electric sources for $\tilde F_1$, 
and magnetic sources for $F_5$. These
are 6-dimensional Hodge duals of each other,
as can be seen from Eq.(\ref{dualhats}).
D3-branes are point-like in 6-dimensions, 
and therefore the tadpole cancellation
condition involves an integral over the
whole 6-dimensional space. It reads
\beq \label{tadpoleB}
 N_{\rm{D3}} - \frac{1}{4} N_{\Ot} +  \frac{1}{(2\pi)^4 \ax'^2} \int 
H_3 \wedge F_3 =0 \ .
\eeq
Using the integral fluxes of Eq.(\ref{quant}),
the number of units of D3-charge induced by
the 3-form fluxes is
\beq \label{Nflux}
N_{\rm{flux}} =  \frac{1}{(2\pi)^4 \ax'^2} \int 
H_3 \wedge F_3 =  \left( e_{K} m^K_{\RR}
- m^K e_{K\, \RR} \right) = N \, \eta \,  N^t_{\RR} \ .
\eeq
where we have used the symplectic vectors (\ref{symplv})
and 
the symplectic matrix $\eta$ is 
$\eta = \left(\begin{array}{cc} 0 & 1 \\ -1 & 0
\end{array} \right)$.

The models of Calabi-Yau orientifolds with D3 and/or D7-branes 
admit a description as F-theory \cite{VafaF} compactified on a Calabi-Yau
four-fold $X_4$ with an elliptic fibration structure $\pi$ over
a three-fold base $M$. This corresponds to a type IIB
compactification on $M$ with a dilaton-axion at a
point $p \in M$ equal to the complex structure
modulus of the fiber $\pi^{-1}(p)$, and 7-branes
at the singularities of the fibration $\pi$.
The tadpole condition for such a construction, which will 
be of use in section \ref{sec:stat}, is
\bea \label{tadpoleF}
N_{\rm{D3}} + N_{\rm{flux}}= \frac{\chi(X_4)}{24} 
\eea
where $N_{\rm{flux}}$ is defined as in (\ref{Nflux})
(the integral being on the base $M$)
and $\chi$ is the Euler number of the four-fold. 
The right hand side arises from the induced D3 charge
of the wrapped D7-branes. 
The orientifold limit \cite{Sen} corresponds to the special case
in which  the singularities 
are $D_4$ singularities, giving an O7-plane
and four coincident D7-branes at each singularity.

The tadpole cancellation condition for NS5 branes is common to IIA and IIB. 
These are magnetic sources for $H_3$, whose Bianchi
identity in the presence of sources is
\beq \label{BianchiNS5}
dH_3 =  \rho_{\rm{NS5}}
\eeq
In type I/heterotic theory the right hand side of (\ref{BianchiNS5}) gets the
higher order correction $\alpha' ({\rm tr} F \wedge F - {\rm tr} R \wedge R)$
(see for example \refcite{JoeBB}).

\subsection{Bianchi identities and special type IIB solutions}

Bianchi identities,  when specialized to special 
type IIB A, B and C backgrounds in Table \ref{ta:IIB}, give a 
particularly simple second order equation.

In type A, the relevant Bianchi identity is that
for NS flux. This flux is related to the fundamental
form $J$ by \cite{Strominger} (we are taking the upper sign 
in Table \ref{ta:IIB})
\beq
H_3 = i (\partial - \bar \partial) J_2
\eeq
where $\partial$ is the holomorphic exterior derivative. 
Bianchi identity (\ref{BianchiNS5}) then gives 
\cite{Strominger} 
\beq
dH_3 = - 2i \partial \bar \partial J_2 =  \rho_{\rm{NS5}} \ .
\eeq

In type C, which is the S-dual of type A, the corresponding 
equation is \cite{gmpt3} 
\beq
dF_3= 2i \partial \bar \partial (e^{-2A} J_2) = 
 H_3 \wedge F_{1} + (2 \pi \sqrt{\alpha'})^{2}  
\, \rho^{\rm{loc}}_{5}
\ .
\eeq 

In type B solutions, the relevant Bianchi 
identity is the one for $F_5$. $F_5$  is
related to the warp factor by
\beq \label{5formB}
F_5^{(10)}=(1+\star)\,  [d(e^{4A}) \wedge
dx^0 \wedge dx^1\wedge
dx^2 \wedge dx^3] \ 
\eeq
or equivalently
\beq
F_5=  e^{-4A} \, * d(e^{4A}) \ .
\eeq
Let us rewrite the metric (\ref{metric}) in the form
\beq \label{metricB}
ds^2=e^{2A} \eta_{\mu\nu} dx^{\mu} dx^{\nu} + e^{-2A} {\hat g}_{mn} dx^m dx^n \ .
\eeq
For the type B solution with constant dilaton (i.e., the one 
not labelled by ``F'' in Table \ref{ta:IIB}), the metric $\hat g_{mn}$
is Calabi-Yau (see Eq. (\ref{confCY})). 
Bianchi identity for $F_5$ is then reduced to 
\beq \label{Laplacpre}
-\hat \nabla^2 e^{-4A} =  \hat * 
\left( H_3 \wedge F_{3} + (2 \pi)^4 \alpha'^{2}  
\, \rho^{\rm{loc}}_{3} \right)
\eeq
where a hat indicates an operation with respect to the metric
$\hat g_{mn}$. Using the ISD property of the complex 3-form flux
in type B solutions, Eq.(\ref{ISD}), we obtain \cite{GP1} 
\beq \label{Laplac}
-\hat \nabla^2 e^{-4A} = (2\pi)^4 \ax'^2 \rho_3(x) + \frac{e^{\phi}}{12} 
G_{mnp}
\bar G^{\widehat{mnp}}
\eeq
where $\rho_3$ is here is a 0-form, equal to $\sum \delta(x-x^i)$
for all D3 and O3 sources. 

Note that to arrive at (\ref{Laplac}) we have used only
 two of the specific features
of type B solutions, namely imaginary self-duality of $G_3$, and the 
relation between $F_5$ and the warp factor (\ref{5formB}). 
On the contrary, we have not used any of properties of $\tilde g_{mn}$.
The two properties that we did use can however
{\it be derived} directly from the
equations of motion if we just demand 
a BPS-type inequality  for the energy
momentum tensor of all localized sources \cite{GKP}:
\beq \label{BPS}
\frac{1}{2} \hat T^{\rm{loc}} \ge T_3 \rho_3 {(x)} \, .
\eeq
where $T_3$ is the tension of a D3-brane, given in (\ref{branetension}).

Giddings-Kachru-Polchinski
(GKP) \cite{GKP} 
(see also Ref.~\refcite{Buchel})
showed that this BPS-type condition
determines  the form of the solution completely.
In order to show it, GKP start from the most general 
five form flux preserving 4-dimensional Poincare symmetry,
which is actually very similar to (\ref{5formB}), but
instead of $e^{4A}$ we should write a general function $f$.
The equation for $f$ coming from the 
Bianchi identity for $F_5^{(10)}$ would be very similar 
to (\ref{Laplac}), but $G_{mnp}$ should be replaced by
 be $i*(G_3)_{mnp}$, since in going from (\ref{Laplacpre}) to
(\ref{Laplac}) we have used ISD of the 3-form flux.
GKP subtract this equation to the trace of
the four-dimensional Einstein's equation, Eq.(\ref{nogoloc})
(with $\tilde R=0$). In (\ref{nogo}), GKP insert
in $\hat T^{\rm{flux}}$
the energy momentum tensor for the NS and RR  
3-form fluxes. The substraction gives the following
equation in the Einstein frame
\bea \label{nogosources}
\hat \nabla^2 (e^{4A} - f) &=& e^{2A+\phi} \frac{1}{6}
\left| *G_3 - iG_3 \right|^2+e^{-6A} \left| \partial (e^{4A} - f) 
\right|^2  \nn\\
&&
+ (2 \pi)^8 \alpha'^4  e^{2A} \left[ \frac{1}{2} \hat T^{\rm{loc}}
-T_3 \rho_3^{\rm{loc}} \right] \, .
\eea
The left hand side of (\ref{nogosources}) integrates 
to zero on a compact manifold,
while assuming (\ref{BPS}), all the terms on the right hand side are 
non-negative. This means that in the case that all the localized sources
obey (\ref{BPS}), a warped compactification to Minkowski space is allowed
only if: i) the warp factor  and the four-form potential are related by
$e^{4A}=f$, i.e., the five-form flux is as in type B, Eq.(\ref{5formB}); 
ii) the complex 3-form flux is imaginary self-dual; iii)
the inequality (\ref{BPS}) is saturated.

D3-branes, O3-planes and D7-branes and O7-planes wrapped on 
the four-cycles arising in F-theory  
saturate the inequality (\ref{BPS}), while D5 and anti-D3-branes 
satisfy (but do not saturate) it. However, D5-branes wrapped 
on collapsed 2-cycles saturate the inequality.
O5 and anti O3-planes, on the contrary, do not satisfy the inequality. 
We are obviously not allowed to
add any number of the sources saturating the inequality
as we want: although  
their contribution
to (\ref{nogosources}) is zero,  (\ref{tadpoleB}) or
(\ref{tadpoleF}) 
 should still be satisfied.

Finally, the internal components of the Ricci tensor and
the axion-dilaton $\tau$ must satisfy, in the Einstein
frame,
\bea \label{Ftheory}
\hat R_{mn} &=& \frac{(2 \pi)^8 \ax^4}{4} e^{2\phi} \partial_{\{m} \tau
 \partial_{n\}} \bar \tau + (2\pi)^7 \left(T^{\rm{D7}}_{mn}
-\frac{1}{8} \hat g_{mn} T^{\rm{D7}} \right) \, ,  \\
\hat \nabla^2 \tau &=& -i e^{\phi} (\hat \nabla \tau)^2 - 4 (2\pi)^7 
e^{-2\phi} \frac{1}{\sqrt {-g}} \frac{\delta \hat S_{\rm{D7}}}{\delta
\hat \tau} \, \nn .
\eea

In summary, if all localized sources satisfy (\ref{BPS}),
the necessary and sufficient conditions for a warped solution
are i) an internal manifold satisfying (\ref{Ftheory}); 
ii) five form flux given by  (\ref{5formB});
iii) ISD complex 3-form flux; iv) the inequality (\ref{BPS})
saturated. 

Note that i), ii) and iii) are the same as in type B
solutions. However, GKP have not
at all imposed supersymmetry. Imaginary self-duality 
of the complex 3-form flux is less restrictive
than the conditions for a supersymmetric type B solution: it allows
for a (1,2) non-primitive component (i.e., a component
of the form $J_2 \wedge w_{(0,1)}$, 
which is in the $\bar 3$ representation
of SU(3)), and a (0,3)--singlet piece. Both pieces should
be zero in a supersymmetric solution. Note that
for an internal manifold with SU(3) structure, 
there are no nontrivial closed 1-forms, which means
that $\bar 3$ representations of the 3-form flux are
not allowed. In the absence of
singlet representations of the 3-form flux, the solution 
is exactly of the type B form.
With no D7-branes, it corresponds to the 
solution without a label ``F'', for which
the internal space is conformal Calabi--Yau ($2 W_5=3W_4$), 
and the dilaton is constant.
When D7--branes are present, the internal space is no longer
conformal Calabi--Yau, but obeys instead
(\ref{Ftheory}) with nontrivial $\nabla \tau$, and
has $W_4=W_5=\partial \phi$. 
From this,  we conclude that we can obtain
a non-supersymmetric solution starting from a type
B solution of Table \ref{ta:IIB}, and turning
on a (0,3) component of 3-form flux.

From now on  we will concentrate on
flux compactifications on Ricci flat manifolds
(mostly SU(3) holonomy, and therefore Calabi-Yau, except 
in sections \ref{sec:modtoriIIB},\ref{sec:modtoriIIA},
where we discuss moduli stabilization for flux compactifications
on tori).
We will come  back briefly to the non Ricci-flat geometries
at the end of sections \ref{sec:fluxpot}, \ref{sec:fluxsup}
and in section \ref{sec:mirror}.

\section{Four Dimensional Effective Theories} \label{sec:4D}

To obtain the four dimensional effective theory for a given
compactification, we should 
perform a Kaluza-Klein (KK)reduction of the ten-dimensional
type II supergravities on a compact internal manifold,
and keep only some finite set of massless modes. As is 
standard in KK reduction, 
massless modes for
each supergravity field (metric, dilaton and B-field 
in the NS sector, and RR potentials in the RR sector)
correspond to harmonic forms on the
internal manifold.

\subsection{Effective theory for compactifications of type II} 
\label{sec:CYspectrum}

When the internal manifold is Calabi--Yau, and no
fluxes are turned on, the four-dimensional 
effective action is well
known: it corresponds to an $\N=2$
ungauged supergravity, whose  matter content
depends on which type II theory we look at \cite{CFG,FS,BCF,BC}.
Let us very briefly review how this effective action. 

A Calabi--Yau
has one harmonic 0-form -a constant-, one (3,0)-form -$\Omega$-, 
one (0,3)-form -$\bar \Omega$, and one (3,3)-form -the volume-. 
Additionally, it has  $h^{(1,1)}$ harmonic (1,1) and (2,2)-forms and $h^{(2,1)}$
harmonic (2,1) and (1,2)-forms. The total number of harmonic 
3-forms is therefore $2 h^{(2,1)}+2$. Finally, there are no harmonic 1 and
5-forms. Table \ref{ta:basis} gives a basis of harmonic forms. The
forms satisfy the normalizations given in (\ref{norm3forms}).

\begin{table}
%\tbl{Basis of harmonic forms in a Calabi--Yau manifold.}
\begin{center}
\begin{tabular}{|c|l|}\hline 
Cohomology group & \hspace{7pt}  basis \\
\hline
$H^{(1,1)}$ & \hspace{10pt} $w_a$ \hspace{40pt} $a=1,..,h^{(1,1)}$ \\
\hline
$H^{(0)} \oplus H^{(1,1)}$ & $w_A=(1, w_a)$ \hspace{3pt} $A=0,..,h^{(1,1)}$ \\
\hline
$H^{(2,2)}$ & \hspace{10pt} $\tilde w^a$ \hspace{40pt} $a=1,..,h^{(1,1)}$ \\
\hline
$H^{(2,1)}$ & \hspace{10pt} $\chi_k$ \hspace{40pt} $ k=1,..,h^{(2,1)}$ \\
\hline
$H^{(3)}$ & $(\alpha_K, \beta^K)$ \hspace{20pt} $K=0,..,h^{(2,1)}$ \\
\hline
\end{tabular}
\caption{  \label{ta:basis}
\text{Basis of harmonic forms in a Calabi--Yau manifold.}}
\end{center}
\end{table}

In the NS sector, the dilaton is ``expanded'' in the only scalar harmonic form.
The B-field can have purely external or internal components. The former
is expanded in the only internal scalar, while the latter
 is expanded in the basis $w_a$. As for the metric, the 4D massless fields
correspond to deformations that respect the Calabi-Yau condition. It was shown in 
Ref.~\refcite{Candelas} that the deformations $\delta g_{i \bj}$ 
correspond to deformations
of the 
fundamental form $J_2$, expanded in the basis of $h^{(1,1)}$ harmonic forms.
$\delta g_{ij}$ correspond on the contrary to deformations of the complex
structure, which are in one to one correspondence with the harmonic (2,1)-forms.
We have therefore the following expansions for the deformations of the fields 
in the NS sector:
\bea \label{expNS}
 \phi(x,y) &=& \phi(x) \, , \nn\\
g_{i \bj} (x,y)&=& i v^a(x) (\omega_a)_{i \bj}(y) \ , \qquad 
 g_{i j} (x,y)= i \bar z^{k}(x) 
\left(\frac{(\bar \chi_{k})_{i \bar k \bar l} \, 
\Ox^{\bar k \bar l}\,_j}{|\Ox|^2}
\right) (y) \ , \\ 
 B_2 (x,y)&=&  B_2 (x) + b^a(x) \omega_a(y) \, . \nn
\eea
Here, all the $x$-dependent fields are the moduli of the 4D theory.
In the NS sector we get a total of $2\,(h^{(1,1)}+1)+h^{(2,1)}$ moduli.

In the RR sector, we perform the following expansions
\bea \label{expRRIIA}
C_1(x,y)&=&  C^0_1 (x) \ , \nn \\
C_3(x,y)&=&  C^a_1(x)\, \omega_a(y) +  \xi^K(x) \alpha_K(y) - \tilde \xi_K(x) \beta^K(y) \, 
\eea
for type IIA, and
\bea \label{expRRIIB}
C_0(x,y)&=& C_0 (x) \ , \nn \\
C_2(x,y)&=&  C_2 (x) + c^a(x) \omega_a(y) \, , \nn \\
C_4(x,y)&=&  V_1^K(x)\, \alpha_K(y) + \rho_a(x) \tilde \omega^a(y) \, 
\eea
for type IIB. In the expansion of $C_4$ we have used the self duality
of $F_5^{(10)}$,  which connects the terms expanded in 
$\alpha_K$ to the ones that would be expanded in 
in the forms $\beta_K$, and similarly for $\rho_a$, which are dual to
$h^{(1,1)}$ tensors
$D_2$ coming from expanding in the basis $\omega_a$.  

These moduli arrange into the $\N=2$ multiplets shown in Tables \ref{ta:IIAmult}
and \ref{ta:IIBmult}, taken from Ref.~\refcite{LGIIA,LGIIB}

\begin{table}
%\tbl{Type IIA moduli arranged in $\N=2$ multiplets.}
\begin{center}
\begin{tabular}{|c|c|c|}
\hline 
gravity multiplet & $1$ & $(g_{\mu\nu},C^0_1)$ \\
\hline
vector multiplets & $h^{(1,1)}$ & $(C_1^a,v^a,b^a)$ \\
\hline
hypermultiplets & $h^{(2,1)}$ & $(z^k,\xi^k, \tilde \xi_k)$ \\
\hline
tensor multiplet & $1$ & $(B_2, \phi, \xi^0,\tilde \xi_0)$ \\
\hline
\end{tabular}
\caption{ \label{ta:IIAmult} \text{Type IIA moduli arranged in $\N=2$ multiplets.}}
\end{center}
\end{table}

\begin{table}
\begin{center}
%\tbl{Type IIB moduli arranged in $\N=2$ multiplets.}
\begin{tabular}{|c|c|c|}
\hline 
gravity multiplet & $1$ & $(g_{\mu\nu},V^0_1)$ \\
\hline
vector multiplets & $h^{(2,1)}$ & $(V_1^k,z^k)$ \\
\hline
hypermultiplets & $h^{(1,1)}$ & $(v^a,b^a, c^a, \rho_a)$ \\
\hline
tensor multiplet & $1$ & $(B_2, C_2, \phi, C_0)$ \\
\hline
\end{tabular} 
\caption{ \label{ta:IIBmult} \text{Type IIB moduli arranged in $\N=2$ multiplets.}}
\end{center}
\end{table}

Inserting the expansions (\ref{expNS}), (\ref{expRRIIA}) 
in the ten-dimensional type IIA action, and 
 (\ref{expNS}), (\ref{expRRIIB}) in the IIB one
and integrating over the Calabi-Yau, one obtains
a standard four-dimensional $\N=2$ ungauged supergravity
action (for a review of $\N=2$ supergravity
see for example Ref.~\refcite{N=2review}), 
whose forms is \cite{Michelson,BCF} (for details, see for example
Refs.~\refcite{DallaH,LouisMicu,LGIIA,LGIIB,Gurrierithesis,halfflat,Grimmthesis})
\bea \label{action4IIA}
  S^{(4)}_{\rm{IIA}} & = &\int_{M_4} -\tfrac12 R * \mathbf{1} 
                      +  \tfrac12 \R \cN_{AB}\, F^{A} \wedge  F^{B} 
		      +  \tfrac12 \I \cN_{AB}\, F^{A} \wedge * F^{B}
\\
                    & & - G_{ab}\, dt^a \wedge *  d\bar t^b 
                        - h_{uv}\, d q^u \wedge * d q^v \ , \nn
\eea
for type IIA, and 
\begin{eqnarray}
  S_{\rm{IIB}}^{(4)} & = & \int_{M_4} -\tfrac{1}{2}R *{\mathbf 1} 
+ \tfrac{1}{2} \R \cM_{KL} \, {F}^K \wedge {F}^L + \tfrac{1}{2} \I
  \cM_{KL} \, {F}^K \wedge * {F}^L \nonumber\\
&&\qquad - G_{kl} \, \dd z^k \wedge *\dd \bar{z}^{l} 
- h_{pq}\, \dd \tilde q^{p} \wedge * \dd \tilde q^{q} \ .
  \label{action4IIB}
\end{eqnarray}
Let us explain these expressions. In the gauge kinetic 
part, the field strengths are  $F^{A} = dC_1^{A}=(dC_1^0,dC_1^a)$
in the IIA action (\ref{action4IIA}), and $F^{K} = dV_1^{A}=(dV_1^0,dV_1^k)$
in the IIB action (\ref{action4IIB}). The gauge kinetic coupling matrices
$\N$ and $\M$, given below in Eq.(\ref{MN}),
 depend on the scalars in the respective vector multiplets. 
In IIA, these are the complex combination of K\"ahler and B-field deformations
$t^a$, called complexified K\"ahler deformations, and defined
\beq \label{tdef} 
B+iJ= (b^a+iv^a)\, \omega_a \equiv t^a \, \omega_a \ .
\eeq
In IIB, the scalars in the vector multiplet moduli space are
the complex structure deformations $z^k$, or the periods, defined as
 \cite{Candelas}
\bea
Z^K=\int \Ox \wedge \beta^K=\int_{A_K} \Ox \, , 
\qquad \F_K=\int \Ox \wedge \alpha_K =  \int_{B^K} \Ox\, .
\eea 
Using these, $\Omega$ can  be expanded as
\beq
\Ox=Z^K \alpha_K -
\F_K \beta^K \ .
\eeq \label{expO}
 It turns out that the Jacobian $\partial_l (Z^k/Z^0)$ 
is invertible, and therefore 
$Z^K$ can actually be viewed as projective coordinates. 
One can introduce special coordinates $z^k=Z^K/Z^0$,
which are the  $h^{(2,1)}$ complex structure deformations
in (\ref{expNS}). These are the scalars in the vector multiplets
in type IIB, Table \ref{ta:IIBmult}, and also part
of the scalars in the IIA hypermultiplets,  Table \ref{ta:IIBmult}.

The scalars in the vector multiplets span a special K\"ahler manifold
of complex dimension $h^{1,1}$ and $h^{2,1}$ in IIA and IIB respectively,
 whose metric
$G_{ab}$ and $ G_{kl}$ will be given shortly. 
The scalars in the hypermultiplets span a quaternionic manifold
whose coordinates are $q^u$, $u=0,...,h^{(2,1)}$ and 
$u=0,...,h^{(1,1)}$ for IIA and IIB respectively.
The explicit expression for the quaternionic metric $h_{uv}$ 
was found in Ref.~\refcite{FS}, and is given for example in 
Refs.~\refcite{Gurrierithesis,Grimmthesis}. 

The metric and K\"ahler potential in the vector multiplet moduli space
will be important later, so let us give their explicit form. 
The K\"ahler potential for the vector multiplet moduli space
in IIA, which is  spanned 
by the complexified K\"ahler deformations $t^a$,
is given by \cite{Candelas,StroYuk}
\beq \label{KIIA}
   K = -\ln \Big[\frac{4}{3} \int J \wedge J \wedge J \Big]=
- \ln \big[\tfrac{i}{6} \cK_{abc}(t-\bar t)^a (t-\bar t)^b (t-\bar t)^c 
\big] = - \ln \tfrac{4}{3} \cK\ ,
\eeq  
where $\cK$ is six times the volume of the Calabi-Yau manifold,
and $\KK_{abc}$ are the intersection numbers defined by 
\bea\label{intnumbers}
  \KK_{abc} &=& \int \omega_a \wedge \omega_b \wedge \omega_c\ ,  \qquad
\qquad
 \qquad 
  \KK_{ab}  = \int\omega_a \wedge \omega_b \wedge J 
= \KK_{abc}v^c \\
  \KK_{a}   &=& \int \omega_a \wedge J \wedge J
=\KK_{abc} v^b v^c \ , \qquad \ 
  \KK = \int J \wedge J \wedge J
 =\KK_{abc}v^a v^b v^c \ .\nonumber
\eea
The metric $G_{ab}$ in the IIA vector multiplet 
moduli space obtained from the K\"ahler potential (\ref{KIIA}) is
\bea \label{metricKIIA} 
  G_{ab} = \partial_{t^a} \partial_{\bar t^b} K 
 = -\frac{3}{2}\left( \frac{\KK_{ab}}{\KK}-
  \frac{3}{2}\frac{\KK_a \KK_b}{\KK^2} \right)= 
\frac{3}{2\KK}
  \int \omega_a \wedge *\omega_b  \ .
\eea

In type IIB, the scalars in the vector multiplet moduli space are
the complex structure deformations $z^k$. They span again a 
K\"ahler manifold, with K\"ahler potential given by
\beq \label{KIIB}
K = -\ln\Big[ i \int \Ox \wedge \bar \Ox\Big] 
      = -\ln i\Big[\bar Z^K {\F}_K - Z^K\bar{\F}_K \Big] \ .
\ ,
\eeq
 The metric derived from this potential is
given by \cite{Candelas}
\beq \label{metricKIIB}
G_{kl} = - \frac{\int \chi_k \wedge \bar \chi_l}{\int \Ox \wedge \bar \Ox} \ .
\eeq

Both K\"ahler potentials for the scalars in the vector
multiplet moduli space (\ref{KIIA}) for IIA and (\ref{KIIB})
for IIB, can be derived from a holomorphic prepotential  $\F$, 
namely
\beq
K(z)=i \left(\bar z^i  \frac{\partial \F}{\partial z^i} -  
z^i  \frac{\partial \bar \F}{\partial \bar z^i} \right) \, .
\eeq
For type IIA, the prepotential is 
\beq \label{prepkahler}
{\mathcal G}=-\frac{1}{6} \KK_{abc} \frac{t^a \, t^b \, t^c}{t^0}
\eeq
where $t^0$ is an extra coordinate set to 1 after differentiation, 
and introduced such that the prepotential
is homogeneous of degree two.   \\
%For type IIB $\F=\frac{1}{2} X^K \F_K$. {\bf no explicit form?
%seems like no, from Candelas}. 

We are now ready to give the expression for the matrices 
$\M$ and $\N$ in (\ref{action4IIA}, \ref{action4IIB}):
\bea \label{MN}
\text{Re} \cN &=& \ \
  \left(\ba{cc}-\frac13 \cK_{abc}b^a b^b b^c &  \frac12 \cK_{abc} b^b b^c \\
              \frac12 \cK_{abc} b^b b^c & - \cK_{abc}b^c  \ea \right)\ , \nn \\
  \text{Im} \cN &=& -\frac{\cK}{6}
  \left(\ba{cc}1 + 4 G_{ab}b^a b^b & -4 G_{ab}b^b  \\
             - 4 G_{ab}b^b &  4 G_{ab}  \ea \right)\ , \\
 \cM_{K L} &=& \overline{ {\F}}_{K L}+2i \frac{(\text{Im}\; 
{\F})_{K M} Z^M
   (\text{Im}\; {\F})_{L N}Z^N }{Z^N(\text{Im}\; \mathcal{\F})_{NM} 
    Z^M}\ . \nn
\eea
where the $(0,0)$, $(0,1)$, $(1,0)$ and $(1,1)$ elements in the
matrix expression for $\N$ give its $(0,0),(0,a), (a,0)$
and $(a,b)$ components, $\F_{KL}=\partial_L \F_K$.

Before moving on to a discussion of Calabi-Yau orientifolds, 
let us make a short pause 
and discuss following Ref.\refcite{GLW} 
the effective four-dimensional theories arising
from compactifications on manifolds of SU(3) structure (or more generally,
\stt\ structure, as discussed in sections \ref{sec:GCG}, \ref{sec:GCY}).

Hitchin showed \cite{GCY,3form} that there is a Special K\"ahler
structure on the space of generalized almost complex structures
(for usual almost complex structures, this bundle is known 
as the twistor bundle \cite{Salo2}).
The space of generalized complex structures is 
the space 
of 
stable\footnote{A real form $\Phi$ is stable if any element in a neighborhood
of $\Phi$ is $GL(6,{\mathbb R})$-equivalent to $\Phi$. 
An equivalent statement is that $\Phi$ lives in an open orbit under
the action of  $GL(6,{\mathbb R})$.},
real,  even or odd forms $\{\R \Phi_+\}$, $\{ \R \Phi_- \}$. 
In the SU(3) structure case, these are the spaces of 
symplectic structures $J$ and complex structures $\rho=\R \Omega$  
(see Eq. (\ref{Fierz})). The spinor $\Phi_+$ can actually
be truly complexified by adding the B-field. 
%Using
%$B \wedge \Omega = 0$, we can also add the B-field
%for free in $\Phi_-$, which will be useful later.
Explicitly, 
\beq \label{puresu3}
\Phi_+ = \frac{i}{8} e^{-(B+iJ)} \ , \qquad
\Phi_- = \frac{1}{8}  \Omega
\eeq
where the relative factors of $i$ 
with respect to (\ref{Fierz}) are introduced for
later convenience.

$\R \Phi_+, \R \Phi_-$ 
belong to irreducible
(``Majorana-Weyl'')  Spin(6,6) representations, as discussed
just above (\ref{Fierz}). 
The imaginary part of the complex spinors $\Phi_{\pm}$ 
is obtained from the real part by $\I \Phi_{\pm}= * \R \Phi_{\pm}$
\footnote{$\I \Phi_{+}$ can be obtained solely from $\R \Phi_{+}$
(i.e. without making use of the metric, constructed out of both
$\R \Phi_+$ and $\R \Phi_-$), by means of the Hithcin function
\cite{3form}. See Refs. \refcite{3form,GLW} for details.}.
The complex \clss\ spinors $\Phi_{\pm} = \R \Phi_\pm + i \, \I \Phi_\pm$
are pure.   

The K\"ahler metric for the space of generalized complex structures
is obtained from the following K\"ahler potential
\beq \label{genKahler}
K_{\pm}= -\ln \Big[ i \int \left< \Phi_{\pm} , \bar \Phi_{\pm} \right> \Big]
\eeq
where the ``Mukai'' pairing $\mukai{\cdot}{\cdot}$ is defined
\begin{equation}
\label{mukai}
\begin{aligned}
  \mukai{\Psi_+}{\Phi_+}
     &\equiv \Psi_6\wedge\Phi_0 -\Psi_4\wedge\Phi_2 
        + \Psi_2\wedge\Phi_4 - \Psi_0\wedge\Phi_6 \ , \\
   \mukai{\Psi_-}{\Phi_-}
     &\equiv \Psi_5\wedge\Phi_1 -\Psi_3\wedge\Phi_3 
        + \Psi_1\wedge\Phi_5\ ,
\end{aligned}
\end{equation}
(the subscripts denote the degree of the component form). 

A straightforward but very important observation is that 
for the case of SU(3) structures, i.e. for $\Phi_{\pm}$ in 
(\ref{puresu3}), the K\"ahler potentials in (\ref{genKahler})
have exactly  the same expressions as their Calabi-Yau counterparts, 
given in Eqs. \ref{KIIA}, \ref{KIIB}. The (big) difference  is that
$\Phi_{\pm}$ in (\ref{genKahler}) need not be
closed, or in other words, need not correspond
to an integrable structure, as it does in the case of
Calabi-Yau structures. Furthermore, the K\"ahler potential
(\ref{genKahler}) applies also to the general
case of \stt\ structures (by using
(\ref{eq:genpureforms}) for $\Phi_{\pm}$), or in other words 
to hybrid complex--symplectic structures (again, not necessarily integrable).

Let us now return to the more familiar case
of Calabi-Yau manifolds, and orientifold them.

\subsection{Effective theory for Calabi-Yau orientifolds} \label{sec:CYOspectrum}

The presence of orientifolds projects out part of the spectra shown in 
Tables \ref{ta:IIAmult} and \ref{ta:IIBmult}. For type IIA, the only
orientifold consistent with supersymmetry in a Calabi-Yau is an O6, extended
along space-time and wrapping a Special Lagrangian internal 3-cycle. 
In type IIB, supersymmetry allows for O3, O5, O7 and O9 planes, the 
O5 and O7 wrapping holomorphic 2- and 4-cycles respectively. 

O6 planes can be included when the Calabi-Yau has a symmetry $\sigma$ which 
is involutive ($\sigma^2=1$), isometric (leaves the metric invariant) 
and antiholomorphic ($\sigma J_n\,^m=-J_n\,^m$). 
The antiholomorphic involution acts on the holomorphic 3-form  
as $\sigma^* \Ox= e^{2i\theta} \bar \Ox$, where $\theta$ is some phase
(and $\sigma^*$ denotes the pull-back of $\sigma$).
One can eliminate $\theta$ by redefining $\Omega$. From now on
we will take $\theta=0$. A consistent truncation of 
the spectrum is obtained when the theory is modded out by 
$(-1)^{F_L}  \Omega_p \sigma$, where $F_L$ is the space-time 
fermion number in the left moving sector, and $\Omega_p$ is the world--sheet 
parity, which exchanges left and right movers \cite{BH,AAHV,LGIIA}.

In type IIB, the involution is also isometric but this time it is
holomorphic. There
are two possible actions on the holomorphic 3-form $\Omega$,
namely $\sigma^* \Omega=\pm \Omega$. 
The plus sign leads to O5/O9 planes, while the minus sign 
to O3/O7. The theory can be consistently modded out respectively by 
$(-1)^{F_L}  \Omega_p \sigma$ and $ \Omega_p \sigma$ \cite{BH,AAHV,LGIIB}.

Table \ref{ta:oraction} gives the transformations of NSNS and RR 
under  the action of $(-1)^{F_L}$, and $\Omega_p$.

\begin{table}
\begin{center}
%\tbl{Parity of the bosonic fields under the actions of $(-1)^{F_L}$, and $\Omega_p$.}
\begin{tabular}{|c|c|c|c|c|c|c|c|c|}
\hline 
& $\phi$ & $g$ & $B_2$ & $C_0$  & $C_1$  & $C_2$  & $C_3$  & $C_4$ \\
\hline 
$(-1)^{F_L}$ & $+$ &  $+$ &  $+$ &  $-$ & $-$ & $-$ & $-$ & $-$ \\   
\hline
$\Ox_p$ & $+$ &  $+$ &  $-$ &  $-$ &  $+$ & $+$ &  $-$ & $-$  \\
\hline
\end{tabular}
\caption{ \label{ta:oraction} \text{Parity of the bosonic fields under the actions of $(-1)^{F_L}$, and $\Omega_p$.}}
\end{center}
\end{table}

The massless states that survive the orientifold projection are
those that are even under the combined action of $(-1)^{F_L} \Omega_p \, \sigma$
for O3/O7 as well as O6, 
and even under $ \Omega_p \, \sigma$ for O5/O9. For example, the 
B-field is odd under $(-1)^{F_L} \Ox_p$ and is also odd under $\Omega_p$ only. 
This implies
that in the presence of any O-plane, the only components of 
B that survive are those that are odd under $\sigma$.

The space of harmonic p-forms, $H^p$, splits into two eigenspaces
under the action of $\sigma$ with eigenvalues plus and minus one.
In IIA, $\sigma$ is antiholomorphic, which implies that the spaces
$H^{(p,q)}$ are not in general eigenspaces of $\sigma$ (i.e. $\sigma$ sends 
$H^{(p,q)}$ into $H^{(q,p)}$). $H^{(1,1)}$ splits into 
$H^{(1,1)}_\pm$ and so does $H^{(2,2)}$, which splits
into $H^{(2,2)}_\pm$. $H^{(2,1)} \oplus H^{(1,2)}$ splits into
halves of dimension $h^{(2,1)}$ each, with positive and negative eigenvalues, 
and $H^{(3,0)} \oplus H^{(0,3)}$ splits into two 1-dimensional
spaces with positive and negative eigenvalues.
As an example, from the massless modes of the B-field given in Table 
\ref{ta:IIAmult}, $b^a$ and $B_2$, only a subset of $b^a$, namely those
multiplying a 2-form in $H^{(1,1)}_-$, survive. The same is true
for $v^a$, the K\"ahler deformations of the metric. Opposite to this, the vector 
$C_1^a$, which is in the same $\N=2$ multiplet than the complexified
K\"ahler deformations and comes from $C_3$, 
should be expanded
in harmonic forms in $H^{(1,1)}_+$. We see that the vector and the scalars
in the $\N=2$ vector multiplet split, building $h^{(1,1)}_-$ $\N=1$ vector
multiplets, and $h^{(1,1)}_+$ $\N=1$ chiral multiplets. The same is true
for the rest: all the original $\N=2$ multiplets break
into  $\N=1$ multiplets. 
Table \ref{ta:IIAOmult}, taken from Ref.~\refcite{Grimmthesis},
shows the surviving IIA multiplets after the orientifold projections.
The moduli $\xi_K$ in Table 
\ref{ta:IIAOmult} correspond
to a combination of the $\N=2$ moduli $(\xi^0,\xi^k,\tilde \xi_k, \tilde
\xi_0)$, namely those in $H^3_+$, of dimension
$h^3_+=h^{(2,1)}+1$. The modulus $Re Z^0$ in Table \ref{ta:IIAOmult} 
corresponds to the dilaton (for details, see Ref.~\refcite{LGIIA,Grimmthesis}).

\begin{table}
\begin{center}
%\tbl{Type IIA moduli arranged in $\N=1$ multiplets for O6 compactifications.}
\begin{tabular}{|c|c|c|}
\hline 
&   \multicolumn{2}{|c|}{ O6} \\
\hline 
gravity multiplet & $1$ & $g_{\mu\nu}$ \\
\hline
vector multiplets & $h^{(1,1)}_+$ & $C_1^\alpha$ \\
\hline
chiral multiplets & $h^{(1,1)}_-$ & $(v^a, b^a) $ \\
\hline
chiral multiplets & $h^{(2,1)}+1 $ & $(Re Z^K, \xi^K)$ \\
\hline
\end{tabular} 
\caption{\label{ta:IIAOmult} \text{Type IIA moduli arranged in $\N=1$ multiplets for O6 compactifications.}}
\end{center}
\end{table}

In type IIB, the involution $\sigma$ is holomorphic. This implies that 
its eigenspaces are inside $H^{(p,q)}$, i.e. $H^{(p,q)}=H^{(p,q)}_+ \oplus
H^{(p,q)}_-$.  Table \ref{ta:IIBOmult} shows the surviving IIB
multiplets after the orientifold projections.

\begin{table}
\begin{center}
%\tbl{Type IIB moduli arranged in $\N=1$ multiplets for O3/O7 and O5/O9 orientifolds.}
\begin{tabular}{|c|c|c||c|c|}
\hline 
 & \multicolumn{2}{|c||}{ O3/O7} & \multicolumn{2}{|c|}{O5/O9} \\
\hline
gravity multiplet & $1$ & $g_{\mu\nu}$ &  $1$ & $g_{\mu\nu}$ \\
\hline
vector multiplets & $h^{(2,1)}_+$ & $V_1^\alpha$ & $h^{(2,1)}_-$ & $V_1^k$ \\
\hline
&  $h^{(2,1)}_-$ & $z^k$ &  $h^{(2,1)}_+$ & $z^\alpha $ \\
\cline{2-5} chiral multiplets &
 $h^{(1,1)}_+$ & $(v^\alpha, \rho_\alpha)$  &
 $h^{(1,1)}_+$ & $(v^\alpha, c^\alpha)$  \\
 \cline{2-5}  &
 $h^{(1,1)}_-$ & $(b^a, c^a)$ &
 $h^{(1,1)}_-$ & $(b^a, \rho_a)$   \\
\cline{2-5}
 & $1$ & $(\phi, C_0)$ &  $1$ & $(\phi, C_2)$ \\
\hline
\end{tabular} 
\caption{\label{ta:IIBOmult} Type IIB moduli arranged in $\N=1$ multiplets for O3/O7 and O5/O9 orientifolds.}
\end{center}
\end{table}

The moduli spaces spanned by the scalars split again into that of
the scalars in the vector multiplets, and that of the hypermultiplets.
They are both K\"ahler, and are appropriate subspaces of the 
special K\"ahler and quaternionic spaces of the $\N=2$ moduli spaces
of the previous section. The effective actions for the type II 
Calabi-Yau orientifold compactifications 
have the standard $\N=1$ form \cite{WessBagger}, namely
\beq\label{N=1action}
  S^{(4)}_{\N=1} = -\int_{M_4} \tfrac{1}{2}R * \mathbf{1} +
  K_{I \bar J} DM^I \wedge * D\bar M^{\bar J}  
  + \tfrac{1}{2}\text{Re}f_{\alpha \beta}\ 
  F^{\alpha} \wedge * F^{\beta}  
  + \tfrac{1}{2}\text{Im} f_{\alpha \beta}\ 
  F^{\alpha} \wedge F^{\beta} + V*\mathbf{1}\ . 
\eeq
Here $M^I$ denote the complex scalars
in the chiral multiplets. The potential $V$ is given in terms of the superpotential
$W$ and the D-terms $D_{\alpha}$ by 
\beq\label{N=1pot}
V=
e^K \big( K^{I\bar J} D_I W {D_{\bar J} \bar W}-3|W|^2 \big)
+\tfrac{1}{2}\, 
(\text{Re}\; f)^{-1\ \alpha \beta} D_{\alpha} D_{\beta}
\ ,
\eeq
where we have used the K\"ahler covariant derivatives, defined as
\beq \label{Kahlercov}
D_I W = \partial_I W + W \partial_I K \,.
\eeq
The expressions for the gauge kinetic coupling matrix
$f_{\alpha \beta}$ are  truncated versions of their $\N=2$ counterparts
$\N_{AB}$ and $\M_{KL}$. These give the couplings of
the $h^{(1,1)}_+$ gauge fields $C^{\alpha}$ for type IIA, and for IIB
the $h^{(2,1)}_+$ gauge fields $V^{\alpha}$ and 
$h^{(2,1)}_-$ $V^k$ for O3/O7 and O5/O9 projections respectively.
 
In order to write the K\"ahler potential for the scalars in the chiral multiplets, 
 one needs to identify the good 
K\"ahler coordinates, i.e. the complex coordinates $M^I$
such that the action is of the form (\ref{N=1action}).  
It turns out that the subset of the scalars in the $\N=2$
vector multiplets that survive the projections, namely
$t^a=b^a + i v^a$ in IIA and $z^k$ for IIB O3/O7 or $z^{\alpha}$ for 
O5/O9 are good K\"ahler coordinates. Their K\"ahler potential 
and 
metric also have the same formal expressions as in $\N=2$, namely
(\ref{metricKIIA}) for type IIA, but where $a$ 
runs only in $h^{(1,1)}_-$, and (\ref{metricKIIB}) in IIB, with
$k$ running up to  $h^{(2,1)}_-$ for O3/O7 and $h^{(2,1)}_+$ for
O5/O9.
For the chiral multiplets, the story is more complicated. 
In type IIA, 
the K\"ahler coordinates (i.e., the coordinates
such that the action takes the form \ref{N=1action})
corresponding to the
complex structure moduli and the dilaton are encoded in
 the expansion of the complex 
3-form field
\bea \label{complexstrO6}
\Omega_c= C_3 + 2i Re(C\Ox) &=&  
\Big(\xi^{\kappa}+ 2 i \,Re(C\,Z^\kappa) \Big) \alpha_\kappa
+ \Big( \tilde \xi_{\lambda}+ 2i  \,Re(C\,\F_{\lambda}) \Big)
\beta^{\lambda} \nn \\
& \equiv& N^{\kappa} \ax_{\kappa} + T_{\lambda} \bx^{\lambda}
\eea
where $C$ is a ``compensator'' field proportional to $e^{-\phi}$
(see Ref.~\refcite{LGIIA} for details), and $(\alpha_{\kappa},\beta^{\lambda})$
is a basis of even 3-forms, $\kappa=0,...,\tilde h$, $\lambda=\tilde h+1,
..., h^{2,1}$ ($\tilde h$, which determines how many of the 
$\alpha$'s are even, is basis dependent, but the total number of 
complex structure moduli is obviously basis independent, equal to
$h^{(2,1)}+1$). 
Finally, the IIA K\"ahler potential is given by \cite{LGIIA}
\beq \label{KIIAO6}
K_{\rm{O6}}=- \, \ln \Big[\tfrac43 \int J \wedge J \wedge J \Big]
-2\, \ln \Big[ \int \R(C \Ox) \wedge * \R (C \Ox) \Big] \, ,
\eeq
where this should be written in terms of the
right K\"ahler coordinates defined in Eqs.(\ref{tdef}) and
(\ref{complexstrO6}).

In type IIB, the K\"ahler coordinates depend on what kind of orientifold
projection is performed. For O3/O7 projections, these are
the complex structure moduli $z^k$ and \cite{LGIIB} 
\begin{eqnarray}\label{tau}
 \tau &=& C_0+i e^{- \phi}\ , \qquad
  G^a = c^a -\tau b^a\ ,\\
  T_\alpha &=& \frac{1}{2}\cK_{\alpha} + i \rho_\alpha 
  - \frac{i}{2(\tau-\bar \tau)}\,  \cK_{\alpha b c}G^b (G- \bar G)^c  \  , \nonumber
\end{eqnarray}
where the intersection numbers $\K_{\alpha b c}$ 
and $\K_{\alpha}$ are defined  in the same way as
the $\N=2$ counterparts (\ref{intnumbers}),
but taking the appropriate basis.
 The K\"ahler potential is 
\beq \label{KIIBO3}
K_{\rm{O3/O7}}= -\ln \Big[-i \int \Ox(z) \wedge \bar \Ox(\bar z) \Big]
- \ln \left[ -i (\tau - \bar \tau) \right] - 2 \, \ln \tfrac16 \K(\tau,G,T) \,
\eeq
where $\K$ has the same formal expression as in (\ref{intnumbers}) as a function
of $v^{\alpha}$, but $v^\alpha$ should be re-expressed
in terms of the K\"ahler coordinates $\tau,G^a,T_{\alpha}$.
This cannot be done explicitly, except for the case in which
there is only one $v$ (and therefore one
$T_{\alpha} \equiv T$), i.e. when $h^{(1,1)}_+=1$. If additionally
$h^{(1,1)}_-=0$, we get a particularly simple and familiar expression
for the K\"ahler potential \cite{GKP}, namely
\beq \label{KIIBsimple}
-2 \, \ln \K= -3 \ln \left[ T + \bar T \right] \ .
\eeq

The K\"ahler potential for the chiral multiplets in type IIB O3/O7 
compactifications coming from $\N=2$ hypermultiplets,
i.e. the last term in Eq. (\ref{KIIBO3}), 
 satisfies a very important property, namely \cite{LGIIB,DFT}
\beq \label{noscale}
\partial_I K \partial_{\bar J} K \, K^{I \bar J} =3 \ ,
\eeq
for $I=(G^a,T_\alpha)$ \footnote{Including the second term 
in (\ref{KIIBO3}) and summing over $I=(\tau, G^a,T_\alpha)$
gives 4 in the right hand side of (\ref{noscale}).}.
This is a no-scale type condition \cite{noscale}. When there is
a nontrivial superpotential (which is the case in the presence of fluxes),
the condition (\ref{noscale}) implies that the positive 
contribution to the potential (\ref{N=1pot}) offsets the negative one $-3|W|^2$, 
and we therefore get $V \ge 0$.
This equality can be easily checked in the simple case of one
K\"ahler modulus, Eq. (\ref{KIIBsimple}).
%Finally, in the absence of fluxes there are no D-terms, so the whole potential
%is zero.

For O5/O9 orientifolds, the right K\"ahler coordinates
are again the complex structure moduli $z^{\alpha}$, and
the combinations \cite{LGIIA}
\bea
t^{\alpha}&=&e^{-\phi} v^{\alpha} - i c^{\alpha} \ , \qquad
A_a = \Theta_{ab} b^b + i \rho_a \ , \nn\\
S&=&\frac{1}{6} e^{-\phi} \K + ih - \tfrac14 (\R 
\Theta^{-1})^{ab} A_a (A+\bar A)_b \ ,
\eea
where we have defined 
\beq \label{kahlero5}
\Theta_{ab}(t)= \K_{ab \ax} t^{\ax}\  , \qquad 
\int C_6 = h+ \tfrac12 \rho_a b^a
\eeq
(i.e. $h$ is an appropriate dual to $C_2$).
The K\"ahler potential for O5/O9 is given by
\beq \label{KIIBO5}
K_{\rm{O5/O9}}=  -\ln \Big[-i \int \Ox \wedge \bar \Ox \Big]
- \ln \left[ \tfrac16 \int e^{-3\phi} J \wedge J \wedge J \right]
 -  \, \ln \tfrac13 e^{-\phi} \K(S,A_a,t^{\alpha}) \ ,
\eeq
where the first term is a function of the complex structure
moduli, the second a function of the moduli $t^{\alpha}$
only, and
in the last one one should solve for $\K$ in terms
of $(S,A_a,t^{\alpha})$ using (\ref{kahlero5}).

\subsection{Flux induced potential and gauged supergravity}
\label{sec:fluxpot}

We showed in section \ref{sec:susy} that fluxes
back react on the geometry, and a Calabi-Yau manifold is no longer
a solution to the equations of motion. If however the typical
energy  scale of the fluxes is much lower than the KK scale, we can
assume that the spectrum of sections \ref{sec:CYspectrum} and
\ref{sec:CYOspectrum} is the same, except that  some of the massless modes
acquire a mass due to the fluxes. The energy scale of, for example,
 3-form fluxes can
be estimated using the quantization condition (\ref{quant}),
and is given by $\frac{N_{\rm flux} \ax'}{R^3}$ (where $N_{\rm flux}$ are the number
of units of 3-form flux). The KK scale
is $\frac{1}{R}$. The former is much lower than the latter
when the radius is much bigger than $\sqrt N_{\rm flux}$ 
times the string scale, which
is in any case needed from the start in order to neglect 
$\alpha'$-corrections to the action. This truncation of
the spectrum to those modes that are massless in the absence of 
fluxes is standard  in  flux compactifications,
and it is shown to yield a consistent $\N=2$ or $\N=1$
gauged supergravity action, depending whether one starts
with a Calabi-Yau 
\cite{PS,Michelson,TaylorVafa,Mayr,germans,HL,DallaH,LouisMicu,BeckerConst,GKTT,FP,KKP} 
 or a Calabi-Yau orientifold \cite{LGIIA,LGIIB}.
A similar argument was used in Ref.~\refcite{halfflat} to show that is it possible
to do a consistent truncation to a set of light modes in the case
of compactifications on half flat manifolds with fluxes, and generalized in
Ref. \refcite{GLW} to the case of any SU(3) structure manifold,
such that the resulting action has the standard form of $\N=2$ gauged   
supergravity. In this section we will concentrate on the case 
of Calabi-Yau compactifications with fluxes, and only say a few comments
about these more general constructions at the end.

Turning on RR fluxes in type IIA and keeping the same light spectrum
of Calabi-Yau compactifications amounts to doing the replacements
\beq
dC_1 \rightarrow dC_1 + m_{\rm{RR}}^a \omega_a \ , \qquad
dC_3 \rightarrow dC_3 - e_{\rm{RR}\,a} \tilde \omega^a  \ .
\eeq
 Reducing the ten dimensional action in the presence
of these fluxes leads to the following new terms in the four dimensional
action \cite{LouisMicu} 
\beq \label{RRtermsIIA}
S_{\rm{RR}}= \int_{M_4} -B_2 \wedge J_2 -\frac{1}{2} M^2\, B_2
\wedge * B_2 - \frac{1}{2} M_T^2 \,B_2 \wedge B_2 - V \ ,
\eeq
where
\bea \label{RRcouplings}
J_2 &=& -e_{\rm{RR}\,A} F^A + m_{\rm{RR}}^A 
\left(\I \N_{AB} *F^B + \R \N_{AB} F^B \right) \nn\\
M^2_{B_2} &=& -m_{\rm{RR}}^A \, \I \N_{AB}\, m_{\rm{RR}}^B \nn \\
M^2_{T\,B_2} &=& -m_{\rm{RR}}^A \, \R \N_{AB} \, m_{\rm{RR}}^B + m_{\rm{RR}}^A \, e_{\rm{RR}\,A}  \\
V_{\rm{IIA \, \RR}} &=& -\frac{e^{4\phi}}{2} (e_{\rm{RR}\,A} - \bar \N_{AC} \,m_{\rm{RR}}^C ) (\I \N)^{AB}
(e_{\rm{RR}\,B} - \bar \N_{BD}\, m_{\rm{RR}}^D ) \nn
\eea 
We see from (\ref{RRtermsIIA}) that RR fluxes induce Green-Schwarz type couplings, 
regular and topological mass terms for the tensor
$B_2$, and a potential that renders massive some of the scalars in the 
vector multiplets.  When magnetic fluxes are present, the tensor $B_2$ 
becomes massive by  a St\"uckelberg-
type mechanism, namely it ``eats''
one combination of the $C_1^a$ gauge vectors, which becomes pure gauge
once the magnetic fluxes are introduced. The potential in (\ref{RRcouplings})
depends on the scalars in the vector multiplets, namely the complexified
K\"ahler deformations. The effect of RR fluxes is summarized in Tables 
\ref{ta:potentials} and \ref{ta:gaugings}.

The extra terms in the action coming from turning on RR fluxes 
where shown to be consistent 
with a standard $\N=2$ gauged supergravity in Ref.~\refcite{LouisMicu} for
the case $m^I=0$, when there are no massive tensors. 
The introduction of magnetic fluxes is also consistent
with an $\N=2$ gauged supergravity with massive tensors
\cite{massivetensors}, in which case the tensor is not
dual to a scalar but rather to a massive vector.

In IIB, RR fluxes are introduced by
\beq \label{expF3}
dC_2 \rightarrow dC_2 + m_{\rm{RR}}^K \alpha_K - e_{\rm{RR}\,K}\, \bx^K  \ .
\eeq
Inserting this in the Lagrangian results in the same new terms
as in IIA, Eq. (\ref{RRtermsIIA}), where the definitions as in IIA,
Eq. (\ref{RRcouplings}), just replacing $\N$ by $\M$, and the indices
$A,B,..$ by $K,L,...$ (i.e. the sums are from $0$ to $h^{(2,1)}$). 
Therefore, RR fluxes in IIB have the same effect as in IIA, namely
induce  Green-Schwarz type couplings, 
regular and topological mass terms for the tensor
$B_2$, and a potential that renders massive some of the scalars in the 
vector multiplets. This is summarized in Tables 
\ref{ta:potentials} and \ref{ta:gaugings}.

NS fluxes are introduced in IIA and IIB by modifying
\beq \label{expH}
d B_2 \rightarrow d B_2 + m^K \alpha_K - e_{K}\, \bx^K  \ .
\eeq
In type IIA, NS fluxes give gauge charges to the 
scalars $a$ (dual to $B_2$) and $(\xi^K,\tilde \xi_K)$ in the 
tensor and hypermultiplets respectively, whose ordinary derivatives 
in (\ref{action4IIA}) are replaced by  covariant derivatives.
The only vector field participating in these gaugings is the graviphoton,
who as a consequence acquires a mass. Fluxes also generate 
a potential for the scalars $(\phi,z^k,\xi^K,\tilde \xi_K)$ 
in the hyper and tensor multiplets. 
%The mass term for the  graviphoton and 
The potential is given by
by \cite{LouisMicu}
\bea \label{NScouplings}
%M^2_{C_1^0} &=& -\frac{e^{2\phi}}{2} (e_{K} + \M_{KM} m^M) \, \I \M^{KL} 
%(e_{L} + \bar \M_{LN} m^N)   \\
V_{\rm{IIA} \, \rm{NS}} &=& -\frac{e^{2\phi}}{4 \K}
 (e_{K} + \M_{KM} \,m^M ) (\I \M)^{KL}
(e_{L} + \bar \M_{LN} \,m^N ) \nn \\
&& + \frac{e^{4\phi}}{2 \K} (m^K \tilde \xi_K
- e_K \xi^K + e_0)^2 \ .
\eea
Note that  the matrix entering the NS flux potential is $\M$, rather than
$\N$ (cf. Eq. (\ref{RRcouplings})), which means that
the scalars that get a potential are all in hypermultiplets.
Note also that among all the axions $(\tilde \xi_K,\xi^K)$, only
a single combination of them, namely $m^K \tilde \xi_K
- e_K \xi^K$, gets a potential. We will come back to this
in section \ref{sec:moduli}.

In IIB, NS electric fluxes gauge the scalars in the tensor multiplet,
namely the (appropriate) dual of $B_2$ (see Ref.~\refcite{BC} for
the redefinitions of the quaternionic coordinates) and 
the dual of $C_2$ \cite{Michelson,DallaH,halfflat}.
Opposite to the case in type IIA, the vectors that gauge these
scalars are those in the vector multiplets, and not the graviphoton.
One combination of these vectors acquires therefore a mass.  
Electric fluxes also generate a potential for the scalars in the vector 
multiplet, and the axion-dilaton.
Finally, magnetic fluxes give a mass to $C_2$. 
The flux generated potential is given by 
\beq \label{NScouplingsIIB}
V_{\rm{IIB} \, \rm{NS}} =
 -\frac{e^{4\phi}}{2 \K} \left(C_0^2+ \frac{e^{-2\phi}}{2 \K}
\right)
 e_{K}  (\I \M)^{KL}
e_{L} \ .
\eeq

The effect of fluxes is summarized in Tables 
\ref{ta:potentials} and \ref{ta:gaugings}. The scalars 
gauged by the fluxes are always ``axions'' in hyper (or tensor) multiplets:
$\xi, \tilde \xi,B_2$ in type IIA, and $B_2$, $C_2$ in type IIB. 
The quaternionic metric in the hypermultiplet
moduli space has translational isometries,  corresponding
to shifts of these scalars. Fluxes gauge these translational
isometries. The partial derivatives in the action turn into
covariant derivatives, namely
\beq
\partial_{\mu} q^u \rightarrow  D_{\mu} q^u = \partial_{\mu} q^u - k_A^u C_{\mu}^A
\eeq
where $k_A^u$  are the Killing vectors, and $C_{\mu}^A$ are the
vectors in the vector multiplet that participating in the gauging.
In type IIA, without magnetic RR flux, the gauging are \cite{LouisMicu}
\beq
k_a^{B}= 2 e_{\RR \, a}  \ , \qquad 
k_0^B= (m^K \tilde \xi_K - e_K \xi^K) \ , 
\qquad k_0^{\xi^K}= m^K \delta^{u\, \xi^K} \ , 
\qquad k_0^{\tilde \xi^K}= e_{K}  \ .
\eeq
where $B$ is the scalar dual to $B_2$ (massless in the absence of magnetic RR flux).
In type IIB, for electric RR and NS flux the Killing vectors are 
\beq
k_K^{B}= 2 e_{\RR \, K} +  e_K \xi^0 \ , \qquad
k_K^{\tilde \xi^0}= e_K \ .
\eeq
Ref. \refcite{KPTflux} showed that the isometries corresponding to these
Killing vectors are not lifted by quantum corrections (instantons)
to the quaternionic metric.

The flux generated potentials can be studied similarly in the case of Calabi-Yau
orientifold compactifications \cite{LGIIB,LGIIA}. In O6 compactifications
of type IIA, the tensor
$B_2$ is projected out of the spectrum, so there are no massive tensors
arising from the introduction of fluxes. The flux term that 
does survive the projection is the potential term in (\ref{RRcouplings}), 
i.e. RR fluxes   
give potential terms to the scalars $v^a, b^a; a=1,...,h^{(1,1)}_-$. NS fluxes
can also be combined in a potential term of the form (\ref{RRcouplings})
which depends on the scalars in the $h^{(2,1)}+1$ chiral multiplets.
Ref.~\refcite{LGIIA} showed that in the language of $\N=1$ supergravity,
this potential arises from a superpotential, 
to be reviewed in the next section, and no D-term.

In type IIB, O3/O7 orientifolds project out the tensors $B_2, C_2$
and the graviphoton. This means that from the combined action of 
RR and NS fluxes, Eq. (\ref{RRcouplings}) --replacing $\N$ by $\M$
and $A,B,...$ by $K,L,...$-- and (\ref{NScouplingsIIB}), only
the potential survives. This potential is a truncated version of 
the sum of the potentials in (\ref{RRcouplings}) and (\ref{NScouplings}),
with indices running from $0$ to $h^{(2,1)}_-$ (i.e. there
is a potential for the complex structure moduli $z^k$ and
the axion-dilaton).
Its explicit expression in terms of the coupling matrix
$\M$ and the fluxes can be found in Ref.~\refcite{LGIIB}.
We will see in next sections that the potential can be derived from 
a superpotential, as in Eq.(\ref{N=1pot}), and no D-term.
In the case of O5/O9 planes, the tensor $C_2$ is not
projected out from the spectrum, and it acquires a mass
when NS magnetic fluxes are present. The potential due to 
the fluxes is again a truncated version of (\ref{RRcouplings})
and (\ref{NScouplingsIIB}). There are two distinguished pieces,
one containing RR fluxes, for which there is a sum 
from $0$ to $h^{(2,1)}_+$ (i.e., it corresponds again a potential
for the complex structure moduli and the dilaton), 
and another one for NS fluxes, whose sum
runs from $1$ to  $h^{(2,1)}_+$, which is also a potential
for complex structure moduli. It was shown in Ref.~\refcite{LGIIA} that
the piece of the potential involving
RR fluxes can be derived from a superpotential as we will review
in the next section, while the NS fluxes give rise to a D-term.

Table \ref{ta:potentials} shows which scalars get a nontrivial potential
due to RR and NS fluxes, for Calabi-Yau, and Calabi-Yau orientifolds
\footnote{The symbols
F and D in Table \ref{ta:potentials}
mean that the potential arises from an F or a D-term respectively (see
Eq.(\ref{N=1pot})). Additionally, $\xi^0$ in the row corresponding to NS 
flux in IIA stands for the combination $m^K \tilde \xi_K
- e_K \xi^K$.}.

\begin{table}
\begin{center}
%\tbl{Scalars that get a potential in the presence of fluxes in CY compactifications.}
\begin{tabular}{|c|c|c|c||c|c|c|c|c|c|}
\hline 
 & \multicolumn{3}{|c||}{ IIA } & \multicolumn{5}{|c|}{IIB} \\
\hline
&  $\N=2$ & \multicolumn{2}{|c||}{$\N=1$ (O6)} & $\N=2$ & 
\multicolumn{2}{|c|}{$\N=1$ (O3/O7)} & \multicolumn{2}{|c|}{$\N=1$ (O5/O9)} \\
\hline
RR flux & $(v^a,b^a)$ & $(v^a,b^a)$ &F & $z^k$ & $z^k$ &F & $\ \ \ \  z^k \ \ \
\  $ &F
\\
\hline
NS flux &   
$(z^k, \xi^0, \phi)$ & $(Re\, Z^k,\xi^0)$ &F & $z^k,\phi, C_0$
& $z^k,\phi, C_0$ &F & $z^k$ &D  
\\
\hline
\end{tabular} 
\caption{\label{ta:potentials} \text{Scalars that get a potential in the presence of fluxes in CY compactifications.}}
\end{center}
\end{table}

Note that in type IIB  the $4 \,h^{(1,1)}$ moduli in hypermultiplets
(or $2 \,h^{(1,1)}$ complexified K\"ahler moduli in the 
orientifolded theory) do not get a potential by fluxes.
In type IIA, the ones that do not get a potential
are $2 h^{(2,1)}$ 
scalars in hypermultiplets (or $h^{(2,1)}$ scalars in chiral multiplets
in the orientifolded theory).
This means that fluxes could potentially be the only ingredient
needed to stabilize all moduli in compactifications of type IIA
on rigid manifolds ($h^{(2,1)}=0$), but in type IIB
there is no way of stabilizing all moduli, as $h^{(1,1)}\ge 1$
(there is at least the volume modulus). 
We will come back to this issue
at length in section \ref{sec:moduli}.

Before discussing the flux induced superpotentials, let us pose
for a moment and discuss the effect of fluxes in the 
case of manifolds with SU(3) structure, not necessarily
Calabi-Yau's.  
As we mention at the beginning of this subsection,
in a similar spirit than the one used to study the effect
of fluxes on Calabi-Yau manifolds,  
Refs.~\refcite{halfflat,GLW} argued that 
in the case of manifolds with SU(3) structure, it is possible
to do a consistent truncation to a finite set of light modes.
The light modes are obtained by expanded in a set of 
p-forms, out of which some are not closed. The non closure
is proportional to the torsion, which as shown in
Refs.~\refcite{halfflat,GLW}, plays a very similar role
as the fluxes (see also Ref. \refcite{CGLM}). We know in fact \cite{VafaM,halfflat} that some of the torsion
classes in Eqs.(\ref{torsion}, \ref{dJdOmega})  are mirror to NS flux, as we will
briefly discuss in section \ref{sec:mirror}.
By this procedure, Ref.\refcite{GLW} showed that the resulting action has the
standard form of $\N=2$ gauged supergravity (we have shown already
in Eq.(\ref{genKahler}) that the moduli space of
complex structures and K\"ahler deformations
is K\"ahler).

To be more precise, torsion is encoded in the non-closure of the forms in the
basis
\begin{equation}
\label{domega}
\begin{aligned}
   d\omega_a &= m_a^K\, \alpha_K - e_{aL}\,\beta^L\ , \\
   d\tilde{\omega}^a &= 0 \ , \\
   d\alpha_K &= -e_{aK} \tilde\omega^a\ , \\
   d\beta^K &=m_a^K \tilde\omega^a\ .
\end{aligned}
\end{equation}
The forms used here are denoted in the same way 
as in the case of Calabi-Yau manifolds, but we should bear in mind
that here they are clearly
not harmonic (or at least some of them are not), and furthermore the indices
$a$ and $K$ run from $1$ to $b_J$ and from $0$ to $b_{\Ox}$, where
$b_J$ and $b_{\Ox}$ are respectively the dimensions
of the finite dimensional set of light 2 and 3-forms respectively.
Furthermore, the ``light'' spectrum is shown to be the same
as in the Calabi-Yau case, Tables \ref{ta:IIAmult}, \ref{ta:IIBmult},
with $h^{(1,1)}$, $h^{(2,1)}$  in IA (Table \ref{ta:IIAmult}) 
replaced respectively by $b_J$, $b_{\Omega}$, and the opposite
in  IIB (Table \ref{ta:IIBmult})
(but differently from the CY case, 
these are not massless in the presence of RR and NS flux).

We will not give the details of the derivation, but just 
quote that the resulting low energy action is consistent with $\N=2$ 
gauged supergravity. Table \ref{ta:gaugings}, taken from Ref.~\refcite{GLW},
summarizes the effect of fluxes and torsion in the four-dimensional action
\footnote{The tensor $D_2^a$  comes from the expansion of $C_4$. In the text
we have used a dual scalar $\rho_a$ instead of $D_2^a$. }

\begin{table}
\begin{center}
%\tbl{Effect of fluxes and torsion in the $\N=2$ four dimensional action.}
\begin{tabular}{| c | c |c|} 
\hline
  & IIA& IIB \\ \hline  
 electric RR-flux $e_{\textrm{RR}}$ & Green--Schwarz coupling 
&Green-Schwarz coupling\\ 
\hline
 magnetic  RR-flux $m_{\textrm{RR}}$ & massive tensor $B_2$ & massive tensor $B_2$\\ \hline
 electric NS-flux $e_0$ & massive $C_1^0$ & 1 massive $V_1^k$ \\ \hline
magnetic NS-flux  $m_0$ &  massive $C_1^0$ &massive tensor $C_2$ \\
\hline  
electric torsion $e_{aK}$ & massive $C_1^a$ & massive $V_1^k$ \\ \hline
magnetic torsion $m_a^K$ & massive $C_1^a$ & massive tensors $D_2^a$ \\ \hline
\end{tabular}
\caption{\label{ta:gaugings} \text{Effect of fluxes and torsion in the $\N=2$ four dimensional action.}}
\end{center}
\end{table}

\subsection{Flux induced superpotential} \label{sec:fluxsup}

In the previous section we showed that fluxes induce a 
potential for certain moduli in type II Calabi-Yau
and type II Calabi-Yau orientifolds (and we also 
briefly discussed the effect of fluxes on manifolds of
SU(3) structure). 
%Table \ref{ta:potentials} summarizes
%which moduli get a non trivial potential for Calabi-Yau 
%compactifications {\bf I can do it for su(3) too...}.
We mentioned that in compactifications yielding
$\N=1$ actions, this potential can be fully derived from a 
superpotential, except in the case of Calabi-Yau O5
compactifications, where the NS-flux generated potential
comes from a D-term. 
%we could have anticipated this by
%the fact that $H=0$ in O5-type vacua, 
%susy says $D=0$

A flux generated superpotential was proposed by
Gukov-Vafa-Witten (GVW) \cite{GVW} for M- theory compactifications 
to three dimensions on Calabi-Yau four-folds.
GVW showed that the  tension for
BPS domain walls (five-branes wrapping a four cycle
of the CY 4-fold) separating vacua
coincided with the jump in the superpotential when going
from one vacua to the other.
Taylor and Vafa \cite{TV}
showed a similar result for type IIB Calabi-Yau 3-folds, where
the BPS domain
walls correspond in that case to five branes wrapped around 
3-cycles in the Calabi-Yau.
They also proposed a type IIA superpotential along
the lines of Gukov's  for Calabi-Yau
four-folds \cite{Gukov}.

The type IIB superpotential for compactifications
of Calabi-Yau 3-folds, and Calabi-Yau O3/O7
generated by the fluxes is
\beq \label{WO3}
W_{O3/O7}=\int G_3 \wedge \Ox 
= (e_{K \,\RR} - i \tau e_K ) Z^K - 
(m^{K}_{\RR} - i \tau m^K ) \F_K 
\eeq
where in the last equality we have used 
Eqs. (\ref{norm3forms}), (\ref{HF}), (\ref{defG}) and (\ref{tdef}). 
This superpotential depends on
the complex structure moduli through $\Ox$, and on the
dilaton-axion, by the definition of $G_3$. 
On the contrary, the K\"ahler moduli, $v^{\alpha},
 \rho_{\alpha}$, as well as the moduli coming from
$B_2$ and $C_2$, $b^a$ and $c^a$
do not appear in the superpotential.

The potential is obtained from this superpotential
by computing the K\"ahler covariant derivatives, Eq.(\ref{N=1pot}).
The K\"ahler potential for the 
chiral multiplets in O3/O7 compactifications
is given in Eqs. (\ref{KIIB}) and (\ref{KIIBO3}).
The K\"ahler covariant derivatives are given by
\bea
  D_{\tau} W&=& \frac{i}{2}e^{\phi}  \int \bar G_3  \wedge \Ox
             +i G_{ab}b^a b^b\  W \ , \qquad 
  D_{T_{\alpha}} W\ =-2\ \frac{v^\alpha}{\cK}\  W \ , \nn\\
  D_{G^a} W&=&2i G_{ab}b^b\  W\ , \qquad \qquad \ \ \quad \qquad \qquad
  D_{z^{k}} W\ = \int G_3 \wedge \chi_{k}  \ ,
  \label{kcov1}
\eea
where we have used 
\beq \label{derox}
\frac{\partial \Ox}{\partial z^{k}}= k_{k}\, \Ox + \chi_k
\eeq
Inserting (\ref{kcov1}) in (\ref{N=1pot}), we get the following potential for 
Calabi-Yau O3/O7 compactifications (the metric $K^{I \bar J}$ is given
explicitly for example in Ref.~\refcite{GGJL}))  
\beq
V_{\rm{O3/O7}}=\frac{ 18 i e^{\phi}}{\cK^2 \int \Ox \wedge \Oxb } 
\left(\int \Gtb \wedge \Ox 
\int G_3 \wedge \Oxb + G^{kl}
\int  G_3 \wedge \chi_{k}  \int \Gtb \wedge \bar{\chi}_{l} \right) \ .
\label{VO3}
\eeq
As anticipated from the no-scale condition (\ref{noscale}),
this potential is positive semi-definite. It contains both RR and NS fluxes,
and depends on the axion-dilaton and complex structure moduli,
in agreement with the gauged supergravity result obtained by
KK reduction, summarized in Table \ref{ta:potentials}. 
Furthermore, inserting the expansions for all the forms given 
in the previous sections, it can be shown that it agrees 
with the sum of (\ref{RRcouplings}) (for type IIB, i.e. with $\N$ 
replaced by $\M$) and (\ref{NScouplings}). 

Another check of GVW superpotential comes from supersymmetry conditions
\cite{GKP}.
A Minkowski vacuum should satisfy $W=0, D_I W=0$. From (\ref{kcov1})
we see that these are satisfied provided 
\beq 
W=0,\  D_I W=0 \ \ \ \ \Rightarrow \ \ \ \
\int G_3  \wedge \Ox =0\ ; \int \bar G_3  \wedge \Ox =0 \ ;\int G_3 \wedge \chi_{k} =0 \ .
\eeq  
These equations imply that in a fixed complex structure, there
are no (0,3), (3,0) or (1,2) pieces in $G_3$, or in other
words, $G_3$ should be (2,1). Since there are non non-trivial
1-forms in a CY, $G_3$ is automatically primitive. 
Then $G_3$ satisfies the type B supersymmetric conditions discussed
in section \ref{sec:N=1} (see Table \ref{ta:IIB}). It is easy to see that when $G_3$
is (2,1), the full potential (\ref{VO3}) is zero.
As we discuss below Table \ref{ta:IIB}, 
the supersymmetries preserved by a type B solution are the same as 
those for O3/O7 planes, so it is expected that supersymmetry
conditions resulting from the GVW superpotential fall into type
B class. Note however that if one adds (2,1) complex 3-form flux 
on a Calabi-Yau manifold, the back--reacted geometry is no longer 
a product, but it is a warped product, and the internal manifold
is no longer Calabi-Yau.
It is nevertheless ``as close as it gets'', namely the 
torsion classes $W_1,\, W_2,\, W_3$ are zero,
and the only non vanishing classes are in the ${\bf 3}$ representation.
Furthermore, for the case $\tau=const$ --no D7-branes-- (corresponding in Table
\ref{ta:IIB} to the row without an ``F''), the internal
manifold is conformally Ricci-flat. As discussed below Table
\ref{ta:IIB}, the conformal factor is the inverse of the warp factor. 
The warp factor behaves at large radius like $e^{2A} \sim 1+ 
O(g_s\,N\,\ax'^2 r^{-4})$,
where $N$ is the number of D3-branes, or the units of 3-form flux.
Therefore, it is a usually argued that in the large radius, weak coupling 
limit, where one trusts the
supergravity approximation, the effect of the warping is negligible. 
An honest computation of supersymmetry conditions
via a superpotential and K\"ahler potential should nevertheless include 
the warping. A first step toward this was put forward in
Refs.~\refcite{DWG,deAlwis,GiMa}, who considered KK-reductions in warped
products with a conformal Calabi--Yau factor.  In particular, it was
claimed in Ref.~\refcite{DWG,GiMa} that the warp factor affects the $N=1$ K\"ahler
potential, but not the superpotential.

In the case of O5/O9 planes, Ref.~\refcite{LGIIB} showed that RR fluxes
generate a superpotential, while NS fluxes generate a D-term. 
The superpotential for this case is the Gukov-Vafa-Witten one,
setting $H_3$ to zero, i.e. 
\beq \label{WO5}
W_{O5/O9}= \int \hat F_3 \wedge \Omega = e_{K \,\RR} Z^K - 
m^{K}_{\RR}  \F_K  \ .
\eeq
This superpotential depends on the complex structure moduli only,
which implies
that all its K\"ahler covariant derivatives except the one
along the complex structure moduli are proportional to $W$ itself.
The derivative along the complex structure moduli, $D_{z^{\alpha}} W$ gives
the same expression as in (\ref{kcov1}), with $G_3$ replaced by
$F_3$. This gives the F-term piece of the potential
\beq
V_{\rm{O5/O9,F}}=\frac{ 18 i e^{\phi}}{\cK^2 \int \Ox \wedge \Oxb } 
\left(\int \hat F_3 \wedge \Ox 
\int \hat F_3 \wedge \Oxb + G^{kl}
\int  \hat F_3 \wedge \chi_{k}  \int \hat F_3 \wedge \bar{\chi}_{l} \right) \ ,
\label{VO5}
\eeq
which agrees with (\ref{RRcouplings}) after truncating appropriately
the sum over moduli. 

Imposing the Minkowski supersymmetric vacuum conditions
$W=0, D_I W=0$, one gets that $F_3$ should have no
(0,3) or (1,2) piece. Since $F_3$ is real, the F-term conditions would 
set all the components of $F_3$ to zero. On the other hand, we argued
that O5/O9 planes preserve type C supersymmetries, and we therefore expect 
to get a type C solution, which, according to Table \ref{ta:IIB}, can have 
non vanishing $F_3$. There is nevertheless no contradiction, because
we see from Table \ref{ta:IIB} that a primitive component of $F_3$
generates the torsion class $W_3$. The internal manifold is then no longer
Calabi-Yau, in a much more drastic way than in type B, namely,
besides the torsion in ${\bf 3}$ representations, 
there is torsion in the ${\bf 6}$ representation.  This means
that in order to get supersymmetry conditions for type C solutions,
one should consider a more general superpotential for 
manifolds of SU(3) structure, not necessarily Calabi-Yau.
We will come back to this issue shortly.

In Type IIA, the potential comes from a superpotential\footnote{In massive 
IIA, there is an additional term in the potential
proportional to $F_0 \int \rm{Im} \, \Ox \wedge H_3$ \cite{DKPZ,WGKT}, which vanishes
when the tadpole cancellation conditions are satisfied.}
of the form \cite{Gukov,TV,LGIIA,DKPZ}
\bea \label{WO6}
W_{O6}&=& \int H_3 \wedge \Ox_c + \int \hat F_A \wedge e^{(B+iJ)} \\
& =& 
T_{\lambda} m^{\lambda} -N^{\kappa} e_{\kappa} + e_{0 \, \RR} + 
e_{a \, \RR} t^a
+ \tfrac12 \K_{abc} m_{\RR}^a t^b t^c + \tfrac16 m_{\RR}^0 \K_{abc} 
t^a t^b t^c
  \ , \nn
\eea
where $F_A=F_0+F_2+F_4+F_6$, 
 $t^a$ is defined in Eq.(\ref{tdef}) and 
$N^{\kappa},T_{\lambda}$ in 
Eq. (\ref{complexstrO6}).

This superpotential depends on all the O6 moduli, and 
so does the corresponding potential, which agrees with the expression
from previous section obtained by doing a KK reduction.
This is a fundamental difference between IIA and IIB flux
superpotential, and will become very important in next
section when we discuss moduli stabilization. 
As in the case of Calabi-Yau O5/O9 compactifications,
variations of this superpotential do not yield the  
$\N=1$ supersymmetry conditions for IIA shown in Table
\ref{ta:IIA}. The reason for this is the same as 
in O5/O9 case, namely $\N=1$ type IIA vacua in manifolds
of $SU(3)$ structure have either NS flux and nonzero
torsion $W_3$, or RR flux and nonzero $W_2$.
In one case the manifold is non-K\"ahler, and in the other 
it is not even complex,
so we do not expect to get the supersymmetry conditions
in Table \ref{ta:IIA} from variations of superpotential
for Calabi-Yau orientifolds \ref{WO6}. What should
be varied instead is the general superpotential for
manifolds of SU(3) structure, to which we turn. \\

The superpotentials given so far, namely Eqs.
(\ref{WO3}, \ref{WO5},\ref{WO6}), can be obtained
from the supersymmetry variation of the gravitino
\cite{germans}, whose
generic form is
\beq
\delta \psi_{\mu} = \nabla_{\mu} \xi + i e^{K/2} 
W \gamma_{\mu} \xi^* \ .
\eeq
where $\xi$ is the (four-dimensional) $\N=1$
supersymmetry parameter and  $K$ and $W$
are the $\N=1$ K\"ahler potential and superpotential. 
Given the ten-dimensional
supersymmetry transformation of the gravitino
(\ref{eq:susyg}), inserting the decomposition of
the ten-dimensional supersymmetry spinor 
(\ref{epsilon}), and the K\"ahler potentials
(\ref{KIIBO3},\ref{KIIBO5},\ref{KIIAO6}),
we obtain the superpotentials
(\ref{WO3}, \ref{WO5},\ref{WO6}) \cite{BeckerConst}.
The O3/O7, O5/O9 and O6 superpotentials
(\ref{WO3}, \ref{WO5},\ref{WO6}) are actually
obtained using respectively $a=ib$, $a=b$ and  $a=b e^{i \beta}$
in (\ref{epsilon}), which are the supersymmetries
preserved by O3, O5 and O6 planes 
(cf. Tables \ref{ta:IIA}, \ref{ta:IIB}). 

The reason why
we mention this is that this procedure allows
us to obtain the superpotential for
SU(3) structure manifolds, not necessarily
Calabi-Yau's. In order to do this, we just
need to insert the appropriate covariant
derivative for the internal spinor.
This was done in Ref.\refcite{GuMi}
for the heterotic theory, in Ref.\refcite{HoMi} for 
M-theory, and in Ref.\refcite{GLW}
for type II theories (see also Ref.\refcite{su3w} for IIA). 
We quote the results
of Ref.\refcite{GLW}, and refer the reader
to the original reference for details.
The general $\N=1$ superpotential 
for unwarped compactifications\footnote{In deriving
this result, the warp factor was set to zero.} 
manifolds of SU(3) structure is
\bea
\label{WSU3}
     {W}_{\text{IIA}} &=& \int 
     \bar a^2 \, e^{-\phi} \langle \Phi_+, d\bar \Phi_-\rangle
      - \int \bar b^2 \, e^{-\phi} \langle \Phi_+, d\Phi_-\rangle 
      + 2 \int \bar a \bar b \,  \revmukai{\hat F_\text{IIA}}{\Phi_+}  , \nn \\
      {W}_{\text{IIB}} &=& \int
    \bar a^2  \, e^{-\phi} \revmukai{d\Phi_+}{\Phi_-}
      + \int \bar b^2 \, e^{-\phi} \revmukai{d\bar\Phi_+}{\Phi_-}
      - 2i \int \bar a \bar b \,  \revmukai{\hat F_\text{IIB}}{\Phi_-} .
\eea
where $\Phi_{\pm}$ are given
in (\ref{puresu3}) and $\hat F_\text{IIA}$ and $\hat F_\text{IIB}$ are the
sum of the RR fluxes, as in (\ref{eq:F}). 
This superpotential is similar to the one proposed in Ref. \refcite{BeMa},
which is expressed in terms of periods of pairs
of Calabi-Yau mirror manifolds.

Inserting $a=ib$ in (\ref{WSU3})
we get the GVW superpotential for O3/O7 orientifolds
of Calabi-Yau \ref{WO3}. 
With $a=b$, we recover the RR part of O5/O9
superpotential \ref{WO5}, which gets modified
by a torsion piece, namely
\beq 
W_{O5/O9, \rm{non CY}} = \int (\, e^{-\phi}\,d J + \hat F_3) \wedge \Ox \ .
\eeq
For $a=0$ in IIA or IIB, we get the heterotic
superpotential \cite{Beck,GuMi} 
\beq
W_{\rm{het}}=\int e^{-\phi}\, (dJ+iH) \wedge \Ox \ .
\eeq
Finally, for $a=ib$ in IIA we get, after integrating
by parts the NS piece,  the ``torsional O6''
superpotential
\beq \label{WO6t}
W_{O6, \rm{non CY}}=\int e^{-\phi} \, (dJ+iH) \wedge \R \, \Ox+ 
i \int {\hat F_{\rm{IIA}}} \wedge e^{B+iJ} \ ,
\eeq
whose RR and NS pieces were proposed respectively 
in Refs.\refcite{Gukov} and \refcite{halfflat}, and checked explicitly
for twisted tori in Ref. \refcite{DKPZ}.
These superpotentials have the right holomorphic
dependence on the respective chiral multiplets.
 
The superpotentials (\ref{WSU3}) were obtained 
for (unwarped) compactifications on manifolds
of SU(3) structure. Nevertheless, Ref.\refcite{GLW}
conjectures that (\ref{WSU3}) is also valid for
manifolds of \stt\ structure on \tts\
(which comprises the cases of SU(3) and 
SU(2) structures on $T$), if we just
replace $\Phi_\pm$ by the appropriate \clss\
spinors, given in (\ref{eq:genpureforms})
(where $\Phi_+$ would have to include a factor of
$e^{-B}$)

\subsection{Mirror symmetry} \label{sec:mirror}

In this section we review the state of the art about 
mirror symmetry for flux backgrounds.

To start with, let us very briefly review the
main ideas of mirror symmetry in Calabi-Yau
compactifications. (See for example \refcite{HKT}
for a review). 
String theory compactified on a Calabi-Yau three-fold gives
a four dimensional $\N=2$ theory. From the
world-sheet point of view, this compactification
yields a 2-dimensional $(2,2)$ superconformal field theory
(SCFT)
whose marginal operators belong to the
(chiral,chiral) and (antichiral,chiral) rings,
of respective dimensions $h^{0,q}(M,\Lambda^p T)=h^{p,q}(M)$
and $h^{0,q}(M,\Lambda^p T^*)=h^{3-p,q}(M)$. From the conformal
field theory point of view, there is a trivial symmetry
corresponding to the exchange
of a relative sign between the two U(1) currents,
by which $(c,c) \leftrightarrow (a,c)$. 
However, on the geometrical level this symmetry is far from
being trivial, as it amounts for example to the exchange
$h^{2,1} \leftrightarrow h^{1,1}$
on the Calabi-Yau cohomologies. It implies that
for a given (2,2) SCFT, there are two interpretations, as
a string theory compactified on two topologically
very different manifolds, $M$ and $\tilde M$ such that
\beq \label{mirrorhodge}
h^{p,q} (M)= h^{3-p,q} \, (\tilde M) \ .  
\eeq
This very nontrivial symmetry is called ``mirror symmetry'', and the
manifolds $M$ and $\tilde M$ are mirror pairs.
IIA compactified on $M$ is identical to a IIB compactification
on $\tilde M$.  
In particular, the complex structure moduli space 
of $M$ is identified to the K\"ahler moduli space of $\tilde M$,
as well as their respective prepotentials, 
i.e. \cite{BC} 
\bea \label{mirrorv}
Z^K=(Z^0, Z^0 z^k) \ &\longleftrightarrow& \ t^A=(1,t^a) \nn \\
\F_K &\longleftrightarrow&  {\cal G}^A
\eea
Furthermore, the whole multiplets are mapped,
as the effective actions resulting from Calabi-Yau
compactifications respect the exchange (see Tables
\ref{ta:IIAmult}, \ref{ta:IIBmult})
\beq \label{mirrorh}
(\xi^K, \tilde \xi_K) \ \ \ \longleftrightarrow \ \ \ 
(c^A, \rho_A)=(C_2, c^a, C_0, \rho_a)
\eeq

Mirror symmetry was also shown to hold at the level of effective actions
in Calabi-Yau orientifolds in the large volume, large complex
structure limit in Ref. \refcite{LGIIA}.

The question we are interested in, is to what extent
mirror symmetry survives when fluxes are present,
or how should this symmetry be modified.
We will try to answer this question from the 
effective supergravity point of view. 
%For a worldsheet
%discussion in the case of NS fluxes, see ?

There are several very important aspects
of this question to take into account.
The first one is that introducing fluxes
generically breaks supersymmetry spontaneously. 
When fluxes are present, we are led to look for
mirror symmetry of the effective actions
rather than on vacua \cite{halfflat,GLW,toma}.
The second aspect is that from (\ref{mirrorhodge})
we expect fluxes in even cohomologies to be mapped to
fluxes in odd cohomologies. For RR fluxes this is 
fine, as IIA contains fluxes in even cohomologies,
while those of IIB are in odd. NS flux however belongs
to an odd cohomology, and its ``mirror''
is an even NS ``flux'': torsion \cite{VafaM,halfflat}.  
Therefore, the right setup to study mirror symmetry
in the presence of fluxes is that of 
compactifications on manifolds with torsion, or more
precisely on manifolds with SU(3) structure. 
This also relates to the first aspect: the effective action
whose vacua are backgrounds with
non zero flux should be those resulting from 
compactifications on  
SU(3) structure manifolds, rather than Calabi-Yau's.

Working on manifolds of SU(3) structure,
Ref. \refcite{halfflat} made precise the conjecture in
Ref. \refcite{VafaM}, showing that the mirror of the
NS flux $H_3$ is the torsion of half-flat manifolds,
namely $\R W_1$, $\R W_2$. Ref. \refcite{FMT}
obtained the mirror symmetry map for general
SU(3) structure manifolds, and Ref. \refcite{toma}
its topological version. 
These results are obtained by extending
the Strominger-Yau-Zaslow (SYZ) procedure 
\cite{SYZ} for Calabi-Yau manifolds to 
manifolds of SU(3) structure. SYZ conjectured
that every \cy\ with a mirror is a $T^3$ Special Lagrangian
fibration 
over a 3-dimensional base, and mirror
symmetry is T-duality along the $T^3$ fiber.
Assuming that the SU(3) structure manifolds in question have
this $T^3$ fibration, Refs. \refcite{halfflat,FMT,toma}
perform three T-dualities along the fiber.

Let us first discuss mirror
symmetry (or 3 T-dualities) in terms of the 
defining objects of the structure, $J$ and $\Omega$
(or equivalently the spinor $\eta$). 
Using the T-duality rules for
the supersymmetry parameter \cite{Hass}, 
one can see that by 3 T-dualities 
there is an exchange 
of $\eta_+$ with $\eta_-$. Using this
in (\ref{eq:genpure}), it is 
natural to conjecture that mirror symmetry
is an exchange of pure spinors, namely \cite{FMT}
\bea \label{mirror}
\Phi_+ \ &\longleftrightarrow& \ \Phi_- \ \\
e^{B+iJ} \ &\longleftrightarrow& \ \Ox \  \nn
\eea
where in the last line we have specialized to manifolds 
of SU(3) structure (see \ref{Fierz}, \ref{puresu3})). 
Ref. \refcite{FMT} checked this conjecture explicitly
by performing 3 T-dualities on a $T^3$--fibered 
metric and a  B-field of fiber-base type. 
Mirror symmetry
exchanges therefore the two \clss\ pure spinors.
We expect (\ref{mirror}) to hold also
for the general case of \stt\ structures
on $T \oplus T^*$.     

We want now to introduce fluxes on the SU(3) structure manifolds. 
The SYZ picture of mirror symmetry makes it clear
that RR fluxes are mapped among themselves (even fluxes
in IIA are mapped to odd fluxes in IIB), while
NS fluxes are mixed with metric components via T-duality 
(see for example Refs. \cite{mirrors,DaHu} for 
constructions relevant to this discussion).
The explicit mirror symmetry map involving
NS fluxes and torsion for manifolds
of SU(3) structure found in Ref. \refcite{FMT} 
is
\bea \label{mirrorsu3}
i (W_3 + i H^{(6)})_{ij} + \Ox_{ijk} (\bar W_4 + i H^{(\bar 3)})^k
& \longleftrightarrow& -2i \bar W^2_{i \bj} - 2 g_{i \bj} 
(\bar W_1 + 3i H^{(\bar 1)}) \nn \\
(W_5-W_4-iH^{(3)})_i
 & \longleftrightarrow& (W_5-W_4-iH^{(3)})_{\bi} \ ,
\eea
or in a more compact version
\beq \label{mirrornabla}
(\nabla J + H)_{ijk} \ \ \ \longleftrightarrow \ \ \ (\nabla J - H)_{i \bj \bar k} \ .
\eeq
This comes from the exchange of $\eta_+$ and $\eta_-$ under
the T-dualities, which results in 
$Q_{ij} \leftrightarrow Q_{i \bj}$ and 
$Q_i \leftrightarrow -\bar Q_{\bi}$, where
the $Q$'s are defined in (\ref{eq:susyRQ}). 
Actually, all the matrices in (\ref{eq:susyRQ}) entering 
the full supersymmetry transformations 
$\delta \lambda, \delta \Psi_M$ where shown
to follow such an exchange \cite{GMPT}, if 
in addition 
\beq \label{mirrorR}
F_{\rm{IIA}} \ \ \ \longleftrightarrow \ \ \ F_{\rm{IIB}} \ .
\eeq

The $\N=1$ supersymmetry equations for the \clss\ pure spinors (\ref{a}, \ref{b}) 
respect the mirror symmetry maps (\ref{mirror}, \ref{mirrorR})\footnote{\label{foot:B}Note that
by using $e^{\pm B} \Phit_{\pm}$ and $H=dB$, 
the twisted covariant derivative 
 $(d+ H \wedge)$ in (\ref{a},\ref{b})
can be replaced by an ordinary derivative. This has 
been used in writing (\ref{WSU3}). }. 
Eqs. (\ref{mirrorsu3}, \ref{mirrorR}) can be understood as the result of 
mirror symmetry exchange
of SU(3) representations 
${\bf 6} + {\bf \bar 3} \leftrightarrow {\bf 8}+ {\bf 1}$ \cite{FMT}.

We turn now to the specific question of 
mirror symmetry of the $\N=2$ effective actions
resulting from compactifications on SU(3) structure
manifolds in the presence of fluxes. 
The K\"ahler potentials for the vector multiplet
moduli space in SU(3) structure
compactifications, spanned by the lines of
pure spinors $\Phi_+$ and $\Phi_-$ for
IIA and IIB respectively, are indeed
mapped to one another, as can be
seen from (\ref{genKahler}). 
The $\N=2$ flux generated potential,
which can be derived from the
mass matrix for the gravitinos, and is an $\N=2$
version of the superpotential, was
also
shown in Ref. \refcite{GLW} to respect the mirror symmetry maps
(\ref{mirror}, \ref{mirrorR}). 
The $\N=1$ superpotential is comprised in 
its $\N=2$ version, namely it is obtained by projecting 
the $\N=2$ on a plane
orthogonal to the $\N=1$ preserved
spinor in the $SU(2)_R$-symmetry space (the parallel projection
gives the D-term). As a result, it also satisfies 
the mirror maps if we do appropriate mirror
projections, as can be seen from (\ref{WSU3}).

Performing
the explicit expansion of $\Phi_{\pm}$ and the
RR potentials in the basis
of ``light modes'' of 
Refs. \refcite{halfflat,GLW}, one can see that
the kinetic terms in the effective actions are symmetric
under the maps (\ref{mirrorv}, \ref{mirrorh})
(remember
nevertheless that these are not ``moduli'',
as some of them are massive due to the torsion
and fluxes).  
In the presence of RR and only electric NS fluxes
(the latter include torsion, see Eq.(\ref{domega})),
the IIA and IIB $\N=2$ superpotential are symmetric
under (\ref{mirrorv}, \ref{mirrorh}) if fluxes 
are mapped via \cite{LouisMicu,halfflat,GLW}
\bea 
(e_{\rm{RR}\, K}, m_{\rm {RR}}^K) \ \ 
&\longleftrightarrow& \ \ (e_{\rm{RR}\, A}, m_{\rm {RR}}^A) \nn \\
e_{A\, K} \ \ &\longleftrightarrow& e_{K \,A} \ .
\eea
where NS flux and torsion have been combined in the flux
$e_{A \, K}\equiv(e_K, e_{a\,K})$. For the magnetic
NS fluxes, Ref. \refcite{GLW} shows that
the $\N=2$ potentials are not mirror symmetric. 
In particular, as summarized in Table \ref{ta:gaugings},
magnetic fluxes give rise to a massive tensor in
IIB, while in IIA this is not the case. This seems
to contradict the conjecture (\ref{mirror}), which
has torsion and H-flux hidden in $d\Phi_{\pm}$ (see footnote
 \ref{foot:B}). However, the expansion in light modes
of Ref.\refcite{GLW} is specialized
to the case of SU(3) structure, where $\Phi_-$ 
contains only a 3-form (see \ref{puresu3}), and
not all odd forms, as in the case of \stt\ structures
(see \ref{eq:genpureforms}). The seeming paradox
is therefore expected to be resolved 
by considering the more general \stt\ structures.

\section{Moduli stabilization by fluxes} \label{sec:moduli}

In the previous section we discussed the flux generated potentials
and superpotentials in Calabi-Yau compactifications and in general SU(3) structure
manifolds. From now on, we will concentrate
on compactifications on Calabi-Yau orientifolds.
The presence of the flux induced potentials implies that some of the moduli
of Calabi-Yau compactifications cease to be moduli, i.e. they
acquire mass due to the fluxes. This is one of the main reasons,
if not the main one, which makes compactifications with fluxes so 
attractive, being a very active domain of research over the past five years. 

Let us first explain why we need to 
understand the ultimate mechanism of moduli stabilization.
Scalars which remain massless lead to long range 
interactions. The coupling of these
scalars to matter (both in standard compactifications and in 
brane-world scenarios) is not universal. This implies
that  different types of matter get different accelerations
from these long range forces, violating the principle
of equivalence. High precision measurements of the principle
of equivalence have tested the ratio of
inertial to gravitational mass up to 1 part in $10^{13}$ \cite{GRtests}.
Such a ``fifth force'' should therefore be very weak, or sufficiently
short ranged not to violate experiments. If moduli remain
massless, we do not
expect their couplings to all types of matter to be 
so much smaller than that of gravity, which implies that in any
realistic theory,  all ``moduli'' should be massive. 
Furthermore, if the vacuum expectation
value of the moduli fields, most notably the volume modulus, 
can be anything, string theory looses any predictability.
  
If the flux generated potential for moduli has  local minima,
moduli will be stabilized at the values where one of these minima lie. 
If the vacua is at a minimum
of the potential which is not the absolute minimum, there is an instability
against tunneling through a barrier to the absolute minimum
(see Ref. \refcite{SilTASI} for a review of moduli and microphysics).
In many cases, notably for the GVW superpotential, the resulting
potential has flat or runaway directions. In such cases, fluxes 
fix only some of the moduli, and in order to obtain realistic theories
one needs to invoke non perturbative effects 
to stabilize the remaining moduli. Once moduli are stabilized,
the overall consistency requires that the dilaton is fixed at small values, i.e. 
$g_s=e^{<\phi>} \ll 1$ and the overall volume, or 
average radius, at large values in string units, i.e. $R
\sim <\int \K_{\ax \bx \gamma} v^\alpha v^\bx v^{\gamma}>^{1/6} \gg \sqrt \ax'$.

We will start by reviewing moduli stabilization 
in orientifolds (O3) of Calabi-Yau manifolds, following
Ref.~\refcite{GKP}, and specializing
on a compact version of the conifold.
Then, we will discuss moduli stabilization on tori,
focusing on the case of O3 orientifolds of $T^6$
and following Ref.~\refcite{KST}

As mentioned in the introduction, we will not cover 
stabilization of moduli by open string fluxes, or
stabilization of open string moduli. We refer the reader
to some of the original references \cite{oms,Madrid,LustMRS}.   
Besides, we will not cover moduli stabilization in M-theory  
\cite{mtm,AsKa}, and in the heterotic theory \cite{hetmod,Beck,CGLM}.

\subsection{Moduli stabilization in type IIB Calabi-Yau orientifolds}
\label{sec:moduliconifold}

In this subsection we review the mechanism of moduli stabilization
in type IIB Calabi-Yau orientifolds, following mainly
 Giddings-Kachru-Polchinski (GKP) \cite{GKP}, who focused on (a compact
version of) the 
conifold. Later developments for other Calabi-Yau manifolds
can be found for example in Refs.~\refcite{GKTT,GKT,CoQ,BBCQ,Quevedo,DWRec}, 
who considered 
Calabi-Yau hypersurfaces
in wighted projective spaces.

The superpotential for Calabi-Yau O3 compactifications 
is given in Eq.(\ref{WO3}). The conditions for a supersymmetric
Minkowski vacuum
\beq \label{susyW}
W=0 \ , \qquad D_{I} W=0 \ ,  
\eeq
result in $2h^{(2,1)}_- + 2$ real
equations, since the covariant derivative of
the superpotential along the K\"ahler moduli
is proportional to $W$ itself (see \ref{kcov1}).
(Note that this implies that 
the condition $W=0$ comes automatically
from demanding supersymmetry, and therefore 
there are no supersymmetric $AdS_4$ vacua
in this construction).
The $2h^{(2,1)}_- + 2$ equations 
are independent of the K\"ahler moduli $(v^\ax, \rho_\ax),
\,(b^a,c^a)$, which remain unfixed. Turning on appropriate
fluxes, it is possible to fix 
$2h^{(2,1)}_- + 2$ real moduli, namely the complex
structure  $z^k$ and the dilaton-axion $\tau=C_0 + i e^{-\phi}$.
The fact that the complex structure moduli and the dilaton
are fixed in type B solutions is easy to understand from
the supersymmetry conditions: given a set of fluxes
$(e_K,m^K,e_{\rm{RR},K},m_{\rm{RR}}^K)$,
there are only some fixed complex structures and 
axion-dilaton which make the complex 3-form flux $G_3$ (2,1) 
(the axion-dilaton enters in the definition of $G_3$).
If there is no complex structure and $\tau$ such that  
$G_3$ is (2,1), then either there is a solution but it is not
supersymmetric (this would be the case if there is
some complex structure and $\tau$ for which $G_3$ 
is (2,1) plus (0,3)), or there is no solution at all
for that set of fluxes.

GKP discuss moduli 
stabilization in a compact version of Klebanov-Strasssler
\cite{KS}, i.e. on a manifold
with a local region with a deformed conifold geometry, embedded
in a compact manifold with an O3 identification. 
GKP assumed $h^{(1,1)}_-=0$, i.e. no $(b^a,c^a)$ moduli,
and $h^{(2,1)}_-=h^{(1,1)}_+=1$, i.e. one complex structure modulus and
one K\"ahler modulus, although the results are easily generalized
to any $h^{(2,1)}_-$, $h^{(1,1)}_+$, as they explain. 

The deformed
conifold is a cone over a space with topology $S^2 \times S^3$.
It is described by complex coordinates $(w_1,w_2,w_3,w_4)$
subject to
\beq
w^2_1+w^2_2+w^2_3+w^2_4= z
\eeq
where the complex parameter $z$ is the complex structure modulus, which
controls the size of the $S^3$. 
Since $h^{(2,1)}_-=1$, there are four non-trivial 3-cycles.
In the vicinity of the conifold, there are two relevant cycles: 
the $S^3$, called A, which intersects once
the dual cycle B \footnote{The cycle
A can be taken to be the $S^3$ on which all $w$'s are real, while 
the the cycle B, which goes off
to infinity in the non compact case, can be constructed by taking 
$w_{1,2,3}$ imaginary and $w_4$ real and positive.}. The Klebanov-Strassler solution
has $M$ units of $F_3$-flux through the A-cycle, and $-K$ units of 
$H_3$ on the B-cycle, i.e.  $e_1=-K$, $m^1_{\rm{RR}}=M$  (cf. Eq.(\ref{quant})). 
Using this in (\ref{WO3}), we get
\beq \label{WKS}
W = -M \F(z) +K \tau z \,
\eeq
$\F(z)$ has an expansion  of
the form $\F(z)=\frac{z}{2\pi i } \ln z +$ analytic terms.
Inserting this in (\ref{WKS}) we get 
from the Minkowski vacuum condition, $W=0$, that
the complex structure is fixed to an 
exponentially small value
\beq
z \sim e^{-\frac{2\pi K}{M g_s}} \ .
\eeq
GKP show that this also satisfies  $D_z W=0$ in the
regime  $\tfrac{K}{g_s} \gg 1, \tfrac{K}{M g_s} \gg 1$.
In order to satisfy the other F-term constrain, $D_\tau W=0$,
GKP showed that in the compact case one needs to turn on additional 
fluxes on the two remaining
3-cycles $A',B'$. 
Calling $-K'$ the number of units of $H_3$ on the B' cycle,
and $z'(z)$ the period of $\Omega$ along the $A'$ cycle
(of order 1), all the susy conditions (\ref{susyW})
stabilize the complex structure modulus and dilaton
at
\beq
\bar \tau = \frac{M \F(0)}{K' z'(0)} \, , \qquad 
z\sim e^{\frac{2\pi K}{K'} Im[\F(0)/z'(0)]} \ . 
\eeq
This implies that by appropriate choices of the fluxes, the dilaton can be
fixed at weak coupling, and $z$ is small. The complex structure 
being stabilized at a small value has interesting phenomenological
consequences, as the warp factor will be very small
close to the end of the throat (located at the points where
the $S^3$ shrinks to zero). This is because the warp factor,
which solves (\ref{Laplac}), goes for D3-branes like
$e^{4A} \sim r^4$, where $r$ is the conical coordinate. 
The resolution of the conifold
cuts this off at $r \propto w^{2/3} \propto z^{1/3}$, which means
that there is a minimum, (but non zero) warp factor
\beq \label{hierarchy}
e^{2A_{\rm{min}}} \sim z^{2/3} \sim e^{-\frac{4 \pi K}{3K'}} \I [\F(0)/z'(0)] 
\eeq
which generates a large hierarchy of scales. This a string realization
of Randall-Sundrum type models \cite{RS} \`a la Verlinde \cite{Ver}, where
the compact region plays the role of IR brane, and 
the size of hierarchy is a function of the fluxes. 
    
As in all type B solutions, the K\"ahler moduli (in this case 
just the overall volume) is not stabilized by the fluxes.

\subsection{Moduli stabilization in type IIB orientifolds of tori} \label{sec:modtoriIIB}

In this section we discuss moduli stabilization in compactifications
of type IIB on tori. We will present one of the first examples
discussed in the literature, that of Kachru-Schulz-Trivedi (KST)
\cite{KST}, consisting compactifications on $T^6$ with
orientifold 3-planes. Other more complicated examples have
been worked out in the literature after KST. In particular,
moduli stabilization on orientifolds (O3) of 
$T^6/({\mathbb Z}_2 \otimes {\mathbb Z}_2)$ 
orbifolds, where there are twisted and untwisted sectors,
 is discussed in Refs.~\refcite{BlTa,CaUr,Mash,CvLi}. 
Stabilization in orientifolds of $T^6/{\mathbb Z}_N$ and in generic 
$T^6/({\mathbb Z}_N \otimes {\mathbb Z}_M)$ is
analyzed reaspecively in Refs. \refcite{Font,LMRS}. In these examples, 
the effect of D9-brane fluxes --which can stabilize
some K\"ahler moduli and also lead to a chiral open string sector 
with potential interest for phenomenology-- is additionally considered.
Moduli Stabilization in
$K3 \times T^2/{\mathbb Z_2}$ is discussed in Refs.~\refcite{KTT,AAFL,LustMRS},
where in the latter D3 and D7-brane moduli and fluxes 
are also taken into account.

KST study $\N=1$ compactifications on a $T^6/{\mathbb Z}_2$ orientifold with
NS flux and RR flux $F_3$. Since there are O3 planes,
we expect the supersymmetries of the solution to be of type
B. We should be careful however in applying
the conditions from supersymmetry outlined Table \ref{ta:IIB},
since a torus is a manifold of trivial structure, while
the results of Table \ref{ta:IIB} concern manifolds of 
SU(3) structure. The fact that the structure group is more
reduced than SU(3) means that there is more than one 
nowhere vanishing spinor that we can use in the decomposition
(\ref{decompepsilon}). This can lead to solutions that are
very different from those of Table \ref{ta:IIB}, as we 
showed in section \ref{sec:GCG} (for example, the 
internal manifold need not be complex. 
Nevertheless, KST showed that
the supersymmetry parameter of their solution 
uses only one internal spinor, as in the case 
of SU(3) structure, Eq. (\ref{decompepsilon}).
The supersymmetry conditions are therefore
those of type B in Table \ref{ta:IIB}, except
that when the structure group is more reduced than
SU(3) there are harmonic 1-forms, and therefore
the condition of $G_3$ being primitive has to
be further imposed (in the SU(3) structure case, the
absence of harmonic 1-forms made this condition automatic,
and we only needed to impose that $G_3$ be (2,1)). 
This will be very important, as it will allow to
fix some of the K\"ahler moduli, which are otherwise
unfixed in the Calabi-Yau case. 

The moduli for O3 compactifications in Calabi-Yau
manifolds are given in Table \ref{ta:IIBOmult}.
In the case of a torus, we have to take into account that the structure
group is trivial. This implies that besides
$h^{(2,1)}_-=9$, $h^{(1,1)}_+=9$, we have to 
consider also the cohomologies $h^{(1,0)}_-=h^{(2,0)}_+=
h^{(3,1)}_+=h^{(3,2)}_-=3$ and their conjugates
 ($h^{(0,1)}_-=3$, etc). Note that in $T^6/\mathbb Z_2$,
all even (odd) forms are even (odd) with respect to
the involution $\sigma$. Therefore, $h^{(2,1)}_+=h^{(1,1)}_-=0$.
The additional cohomologies add 12 extra vectors 
coming from $B_{\mu m}$ and $C_{\mu m}$, and 
six extra scalars from $C_{(3,1)}$ and
$C_{(1,3)}$. This gives a total of 12 
vectors, 21 scalars from the metric (9 from
the K\"ahler moduli $v^{\alpha}$ and 12
(real) from complex structure deformations 
\footnote{Since $h^{(2,1)}_-=9$, 
one would expect 9 complex structure deformations
instead of 6. The difference appears because 
%the fact that differently from a Calabi-Yau, 
in the torus (as well as in any 
manifold with no-where vanishing vectors)
not all complex structure deformations correspond
to deformations of the metric. There are 
six real deformations of the complex structure that 
leave the metric invariant \cite{KST,GLW}.}, 15 scalars 
from $C_4$ (9 scalars $\rho_{\alpha}$ plus 6 extra  
from  $C_{(3,1)}$ and
$C_{(1,3)}$) and 2 from the axion-dilaton. These
arrange into multiplets of $\N=4$, as the 
existence of 4 nowhere vanishing vectors
plus the orientifold projection implies that the effective four-dimensional 
action is $\N=4$. The graviton, 6 gauge bosons and the axion dilaton
are in the $\N=4$ supergravity multiplet, and the 
others build, together with their fermionic partners, 
six vector multiplets with one vector and six scalars each. 
The scalars span the manifold
\beq
\M_{\N=4} = \frac{SU(1,1)}{U(1)} \times \frac{SO(6,6)}{SO(6) \times SO(6)} \ ,
\eeq
where the first factor corresponds to the dilaton-axion, and the second 
to the scalars in the vector multiplets.

The explicit solution is constructed as follows. First,
let $x^i,y^i$, $i=1,2,3$ be six real coordinates on the
torus, with periodicities $x^i\equiv x^i+1$, $y^i \equiv y^i+1$,
and take the holomorphic 1-forms to be
\beq \label{torus}
dz^i=dx^i+ \kappa^{ij} dy^j \ .
\eeq
The matrix $\kappa^{ij}$ specifies the complex structure. The holomorphic
3-form is
\beq \label{oxsimple}
\Ox = dz^1 \wedge dz^2 \wedge dz^3 \, .
\eeq
The basis
$(\alpha_K,\beta^L)$ from Eq. (\ref{norm3forms}), where
$K=0,..,9$ is taken to be
\bea
\alpha_0 &=& dx^1 \wedge dx^2 \wedge dx^3 \ , \nn\\
 \alpha_{ij} &=& \frac{1}{2} \epsilon_{ilm}
dx^l \wedge dx^m \wedge dy^j \ ,  \\
\beta^{ij} &=& -\frac{1}{2} \epsilon_{jlm}
dy^l \wedge dy^m \wedge dx^j \, \nn \\ 
\beta^0 &=& dy^1 \wedge dy^2 \wedge dy^3 \ . \nn
\eea
The holomorphic 3-form $\Ox$ in (\ref{oxsimple})
is given in this basis by
\beq 
\Omega=\alpha_0 + \alpha_{ij}\ka^{ij}  - \beta^{ij}
(\cof \ka)_{ij} + \beta^0(\det \ka) \ ,
\eeq
where
\beq
(\cof\ka)_{ij}\equiv (\det\ka) \ka^{-1,{\rm T}} = {1 \over
2} \,
\epsilon_{ikm} \epsilon_{jpq} \ka^{kp} \ka^{mq} \ .
\eeq
The NS 3-form fluxes along these 3-cycles  are denoted
$e_0$, $e_{ij}$, $m^0$ and $m^{ij}$ (see Eq. \ref{HF}),
and similarly for the RR fluxes, adding a subindex ``RR''.
The number of units of flux is constrained by 
the tadpole cancellation condition (\ref{tadpoleB}).
In $T^6/{\mathbb Z}_2$, there are $2^6$
O3-planes, giving a negative contribution
of $-16$ to the tadpole (\ref{tadpoleB}), leading
to the following condition 
\beq \label{tadpoleT6}
N_{\Dt} + \frac{1}{2}  \frac{1}{(2\pi)^4 \ax'^2} \int_{T^6} 
H_3 \wedge F_3 = N_{\Dt} + \frac{1}{2}
\left( e_{K} m^K_{\RR}
- m^K e_{K\, \RR} \right)  = 16
\eeq
where the factor $1/2$ comes from the volume of  $T^6/{\mathbb Z}_2$,
which is half the volume of $T^6$.

The type B, KST solution has additionally 5-form flux, and nontrivial
warp factor whose Laplacian is given by (\ref{Laplac}).
The internal manifold is not a torus, but a conformally
rescaled torus. 
However, as usually done in this class of
examples \cite{GKP,FP}, it is argued that at large radius and weak 
coupling one can neglect
the warping (and five-form flux)
and treat the moduli space as if it was that of a Calabi-Yau.  
One should nevertheless demand that the integrated Bianchi identity
for $F_5$ is satisfied, or in other words, satisfy Eq.(\ref{tadpoleT6}).

For the given setup, the GVW superpotential (\ref{WO3})
is 
\beq
W= (m^0_{\RR} - \tau m^0) \det \ka - (m^{ij}_{\RR} - \tau m^{ij})
(\cof \kappa)_{ij} - (e_{ij \, \RR} - \tau e_{ij}) \ka^{ij}
- (e_{0 \, \RR} - \tau e_{0}) \  .
\eeq
The supersymmetry conditions reduce to eleven equations, namely 
\bea \label{susyugly}
W- \partial_\tau W &=& 0 \ \Rightarrow \ 
m^0_{\RR} \det\tau - m^{ij}_{\RR}(\cof\ka)_{ij}
- e_{ij \, \RR}\ka^{ij} - e_{0 \, \RR} = 0  \\
\partial_\ka W &=& 0  \  \Rightarrow  \ m^0 \det\ka - m^{ij}(\cof\ka)_{ij}
- e_{ij}\ka^{ij} - e_0 =0\\
\partial_{\ka^{ij}} W &=& 0  \  \Rightarrow \ (m^0_{\RR}-\tau \,m^0)(\cof\ka)_{kl}
- ( m^{ij}_{\RR} -\tau \,m^{ij})\epsilon_{ikm}\epsilon_{jln}\ka^{mn}
 \nn \\
&& \ \ \ \, \qquad \quad \qquad \qquad \qquad\ \qquad - (e_{ij \, \RR}-\tau\,e_{ij})\delta^i_k\delta^j_l = 0 \ .
\eea
In these equations, the condition $W=0$ is a consequence of demanding 
vanishing F-terms along the K\"ahler moduli, $D_{T_{\alpha}}W=0$
 (see (\ref{kcov1})). When $W=0$, the K\"ahler 
covariant derivatives reduce to ordinary derivatives. 

Eqs(\ref{susyugly}) are eleven complex coupled non linear 
equations for 
ten complex variables, namely the axion-dilaton and
the 9 complex structure moduli $\ka^{ij}$. Generically,
these cannot be solved, and supersymmetry is broken,
or even more, for a given set of fluxes, there might be no solution
at all to the equations of motion. 
Besides (\ref{susyugly}), we should additionally impose the 3-form flux 
to be primitive, i.e.
\beq \label{primit}
J_2 \wedge  G_3 = 0 \, .
\eeq
These are six real equations ($J \wedge G$ is a five-form, 
with six different components) for the nine K\"ahler moduli
$v^{\alpha}$, which means that generically, and
differently from the Calabi-Yau case,  only
three of them remain unfixed. 

KST work out several $\N=1$ vacua. Let us briefly discuss 
one of them. Taking the flux matrices to be diagonal,
namely
\beq \label{flll}
(e_{ij},m^{ij}, e_{ij \, \RR}, m^{ij}_{\RR}) = (e,m,e_{\RR},m_{\RR}) \, \delta_{ij} \ ,
\eeq
the complex structure matrix that solves (\ref{susyugly})
should be proportional to the identity
\beq
\ka^{ij} = \ka \, \delta^{ij} \ .
\eeq
This means that the torus factorizes as $T^6=T^2 \times T^2 \times T^2$
with respect to the complex structure.
$\ka$ turns out to be the root of a third degree polynomial equation where
the coefficients are given by the fluxes. Additionally, 
$\tau$ is fixed by he third
equation in (\ref{susyugly}). Let us take for example the set of fluxes
\beq \label{setf}
(e_0,e_{ij},m^{ij},m^0,  e_{0 \, \RR},
e_{ij \, \RR}, m^{ij}_{\RR}, m^{0}_{\RR} ) = (2,-2 \delta_{ij},-2 \delta^{ij},
-4, 2,0,0,2) \ .
\eeq
These fluxes contribute $N_{\rm{flux}}=12$ to the tadpole cancellation
condition (\ref{tadpoleB}, \ref{Nflux}), and therefore for consistency 
we should add 4 D3-branes.
KST showed that the complex structure modulus and the axion-dilaton moduli
are fixed for these fluxes at $\ka=\tfrac12 \tau =e^{\frac{2\pi i }{3}}$. 

Generically six of the K\"ahler moduli are fixed 
by the primitivity condition (\ref{primit}).
However, in the example
at hand, KST showed that only 3 are fixed, and we are left with
6 flat directions given by the 3 radii of the $T^2$'s ($J_{i \bi}$), plus 
the components $J_{1 \bar 2}+J_{2 \bar 1}$, and the same
for ${1,3}$ and ${2,3}$.

Note that the coupling constant $g_s$ in this example is fixed 
at a value where perturbative corrections are important,
namely $g_s=1/\sqrt 3$. It is possible however to chose fluxes such that
$g_s$ is fixed in the perturbative regime.

\subsection{Moduli stabilization in type IIA Calabi-Yau orientifolds}
 \label{sec:modCYIIA}

In this section we discuss stabilization of moduli in 
type IIA Calabi-Yau (O6) orientifolds.
As already mentioned, there are two main differences
between 
the mechanisms of moduli stabilization by fluxes in type IIA
and type IIB. The first difference is that the potential
in type IIA depends on complex structure as
well as K\"ahler moduli, while its IIB counterpart
depends on the complex structure moduli, 
but not the K\"ahler moduli. This means that a priori fluxes alone
could fix all the moduli in IIA. But most importantly,
in IIA there are no vacua (neither Minkowski,
nor Anti-de-Sitter) that involve a 
Calabi-Yau manifold (or better a conformal Calabi-Yau,
as is the case of type B solutions in IIB). Even more,
one can see from Table \ref{ta:IIA} that Minkowski
vacua in IIA involving manifolds of SU(3) structure
have either $W_3$ or $W_2$ non-zero, or in other
words, are either non-K\"ahler, or non-complex.
Besides, Ref. \refcite{BeCvIIA} showed that Anti-de-Sitter
vacua with internal manifolds of 
SU(3) structure are only possible with ``type BC'' supersymmetries
(i.e., those in the second column of Table \ref{ta:IIA}), with $a=b e^{i \beta}$),
and have non-vanishing $W_1$ (besides non-vanishing $W_2^+$ 
when there is 2-form flux in the {\bf 8} representation),
which means that they are not complex.
$W_1$, as well as the mass parameter of type IIA and the singlets
in 4-form flux and NS flux, are all proportional
 to
the cosmological constant.
In summary, Minkowski or AdS vacua involving
manifolds of SU(3) structure are far from being
Calabi-Yau: they are either non-symplectic
($W_3 \neq 0$) or non complex 
($W_1\neq 0$ or $W_2 \neq 0 $ or both). This means
that using the Calabi-Yau orientifold superpotential
\ref{WO6} to determine vacua is just not correct\footnote{We remind the reader 
that in IIB, one of the possible Minkowski
vacua
(type B) involves a conformal CY, and it was argued in
Refs.\refcite{DWG,GiMa} that the conformal factor does not
enter the GVW superpotential, which remains that of Eq. \ref{WO3}.}.
One should consider instead the general superpotential for
SU(3) structure manifolds, Eq. (\ref{WSU3}). 
Nevertheless, as Ref.~\refcite{WGKT}
argues, one can use the Calabi-Yau superpotentials and K\"ahler 
potential to attempt to
determine the vacua and dynamics in terms of the properly corrected
superpotential and K\"ahler potential.
In any case, the reason why we review moduli stabilization
mechanism in type IIA Calabi-Yau orientifolds is to
understand the next subsection, which deals with tori:
when the SU(3) structure is broken to SU(2) or further\footnote{We could also say that SU(3) structure is ``enlarged''
to SU(2), depending on the point of view, namely whether one 
looks at the number of generators
of the group, or the number of invariant spinors.},
a conformally Ricci-flat space 
is a possible vacuum of IIA \cite{BeCvIIA}. Furthermore, when
there are no non-trivial 1-forms on the manifold, i.e.
when $h^1=0$, their moduli spaces 
are the same as those
of a Calabi-Yau (or Calabi-Yau orientifold), and
moduli stabilization works in the same way as for 
Calabi-Yau's (with the caveat that again for discussing
moduli, we will have to 
neglect the effect of the warping --or conformal factor--).
The example studied in Ref.~\refcite{WGKT}, namely a 
 $T^6/({\mathbb Z}_3 \otimes {\mathbb Z}_3)$, is 
precisely of this type.

After this long discussion of the differences between
the IIA and IIB flux induced superpotentials and
moduli stabilization mechanisms, let us review
the technicalities of IIA moduli stabilization
on Calabi-Yau orientifolds, following Ref.~\refcite{WGKT}.

The superpotential for 
Calabi-Yau O6 is given in (\ref{WO6}), and the proper
K\"ahler coordinates are given in Eqs. (\ref{tdef}) and (\ref{complexstrO6}). 
Supersymmetric
vacua are given by the conditions 
$ D_I W=0$, for $I=(N^{\kappa},T_{\lambda}, t^a)$.  
The conditions $D_{N^{\kappa}} W= 0, D_{T_{\lambda}}W=0 $ give
\bea \label{dercs}
e_{\lambda} + 2 i e^{2D} W \I(C \F_{\lambda}) &=&0 \ , \nn\\
m^{\kappa} + 2 i e^{2D} W \I(C Z^{\kappa}) &=&0
\eea
where $e^{2D}=6 e^{2\phi}/(\int J \wedge J \wedge J)$,  
is a real function of the dilaton and 
K\"ahler moduli $t^a$. Given that $C$ and $D$ are
real, vanishing of the imaginary part of (\ref{dercs}) 
implies that the real part of the superpotential 
is zero:
\bea \label{idercs}
\R W&=&\tilde \xi_{\lambda} m^{\lambda} - \xi^{\kappa} e_{\kappa}  +
 \nn \\
&& \R \left(e_{0 \, \RR} + 
e_{a \, \RR} t^a+ \tfrac12 \K_{abc} m_{\RR}^a t^b t^c + \tfrac16 m_{\RR}^0 \K_{abc} 
t^a t^b t^c
\right)=0  \ .
\eea
This single equation is the only condition involving the axions,
which means that only one combination of them is fixed by
the fluxes. Therefore, the remaining $h^{(2,1)}$ axions are not fixed
by the fluxes, and have to be stabilized by 
other (non perturbative) mechanisms. 
These are nevertheless the only moduli
that cannot be stabilized by fluxes in type IIA
Calabi-Yau orientifolds.
The real parts of (\ref{dercs}) say that if any NS flux
is non-zero, then $Im W$ has to be nonzero. Given that
$Im W \neq 0$, the real parts of (\ref{dercs}) are
$h^{(2,1)}$ real equations that generically fix the
$h^{(2,1)}$ (real) complex structure moduli in terms of the dilaton. 
Note that there are $h^{(2,1)}+1$ real NS fluxes, so
we did not expect to have the $ h^{(2,1)} + 1$
(real) complex structure at the same time as the
$ h^{(2,1)} + 1$ axions fixed by the NS fluxes.

The K\"ahler moduli $t^a$ appear in the RR piece
of the superpotential. Its K\"ahler covariant derivative
$D_{t^a} W$ also splits into a real and imaginary part.
Since the K\"ahler potential for $t^a$ depends only on their
real part, $v^a$, the imaginary part of the K\"ahler covariant
derivative contains just the regular derivative. To be more explicit,
we have to impose  
\beq \label{kastaIIA}
\I (D_{t^a} W) = \I (\partial_{t^a} W)=0 \ \Rightarrow b^a=- \frac{m^a_{\RR}}{m^0_{\RR}} \ .
\eeq
All the moduli $b^a$ are therefore fixed\footnote{The case  
$m^0_{\RR}=0$ is not interesting as 
in that case all other RR fluxes must vanish as well,
therefore leaving all K\"ahler moduli unfixed, or $v^a$ are 
driven to zero.}. The real part of the
K\"ahler covariant derivative gives, after some algebraic manipulations,
$h^{(1,1)}_-$ simple quadratic equations for the $h^{(1,1)}_-$
moduli $v^a$, which are therefore generically fixed. Finally
Ref. \refcite{WGKT} showed that the dilaton also gets
stabilized. 

In summary, in massive Type IIA Calabi-Yau O6 compactifications with
fluxes, enforcing $D_I W =0$, for $W$ given by 
(\ref{WO6}), leads to an $AdS_4$ supersymmetric
vacuum with all K\"ahler moduli
$(v^a,b^a)$ generically stabilized; all complex structure moduli
and dilaton $(Re (Z^k), Im (\F_{\lambda}))$ stabilized, but 
only one combination  of the axions $(\xi^{\kappa},\tilde \xi_{\lambda})$
fixed, while the remaining $h^{(2,1)}$ stay massless.

\subsection{Moduli stabilization in type IIA orientifolds of tori} 
\label{sec:modtoriIIA}

In this section, we illustrate the mechanism of moduli stabilization 
in type IIA orientifolds of tori with
a specific example: an orientifold (O6) 
of the orbifold $T^6/({\mathbb Z}_3 \otimes {\mathbb Z}_3)$,
constructed in Ref. \refcite{WGKT}. 
We will concentrate
on supersymmetric vacua, but we note that Ref.~\refcite{WGKT}
considers additionally stabilization of moduli in non supersymmetric
cases, by inspecting the minima of the potential. At the end of the section,
we mention briefly other constructions of O6 orientifolds of twisted tori
(i.e., manifolds of trivial structure but not trivial holonomy).

As stressed in the previous section, the back reaction of the fluxes
in IIA allows for vacua involving  conformally Ricci-flat manifolds 
(or orbifolds/orientifolds thereof)
only when the structure group is more reduced than SU(3). 
This is the case of tori, whose structure group is trivial,
and, as we will see, can support supersymmetric fluxes.
Furthermore, as also stressed in previous sections,
if $h^1=0$, as is the case for  
the orbifold $T^6/({\mathbb Z}_3 \otimes {\mathbb Z}_3)$,
the moduli spaces and moduli fixing mechanisms 
work as in Calabi-Yau manifolds.

We will show, following Ref.~\refcite{WGKT}, that differently
from the case of type B compactifications, 
fluxes fix all moduli (since $h^{(2,1)}=0$
for this orbifold), and they can do it at arbitrarily large 
volume and weak coupling. 
Let us review how this magic works.

The orbifold $T^6/({\mathbb Z}_3 \otimes {\mathbb Z}_3)$ is constructed
as follows: the torus is parameterized by 3 complex coordinates $dz^i=
dx^i+i\, dy^i$ (the ${\mathbb Z}_3$ action does
not leave any freedom in the choice of complex structure),
with the periodicity condition
\beq
z^i \equiv z^i+1 \equiv z^i + \alpha
\eeq
with $\alpha=e^{\pi \, i /3}$. The two $\mathbb Z_3$ actions 
are given by
\bea
T: (z^1,z^2,z^3) & \rightarrow & \alpha^2 (z^1,z^2,z^3) \nn\\
Q: (z^1,z^2,z^3) & \rightarrow &
(\alpha^2 z^1 +\frac{1+\ax}{3},\alpha^4 z^2 +\frac{1+\ax}{3},
 z^3+ \frac{1+\ax}{3} ) \ .
\eea
$T$ 
has 27 fixed points, while $Q$, is freely acting.
The result is a Ricci-flat manifold (with curvature concentrated
at 9  ${\mathbb Z}_3$ singularities), and Euler number
$\chi=24$. The Euler number being non zero implies that there are no
nowhere vanishing vectors,
i.e. $h^1=0$, which
in turn makes the whole Calabi-Yau moduli business work 
for this case.  The other Hodge numbers are $h^{(2,1)}=0$,
$h^{(1,1)}=3_u + 9_t$ (meaning 3 untwisted moduli and 9 twisted moduli).
The twisted 
moduli are localized at the singularities, and correspond to 
blow-up modes. In this review article we will not discuss
twisted moduli and twisted moduli stabilization. We refer
the reader to Ref.~\refcite{WGKT}
for the discussion of blow up mode stabilization (see also
Refs. \refcite{CU,LMRS}).

The O6 projection is given by $(-1)^{F_L} \Ox_p \sigma$, 
for the involution $\sigma$
\beq
\sigma : z^i \rightarrow -\bar z^i \ \Rightarrow \ x^i \rightarrow -x^i \ , \ 
y^i \rightarrow y^i 
\eeq
The $h^{(1,1)}_u=3$ untwisted normalized two- and four-forms 
and the $h^3= 2 h^{(2,1)}+ 2=2$ normalized three-forms are
\bea
w_i &=& 2 (3)^{1/6} dx^i \wedge dy^i = i (3)^{1/6} dz^i \wedge d\bar z^i
 \qquad \qquad  \quad   \qquad \qquad  \quad \ \ \, \, \ \, (-) \nn \\
\alpha_0&=& (12)^{1/4} \left(dy^1 \wedge dy^2 \wedge dy^3 - \frac{1}{2} 
 \epsilon_{ijk} dx^i \wedge
dx^j \wedge dy^k\right) \ \  \qquad \qquad   (+)  \nn\\
\beta^0 &=& (12)^{1/4} \left(dx^1 \wedge dx^2 \wedge dx^3 - \frac{1}{2}\epsilon_{ijk} dy^i \wedge
dy^j \wedge dx^k \right) \ \  \qquad \qquad  (-) \nn\\
\tilde w^i&=& \frac{4}{3}(dx^j \wedge dy^j) \wedge (dx^k \wedge dy^k)=
-\frac{1}{3}  (dz^j \wedge d\bar z^j) \wedge (dz^k \wedge d\bar z^k) 
  \ \ \ \, (+)
\eea
where in parenthesis we have indicated the parity under $\sigma$,
and we have used $\int_{T^6/({\mathbb Z}_3)^2} dx^1 \wedge 
 dx^2 \wedge  dx^3 \wedge  dy^1 \wedge  dy^2 \wedge  dy^3 =\tfrac{1}{8 \sqrt 3}$.
The O6 is wrapped along $A_0$, the cycle dual to $\beta^0$.   
The holomorphic 3-form $\Ox$ is given by 
\beq
\Ox= \frac{1}{\sqrt 2} (\ax_0 + i \beta_0 )= i (3)^{1/4} dz^i \wedge dz^2 \wedge
dz^3 \ .
\eeq

For this orientifold, Table \ref{ta:IIAOmult} tells
us that the untwisted moduli are a total of 8 real scalars:
6 of them are in 3 chiral 
multiplets $t^i=b^i+iv^i$, $i=1,2,3$
(cf. Eq.(\ref{tdef})), and the remaining two 
are the dilaton (written in the table as $Re Z^0$)
and an axion $\xi^0$ coming from $C_3$
along $\alpha_0$ (cf. Eq.(\ref{complexstrO6})).

Now we want to turn on fluxes on this orientifold. 
The NS flux $H_3$ should be odd under $\sigma$,
which implies that it should be along $\beta^0$. The RR fluxes
$F_0$ and $F_4$ should be even, while $F_2$ and
$F_6$ are odd. We can therefore turn on the following fluxes
\beq
H_3 = -e_0 \beta^0 \ , \quad
F_0 = m^0_{\RR} \ , \quad F_2=m^i_{\RR} w_i \ , \quad
F_4=-e_{\RR \, i}\tilde w^i \ , \quad
F_6=e_{\RR \, 0} \ .
\eeq
The tadpole cancellation condition (\ref{tadpole6}) 
enforces
\beq \label{tadi}
 m^0_{\RR} e_0 = -2 \ .
\eeq  
This means that the NS flux and the mass parameter
are basically fixed by the tadpole (up to four choices
$\pm(1,-2)$ or $\pm(2,-1)$), but we are free to add
any number of $F_2$, $F_4$ and $F_6$ fluxes.
We can use (\ref{tadi}) to write the full solution in terms
of RR fluxes only. 
%Ref.~\citeref{WGKT} argue that vacua with nonzero 
%$F_2$ have qualitatively the same behavior as those
%with $F_2=0$, and therefore study this latter case.

From  $\I(D_{t^a} W)=0$, Eq.(\ref{kastaIIA}) we know that the K\"ahler moduli 
$b^a=Re(t^a)$ are stabilized at the ratio of the mass parameter
and the two-form flux, namely
\beq \label{b^i}
b^i= -\frac{m^i_{\RR}}{m^0_{\RR}} \ .
\eeq
We said that the condition $\R(D_{t^a} W)=0$ gives $h^{(1,1)}_-$  
quadratic equations for the moduli $v^a=Im(t^a)$. Ref.\refcite{WGKT}
showed that the solution to the $h^{(1,1)}_-=3$  equations 
that one gets for the  $T^6/{\mathbb Z}_3^2$ are \footnote{The factor
 81 comes from the triple intersection number $\K_{123}=81$.}
\beq \label{v^i}
v^i= \frac{1}{9 |\hat e_{\RR \, i}|} 
\left(5\frac{\hat e_{\RR \, 1} \, \hat e_{\RR \, 2}
 \, \hat e_{\RR \,3}}{m^0_{\RR}}\right)^{1/2} \ , \quad 
\hat e_{\RR \, i} \equiv e_{\RR \, i} - 81 \frac{m^j_{\RR} m^k_{\RR}}{m^0_{\RR}} \ .
\eeq

We showed that the complex structure equation $D_{N_{\kappa}}W=0$ 
splits into a real and an imaginary part. The imaginary part
stabilizes the dilaton at \footnote{We are setting here and in the
next equation the twisted fluxes to zero}
\beq \label{dilf}
e^{-\phi}=\frac{12 \sqrt 3}{5}  (v^1 v^2 v^3)^{1/2} = 
\frac{4 \sqrt 3}{241  |m^0_{\RR}|} \left(5\frac{\hat e_{\RR \, 1} \, \hat e_{\RR \, 2}
 \, \hat e_{\RR \,3}}{m^0_{\RR}}\right)^{1/2}\ .
\eeq 
We argued that the real part of $D_{N_{\kappa}}W=0$, 
Eq. (\ref{idercs}) implies that only one axion $\xi$
is fixed by the fluxes. In the example at hand 
there is only one axion, bingo!. It is stabilized at
\beq \label{axf}
\xi^0= - 2  \left( \frac{e_{\RR \, 0}}{m^0_{\RR}}
+ \frac{e_{\RR \, i} m_{\RR}^i}{(m^0_{\RR})^2}
- 162 \frac{m^1_{\RR} m_{\RR}^2 m_{\RR}^2 }{(m^0_{\RR})^3} \right) \ .
\eeq

In summary, the orientifold of the
$T^6/({\mathbb Z}_3 \otimes {\mathbb Z}_3)$ orbifold
worked out in Ref.~\cite{WGKT} has all moduli stabilized. 
The K\"ahler moduli
$t^i=b^i+iv^i$ are stabilized at the values given by
Eqs. (\ref{b^i})-(\ref{v^i}), and the fixed dilaton-axion
is given in (\ref{dilf})-(\ref{axf}). Let us make a 
few very important comments about the solution.
First, this example with trivial structure has the same
moduli as a Calabi-Yau because $h^1=0$. Second,
all moduli can be fixed thanks to the property
$h^{(2,1)}=0$, otherwise there would be
 $h^{(2,1)}$ axions unfixed. 
Third, all the vacuum expectation values of the scalars depend
on the mass parameter of type IIA $m^0_{\RR}$, which by the
tadpole cancellation condition (\ref{tadi}) is fixed to
be $\pm 1$ or $\pm 2$. Last but not least, 
the flux parameters $e_{\RR \, i}$ can be anything,
since they do not enter the tadpole cancellation
condition (neither do the $F_2$  fluxes $m^i_{\RR}$,
which give the same qualitative behavior).
If we take all $\hat e_{\RR \, i} \sim N$, the radii
and the dilaton are stabilized at 
\beq
\hat e_{\RR \, i} \sim N \ \Rightarrow \ \sqrt{v^i} \sim R \sim N^{1/4}
 \ , \ g_s \sim N^{-3/2} \ .
\eeq
Therefore, the stabilization can be done at
arbitrarily large radius of compactification and 
weak coupling.
Furthermore, inserting the stabilized moduli 
in the superpotential (\ref{WO6}), we get
that the cosmological constant is parametrically 
small, namely
\beq
\Lambda = -3 e^{K} |W|^2 \sim N^{-9/2} \ ,
\eeq
where we have used 
$\rm{exp}(-2\, \ln [\int \R(C \Ox) \wedge * \R (C \Ox)])=
e^{4 \phi}/(\rm{vol})^2$ (see Refs.\refcite{LGIIA,WGKT}
for details).

Let us analyze these vacua from the point of view of
supersymmetry conditions discussed in section \ref{sec:N=1}.
First of all,  
the supersymmetric vacua are all AdS, since the
superpotential is not zero. The NS flux has obviously
only a singlet component, proportional to $i (\Ox - \bar \Omega)$.
Then, setting the
two-form RR fluxes $m^i_{\RR}=0$ for convenience,
it is not hard to show using (\ref{v^i}) that
the four-form flux $F_4$ 
is proportional to 
$J^2 \propto \epsilon^{ijk} e_{i \, RR}\, w_j \wedge w_k $, 
namely it is also
in the singlet representation. Therefore, this
solution contains only singlets in the fluxes, 
which give a parametrically small cosmological constant, but not
a torsion class $W_1$. This singlets are nevertheless
not that innocent, as they allow us to fix all moduli, and
in a region were supergravity approximation holds.

Needless to say again how differently moduli stabilization
 works in type IIA and type IIB! Even if needless, let us stress
again the two main differences:
first, in IIB, the fluxes that stabilize moduli enter the tadpole cancellation,
so we are not free to make them as large as we want. Secondly,
in IIB Calabi-Yau orientifolds, no K\"ahler moduli are fixed.
This is better in the case of tori, where some K\"ahler moduli are
fixed by the primitivity condition $J \wedge G=0$.   
However, there will always be at least one unfixed K\"ahler modulus, 
the overall volume, since
as GKP showed, all the type B conditions are invariant 
under rescalings of the metric. In type IIA the unfixed
moduli are all but one axionic partners of the complex structure moduli
and dilaton \footnote{This fact has even been regarded as a ``blessing'' 
in Ref. \refcite{IFC}, since the unfixed axions give masses to potentially
anomalous U(1) brane fields in SM-like constructions.}
In manifolds with a rigid complex structure, as the one discussed
in this section, there is only one axion, and therefore all
the moduli are fixed. All this agrees with the results from section
\ref{sec:fluxpot}, summarized in Table \ref{ta:potentials}.

Other discussions of closed moduli stabilization in IIA tori 
are given in Refs. \refcite{DKPZ,IFC}. Ref. \refcite{DKPZ} 
studied moduli stabilization in O6 orientifolds of the
orbifold 
$T^6/({\mathbb Z}_2 \otimes {\mathbb Z}_2)$
including torsion (usually called metric fluxes), and found
by exploiting the underlying gauged supergravities, 
that all untwisted moduli can in principle 
be stabilized by fluxes and torsion. 
Ref. \refcite{IFC} studied moduli stabilization in general
orientifolds of Calabi-Yau twisted tori (with torsion, or ``metric
fluxes''), by analyzing the superpotential (\ref{WO6t}).  
They found that some axions remain unfixed in Minkowski vacua, while
all moduli can be stabilized in some AdS ones.

\section{Moduli stabilization including non perturbative effects and De Sitter vacua} \label{sec:nonpert}

We saw in the previous sections that fluxes are usually
not enough to stabilize all moduli. In particular, we
reviewed in section \ref{sec:moduliconifold} that in
type IIB compactifications on Calabi-Yau orientifolds, 
fluxes stabilize the complex structure moduli and the dilaton, but
leave K\"ahler moduli unfixed. There are nevertheless perturbative 
and non 
perturbative corrections to the leading order K\"ahler potential and
superpotential considered in previous sections 
that can help in stabilizing the remaining
moduli.  
In this section we discuss these
corrections, concentrating on their effect on 
moduli stabilization, and 
whether they lead to de Sitter vacua.
We will mainly focus on type IIB compactifications
on Calabi-Yau orientifolds.

\subsection{Corrections to the low energy action} \label{sec:corrections}

Corrections to the low energy effective
supergravity action are governed by the Planck scale,
which in string theory is given by $M_P^8=\frac{1}{g_s^2 (\alpha')^4}$.
In the low energy limit, the dimensionless parameter 
$\tfrac{l_P}{R}$, where $R$ is a characteristic length of the solution,
controls the corrections. One then thinks of the corrections
as a double series expansion 
in $g_s$ and $\alpha'$. There are perturbative and 
non perturbative corrections to the supergravity action.
The non perturbative arise from world-sheet or brane instantons.
A world-sheet or a p-brane
wrapping a topologically non-trivial
space-like 2-cycle or p-cycle $\Sigma$ on the internal manifold gives
instanton corrections which are 
suppressed by $e^{-\frac{Vol(\Sigma)}{2 \pi \alpha'}}$.
This will be the main effect stabilizing the K\"ahler moduli.
As we will briefly discuss, 
the number of fermion zero modes on the instanton world-volumes  
dictates whether these corrections are there or not.

Let us discuss the perturbative corrections in
the case of $\N=1$ compactifications, concentrating on
compactifications of IIB on Calabi-Yau orientifolds.
The ten dimensional supergravity action 
is corrected by a series of $\ax'$ terms,
coming from higher derivative terms in the action:
\beq \label{alphaexp}
S= S_{(0)} + \alpha'^3 S_{(3)}
+ ...+ \alpha'^n S_{(n)} + ... +S^{CS}_{(0) } + S_{(0)}^{\rm{loc}} +  
\alpha'^2 S_{(2)}^{\rm{loc}}  \ ,
\eeq
where $S^{CS}$ are the Chern-Simons terms, and 
$S^{\rm{loc}}$ the localized p-brane actions.
In addition to higher derivative corrections
contributing to (\ref{alphaexp}),
there are string loop corrections to the action, 
both for the bulk and for the localized 
pieces, whose effect is less known.  They are suppressed
by powers of $g_s$. Some of these corrections were
computed for IIB orientifolds of various tori  with
D5/D9 and D3/D7 branes in Refs.~\refcite{BHK,strloop} (see also
Ref.~\refcite{AMTV} for the case of $\N=2$ compactifications).
String loop corrections to the bulk effective
action appear nevertheless at 
order $\alpha'^3$, so their effects 
are subsumed in the expansion (\ref{alphaexp}),
as a further $g_s$ expansion of each term.

The term $S_{(3)}$ contains ${\mathcal R}^4$ corrections to the
action (where ${\mathcal R}$ is the Ricci scalar), 
as well as ${\mathcal R}-F_p$ terms mixing flux and
curvature.
These corrections break the no-scale structure
of the flux generated potential, both at string tree level \cite{BBHL}
and at one loop \cite{BHK}. 
Considering the scalings
of all possible $\alpha'^3$ correction to the bulk type IIB
action compactified on a Calabi-Yau orientifold 
in the presence of fluxes, Refs.~\refcite{BBCQ,Quevedo} conclude
that the leading term
has a scaling ${\mathcal O}(\alpha'^3/R^6)$
relative to the zeroth order term, in agreement with the result
of Ref.~ \refcite{BBHL}.

Higher derivative corrections to the localized sources,
$\alpha'^n S_{(n)}^{\rm{loc}}$  
lead to a potential energy in the case of D7-branes,
but not D3-branes \cite{Quevedo}.
The $\ax'^2$ correction to the D7-brane action
gives their effective D3-brane charge and tension \cite{GKP}.
In F-theory this effective charge is given in terms of the
Euler number of the fourfold by $Q_3^{D7} = - \frac{\chi}{24}$.
This adds the constant term in the 
F-theory version of the tadpole cancellation condition,
 Eq.(\ref{tadpoleF}).
 Higher $\ax'$ corrections to the D7-brane action do not lead to
potential energy.
%Finally,  non perturbative corrections to the localized action
%are the main cause
%of K\"ahler moduli stabilisation, and will therefore be discussed
%in more detail in next section.

The corrections just outlined lead to corrections to the 
four dimensional K\"ahler potentials and 
superpotentials. 
The $\N=1$  K\"ahler potential receives corrections at every order
in perturbation theory, while the superpotential
receives non-perturbative corrections only. Considering
leading $\alpha'$ corrections, the K\"ahler potential
and superpotential can be written
\bea \label{corr}
K&=&K_{0}+K_{\rm{p}}+K_{\rm{np}} \nn \\
W&=&W_{0}+W_{\rm{np}} \ .
\eea
where $K_{\rm{p}}$ comes from the corrections discussed above, 
$K_{\rm{np}}$  comes mainly from fundamental string
wordlsheet instantons, and $W_{\rm{np}}$
comes from string non perturbative effects
such as  D-brane instantons (or similarly, from
gaugino condensation on D-branes).

Refs.~\refcite{BaBe,BBCQ,Quevedo}
analyze the effect of of these corrections on the 
potential (\ref{N=1pot}), namely  
\beq
V=V_0 + V_{K_{\rm{p}}} + V_{W_{\rm{np}}}
+...
\eeq
where
\beq
V_0 \sim W_0^2 \ , \qquad  V_{K_{\rm{p}}} \sim W_0^2 K_{\rm{p}} \ , 
\qquad  V_{W_{\rm{np}}} \sim  W_{\rm{np}}^2+ W_0  W_{\rm{np}} \ .
\eeq
As discussed above, in four dimensional supergravities arising from
compactifications of type II theories on Calabi-Yau orientifolds
with D-branes, we have more information about
$W_{\rm{np}}$ (coming from field theoretic considerations, 
though) than about the perturbative corrections 
to the K\"ahler potential, $K_{\rm{p}}$.
We wish therefore to see whether there is any regime
in which the latter can be neglected,
and moduli stabilization can be studied
 just including the non perturbative corrections to the
superpotential, as is the case in the KKLT scenario \cite{KKLT}
to be discussed in next section.
If the tree level superpotential $W_0$ vanishes, then the first
contribution to the potential is proportional to
$W_{\rm{np}}^2$, and we can safely ignore $K_{\rm{p}}$.
Similarly, if $W_0  \ll 1$ 
in suitable units, then the tree level superpotential
can have similar magnitude than the non-perturbative
superpotential, leading to 
\beq 
W_0 \sim W_{\rm{np}} \ \Rightarrow
  W_{\rm{np}}^2 \sim V_{W_{\rm{np}}} 
\gg V_{K_{\rm{p}}} \sim  W_{\rm{np}}^2 K_{\rm{p}} \ .
\eeq
This is the relevant behavior for the KKLT scenario to be reviewed
in section \ref{sec:KKLT}.
Finally, when $  \tfrac{W_{\rm{np}}}{K_{\rm{p}}} < W_0 \ll
1$, the perturbative effects dominate, and it is not consistent to neglect
them.
 Considering non perturbative corrections to the 
superpotential while at the same time neglecting 
perturbative corrections
to the K\"ahler potential is therefore justified only
when the flux generated superpotential $W_0$ is zero, or
of the same order of magnitude than the non-perturbative ``correction''. 

\subsection{Non perturbative corrections to the superpotential}

There are two classes of effects that lead to corrections to the 
superpotential depending on the K\"ahler moduli of IIB compactifications: 
gaugino condensation
and D-brane instantons. 

Gaugino condensation arises in D7-branes: in the presence of 3-form flux, 
many or all of the world-volume matter fields acquire masses 
\cite{LustMRS,soft7}. If at the same time fluxes stabilize 
the D7's at coincident locus, gaugino condensation is expected to 
occur at energy scales much lower than
the mass scale, where the low energy dynamics is that of pure
$\N=1$ SYM.

Euclidean D3 branes wrapping 4-cycles 
can also lead to non-perturbative potentials. 
Non-perturbative superpotentials will arise
when the D3-branes in question lift to 
M5-branes wrapping a ``vertical'' divisor $D$
(where vertical means that it wraps the fiber
directions that shrink in the F-theory limit)
and that supports two fermionic zero modes \cite{Witten0}.

The behavior of the non-perturbative 
superpotential with the  K\"ahler moduli
is similar in both cases, gaugino condensation and 
D-brane instantons. Here we will discuss 
the brane instanton case.
For euclidean D3-branes  
wrapping  a four-cycle
dual to $n^{\ax}w_{\ax}$, the non-perturbative superpotential is
is given by \cite{DDF,Kanp,BeMa}
\beq \label{Wnp}
W_{\rm{np}}=B_{n} e^{-2 \pi n^{\alpha}T_{\alpha} } 
\eeq
where $B_{n}$ are one loop determinants that depend on the
expectation values of the complex structure moduli, and
$T_{\alpha}$ are the K\"ahler moduli defined in  (\ref{tau}),
whose real part gives the volume of the cycle $[w_{\ax}]$
(we are taking here $h^{(1,1)}_-=0$). The non-perturbative
superpotential depends therefore on K\"ahler moduli,
which were absent in the flux induced GVW superpotential 
(\ref{WO3}). We will see in next section that
taking into account this non perturbative correction
to the superpotential can lead
to Calabi-Yau O3/O7 compactifications with
all moduli stabilized. 

The one loop determinants 
$B_n$ are non-vanishing whenever
the instantons support two fermionic zero modes. 
In the absence of background flux, this is
translated into a condition on the M-theory divisor,
which should have holomorphic Euler characteristic 
(called sometimes arithmetic genus) one,
$\chi(D) = \sum_p (-1)^p h^{(0,p)} (D) =1$. 
It has been argued
recently \cite{TT2,Saulina,fag} 
that this requirement is relaxed in the presence 
of background flux, and branes wrapping divisors 
of holomorphic Euler characteristic $\chi(D) \ge 1$ can contribute
to the non-perturbative superpotential.
For example, Ref.~\refcite{fag} showed
that in type IIB string theory compactified on a
$K3 \times T^2/{\mathbb Z}_2$ orientifold,
 a D3-brane wrapped on $K3$ sitting at the
O7 fixed point on the $T^2$ would have $\chi(D)=2$ in the absence
of fluxes. The presence of  
complex 3-form flux with two legs along
the holomorphic 2-form on $K3$ and 
the third leg along the antiholomorphic direction on $T^2$
(making it overall (2,1) and primitive)
lifts some zero modes,
leaving $\chi (D_{\rm{flux}})=1$, thus allowing
for instanton contributions to the superpotential.

\subsection{Moduli stabilization including non perturbative effects} 
\label{sec:KKLT}

Considering the non-perturbative corrections to the superpotential
discussed in the previous section, all moduli can
be stabilized in type IIB compactifications on Calabi-Yau
orientifolds. The idea was first put up in the seminal paper of 
Kachru-Kallosh-Linde-Trivedi (KKLT) \cite{KKLT}, where it was shown that 
besides having all moduli stabilized in this type of compactifications, 
it is possible to obtain de Sitter vacua by adding a small number of
anti-D3-branes. In this section we will first review how 
the inclusion of a non-perturbative superpotential can lead to 
moduli stabilization,
following KKLT. We will then mention the explicit 
supersymmetric and non supersymmetric examples 
constructed in the literature, and finally, in the next section,
we will review how de Sitter spaces arise in these construction. 

KKLT study IIB flux compactifications on Calabi-Yau O3 orientifolds
with $h^{(2,1)}$ arbitrary, and $h^{(1,1)}_+=1$, $h^{(1,1)}_-=0$, 
i.e. with any number of complex structure moduli but with a single 
K\"ahler modulus
$T$, whose real part is $\R T =\tfrac{1}{2}\cK^{2/3}\sim R^4$ 
(where $\cK$ is the overall volume, $R$ the radius of compactification),
as in Eq.(\ref{KIIBsimple}). 
%\footnote{The Einstein frame volume
%$\cK_E^{2/3}$ has a factor of $e^{-\phi}$ relative to the string frame volume,
%which accounts for a difference of $e^{-\phi}$ in the definition of $T$ 
%given in Ref.~\refcite{KKLT} relative to that in Eq. (\ref{tau}).}. 

3-form fluxes
generate a superpotential for the complex structure moduli and dilaton-
axion. 
Using the quantization conditions
(\ref{quant}) and the fact that the 3-cycles have volumes
$R^3$, the complex structure moduli and dilaton have masses of order 
$m \sim \tfrac{\ax'}{R^3}$. 
KKLT set the complex structure moduli and the dilaton
axion equal to their VEVs, and concentrate then on the effective field theory
for the volume modulus $T$.
In the presence of D3-brane instantons there is a non-perturbative superpotential
for the K\"ahler modulus $T$, Eq.(\ref{Wnp}). A single instanton
will lead to the a total superpotential
of the form (\ref{corr}), namely
\beq \label{Wtot}
W=W_0 + B e^{-2\pi T}
\eeq
where $W_0$ is the contribution coming from the fluxes. Setting
the axion $\rho=0$, and defining $\sigma\equiv Re T= \frac{1}{2} \K^{2/3}$
a supersymmetric minimum satisfying $D_{T}W=0$ is attained at
\beq \label{W0KKLT}
W_0 = -B e^{-2 \pi \sigma_{cr}} \left( 1+ \frac{4\pi}{3} \sigma_{cr}\right) \ ,
\eeq
(to perform the K\"ahler covariant derivative we have used the 
K\"ahler potential (\ref{KIIBsimple})). The volume is therefore 
stabilized at $\cK_{cr}=(2 \sigma_{cr})^{3/2}$, which can take 
reasonably large values for sufficiently small $|W_0|$.

Inserting this in the potential (\ref{N=1pot}), 
we get that the minimum  
leads to an AdS vacuum 
\beq \label{VAdS}
V_{\rm{min}}=-3 (e^K |W|^2)_{\rm{min}}= - 
\frac{2 \pi^2 B^2 e^{-4\pi \sigma_{cr}}}{3 \sigma_{cr}} \ .
\eeq

A few comments are in order. First, note that we need
$W_0 \neq 0 $  for the complete stabilization
to work. In section \ref{sec:moduliconifold}, when we discussed
the supersymmetry conditions, we set
$D_I W_0=0$, as well as $W_0=0$. Looking at Eqs.(\ref{kcov1}),
we see however that if we do not consider derivatives along
the K\"ahler moduli $T_{\ax}$, and take $h^{(1,1)}_-=0$
(i.e., no $G^a$ moduli), the supersymmetry conditions
are just $\int \bar G_3 \wedge \Omega = \int G_3 \wedge
\chi_k = 0$, or in other words, $G_3$ must be
ISD -(2,1) primitive plus (0,3)-. If we consider
a (0,3) piece of the complex 3-form flux, the 
superpotential does not vanish. If we had just 
the flux superpotential, this will 
break supersymmetry, as $D_{T_{\alpha}}W_0 \neq 0$.
However, KKLT have shown that taking into account the non 
perturbative corrections to the superpotential, (0,3)
pieces of the 3-form flux can lead to supersymmetric
(AdS) vacua. This piece has to be fine tuned, though,
to give $e^K |W_0|^2 \ll 1$, otherwise there is no large
radius minimum of the potential. Examples of
flux vacua with $e^K |W_0|^2 \sim 10^{-3}$
and less were constructed in Refs.~\refcite{GKTT,DDF}.
The statistical results,
as we will review in section \ref{sec:stat}, 
suggest that even smaller values are possible,
and conclude that the fraction of vacua having 
$e^{K_0} |W_0|^2 \le \epsilon$ scales like $\epsilon$ \cite{DeDo}.

There are several critiques to the KKLT procedure,
to be discussed shortly. 
These critiques do not affect the main results, but they
do affect the detailed physics, and therefore 
tell us that KKLT should be taken only as a toy model
of complete moduli stabilization in compactifications of IIB 
Calabi-Yau orientifolds.

The first critique is that
the procedure of obtaining an effective potential for 
light moduli via non-perturbative corrections
after integrating 
 out moduli that are assumed to be heavy at the classical level
is in general not correct (see for example 
Refs.\refcite{CFNO,deAl,Quevedo}).
In some cases, this two step procedure can fail, giving rise to 
tachyonic directions \cite{CFNO}.
One should instead minimize the full potential, which has
additional terms (mixing the light and heavy modes). 
This is a highly involved procedure, which has not been 
carried out in the explicit examples
of full stabilization worked out in the literature 
\cite{DDF,DDFGK,AsKa}.

Another critique, outlined in section \ref{sec:corrections},
 is that the corrections to the K\"ahler potential, both 
perturbative and non-perturbative in nature,
have not been taken into account. As reviewed in that section,
Refs.~\refcite{BaBe,BBCQ,Quevedo} show that the $\ax'$ corrections to the
K\"ahler potential are subleading whenever $W_0 \sim W_{np}$.
Otherwise, the perturbative corrections to the K\"ahler
potential dominate, and one should include them in order
to analyze the details of the potential. 
At large volumes, the vev of the GVW superpotential, 
Eq. (\ref{W0KKLT}),
is indeed larger than the non-perturbative one, and
according to Refs.~\refcite{BaBe,BBCQ,Quevedo}, perturbative 
$\ax'$ corrections start to take over. 
Taking into account
the known $\ax'$ correction to the K\"ahler potential
computed in Ref.~\refcite{BBHL}, Refs. \refcite{BaBe,BBCQ,Quevedo}
show that there is a large volume minimum which
for sufficiently small values of $W_0$ coexists 
with the KKLT minimum.

Finally, Ref.~\refcite{DDF} argues that in order to stabilize
the K\"ahler moduli at strictly positive radii, one needs
a sufficient number of distinct divisors, which
excludes the case of internal manifolds with $h^{(1,1)}=1$,
as in the toy model of KKLT.

In the past year, explicit examples of type IIB orientifold 
compactifications were constructed with
all moduli (also twisted and open string moduli) stabilized 
\cite{DDF,DDFGK,AsKa}. 
Ref.~\refcite{DDF} studied Calabi-Yau four-folds with Fano
and ${\mathbb P}^1$ fibered base. 
Out of 92 models with fano base, they found 29 in which
all K\"ahler moduli can be stabilized by arithmetic genus
one divisors. The simplest one 
is ${\mathcal F}_{18}$, which has 89 complex structure
moduli.   
Ref.~\refcite{DDFGK} discusses an orientifold of the
Calabi-Yau resolved orbifold $T^6/\mathbb Z_2^2$,
while Ref.~\refcite{AsKa} studies M-theory on $K3 \times K3$, dual 
to type IIB on $K3 \times T^2/{\mathbb Z_2}$, where the
flux modification to the condition $\chi(D)=1$ plays a prominent role.
We will not review these explicit constructions, but just
remark that 
Ref. ~\refcite{DDFGK} argues that
moduli can be stabilized supersymmetrically 
in the perturbative regime
for reasonable values of $W_0$. 
Furthermore, the corrections to the K\"ahler potential
$K_{p}$, $K_{np}$ coming respectively from
the tree level $\ax'^3$ correction of Ref.~\refcite{BBHL}
and string worlsheet instantons are estimated to be
small. There is no quantitative analysis of string loop corrections,
but Ref.~\refcite{DDFGK}  argues that there are no expected reasons to 
believe that these will destabilize
the solution for the values of $g_s$ found.

Refs. ~\refcite{BBCQ,Quevedo} find that if in addition
to the non perturbative superpotential one considers
the $\ax'^3$ corrections to the K\"ahler potential
of \cite{BBHL} in generic Calabi-Yau
orientifolds with 
$h^{(2,1)}>h^{(1,1)}>1$,  
all geometric moduli can be stabilized, 
and there are non-supersymmetric AdS 
minima at exponentially large volume. 
Taking the example of the orientifold of $\mathbb P^4_{[1,1,1,6,9]}$,
Refs. ~\refcite{BBCQ,Quevedo} show that
as $|W_0|$ increases, the perturbative corrections
dominate the non-perturbative ones, and
including these corrections there is
a large volume minimum which for small values
of $W_0$ coexists
with the KKLT minimum.
Very recently, Ref.\refcite{BeHaCo} argued,
along the lines of Refs. \refcite{strloop,Bobkov},
that 
$\ax'$ corrections to the K\"ahler potential
(tree level and string loop) should be enough
to stabilize all moduli in a IIB orientifold compactification,
without the need of non-perturbative corrections. 
Explicit models with the volume stabilized at large radius
are not yet constructed, though.

\subsection{De Sitter vacua} \label{sec:deSitter}

After having fixed all moduli, KKLT outline the construction of de Sitter vacua.
In order to get de Sitter solutions
from IIB flux compactifications, one should uplift the
AdS vacuum found after having fixed all moduli, as discussed
in the previous section. 
KKLT do this by adding a small number of anti-D3-branes 
at the bottom of the warped throat\footnote{The throat is the 
highly warped region around the D3-brane sources 
where the warp factor $e^{2A}$ is very small, or equivalently
where the conformal
factor multiplying the Calabi-Yau metric, $e^{-2A}$
becomes very large.}. 
Other
uplifting mechanisms involve adding D7-brane fluxes
\cite{BKQ}; starting with a local nonzero minimum of the no-scale
potential (which does depend on the overall volume
through the factor $e^K$)
and expanding around it \cite{SS};
Let us review here the KKLT uplifting procedure.

\begin{figure}[h]
\begin{center}
\includegraphics[height=5cm]{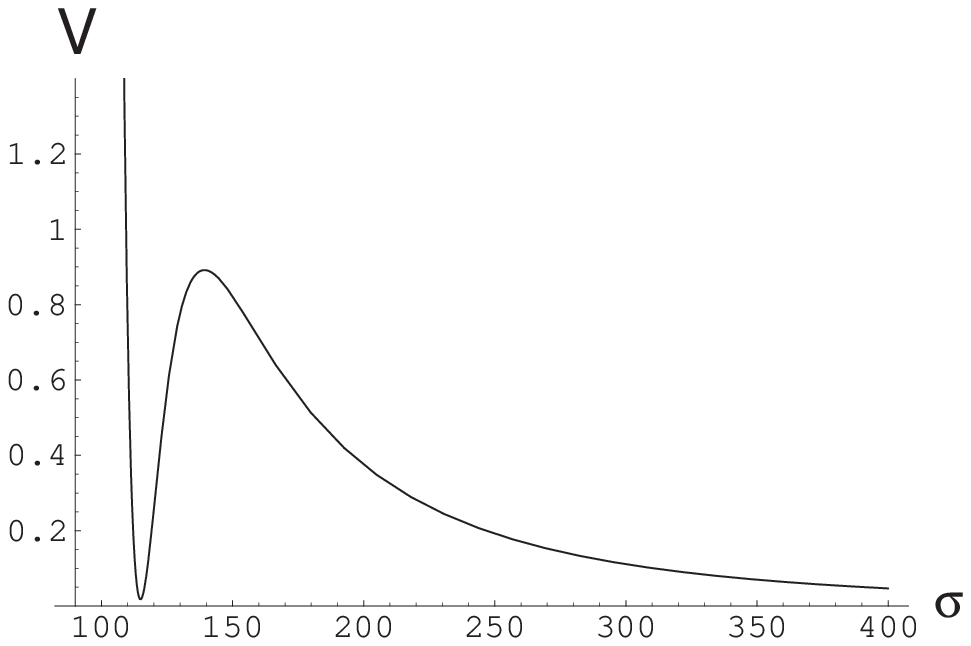}
\caption{\label{fig:ds} 
\text{Potential (\ref{VdS}) 
multiplied by $10^{15}$, taken from Ref.\refcite{KKLT}.}}
\end{center}
\end{figure}

We start by cooking up together all the ingredients of the previous sections,
namely ISD fluxes, D3-branes, orientifold planes and instantons
(or gaugino condensation on D7-branes). Let us assume nevertheless
that the tadpole cancellation condition  (\ref{tadpoleB}) is not satisfied,
and we need a small number of anti-D3-branes to satisfy it.
Anti-D3-branes in the warped geometry created by ISD fluxes,
break supersymmetry explicitly and
do not have translational moduli (see for example Refs.~\refcite{antiD3}),
and are driven to the end of the throat,  where the warp
factor is minimized. The potential energy of such
$\overline{\rm{D3}}$-brane is proportional to $e^{4A}$ at 
the location of the brane,
and 
inversely proportional to the square of the
volume (see for example Refs.~\refcite{antiD3,GGJL}). 
Adding a small number $n$ of $\overline{\rm{D3}}$-branes, there is an
extra contribution to the potential of the previous section,
given by 
%\footnote{The original power of $\R T$ in 
%Ref.\refcite{KKLT} was 3. This was corrected in Ref.\refcite{KKLMMT}
%to be 2 due to an extra power of $\R T$ in the warp factor.}
\beq \label{Vanti}
V_{{\overline D3}}= \frac{D}{\sigma^3}= \frac{D}{(\R T)^3} \ ,
\eeq
where the coefficient $D$ is proportional to
$n$ and $e^{4A}$ at the position of the branes. 
Adding this to the potential 
from the previous section, obtained by inserting the superpotential
(\ref{Wtot}) and K\"ahler potential (\ref{KIIBsimple}) into
the $\N=1$ potential (\ref{N=1pot}), we get
\beq \label{VdS}
V=\frac{B \pi e^{-2\pi \sigma}}{\sigma^2} 
\left( B e^{-2\pi \sigma} \left(1+ \frac{2  \pi}{3} \sigma \right)
+W_0 \right) + \frac{D}{\sigma^3} \ .
\eeq  
This potential has few terms because the no scale structure of
the K\"ahler potential, Eq. (\ref{noscale}), gives the cancellation
$K^{T \bar T} \partial_T K \partial_{\bar T} K |W|^2 - 3 |W|^2=0$. 
There are two extrema of the potential, a local minimum at positive
energy and a maximum separating the de Sitter minimum
from the  vanishing potential at infinity.
By fine tuning $D$, it is easy to get very small positive energy,
at large values of $\sigma$, i.e. at large volume. Figure
\ref{fig:ds}, taken from \refcite{KKLT}
 shows the potential as a function of $\sigma$ 
for $W_0=-10^{-4}$, $B=1$, $D=3 \times 10^{-9}$, and
KKLT take $2\pi=0.1$ in the non-perturbative superpotential (\ref{Wnp}).

%\begin{figure}[pb]
%\centerline{\psfig{file=2.ps,width=5cm}}
%\vspace*{8pt}
%\caption{Potential (\ref{VdS}) (multiplied by $10^{15}$) 
%for $W_0=-10^{-4}$, $B=1$, $D=3 \times 10^{-9}$ and $2\pi=0.1$ 
%in (\ref{Wnp}), 
%taken from Ref.\refcite{KKLT}.
% \label{fig:ds}}
%\end{figure}

Actually, the value of $\sigma$ that minimizes the potential, $\sigma_{cr}$, 
is just slightly shifted from the one that gives the AdS minimum
of the previous section, Eq.(\ref{VAdS}). The effect of  
the last term in (\ref{VdS}), coming from the $\overline{\rm{D3}}$-branes,
 is therefore to lift the AdS minimum without changing too much 
the shape around the minimum.

 The different versions
of the uplifting mechanisms (Refs. \cite{BKQ,SS}) 
differ in the precise shape of the uplifting potential
(\ref{Vanti}), 
but have the same overall behavior.
The breaking of supersymmetry is spontaneous instead of 
explicit, as it is in KKLT. However, there are consistency requirements to combine D-term breaking with a non-perturbative superpotential (see Refs. \refcite{ViZw,Dudas} for details). 

Note that the de Sitter vacuum just obtained is metastable, as there
is a runaway behavior to infinite volume. This is expected
for many reasons. On one hand, it has become clear
on entropy grounds that de Sitter space cannot be
a stable state in any theory of quantum gravity \cite{dSstable}.
On the other hand, 
the runaway behavior
is a standard feature of all string theories \cite{DiSe}. 
Ref. \refcite{GiMy} argues very generically that a positive vacuum energy 
in a space with extra dimensions implies an unstable universe 
toward decompactification.
KKLT showed nevertheless that the lifetime of the dS vacuum
is large in Planck times (it can be longer than the cosmological
time scale $\sim 10^{10}$ years), and shorter than the 
recurrence time $t_r \sim e^S$, where
$S$ is the dS entropy \cite{dSstable,GiMy}.    
Ref.~\refcite{FLW} explored the possible decay channels of the KKLT
de Sitter vacuum, finding, in agreement with KKLT, that even the fastest
decays have decay times much greater than the age of our universe.

Let us note again that due to the critiques discussed
in the previous section, KKLT is a toy model for
getting de Sitter vacua in IIB compactifications.
Differently from the case of AdS vacua,
no explicit models with dS vacua were constructed so far
(some partial constructions can be found for example
in Ref.\refcite{BdA}).
On top of the difficulties already discussed
in stabilizing all moduli
in a controlled way, there is an extra fine tuning needed in order 
to make the constant $D$ sufficiently small, and get at
the same a long lived vacuum.
There are however no fundamental reasons to doubt the
existence of such explicit dS vacua.

Uplifts to de Sitter of IIA rigid Calabi-Yau orientifold flux 
vacua of the type
discussed in section \ref{sec:modtoriIIA} were considered 
in Ref.\refcite{STV} (see also Ref.\refcite{KKP} 
for moduli stabilization in IIA including 
corrections, and possible de Sitter vacua). De Sitter vacua are possible after
taking into account non-perturbative corrections
to the superpotential, and
perturbative corrections to the
K\"ahler potential. The latter give rise to a positive
contribution analogous to the anti-D3-brane one of KKLT. 
Similarly, de Sitter vacua can be obtained
from heterotic M-theory \cite{hetdS}, using background 
fluxes and membrane
instantons / gaugino condensation on the
hidden boundary.
In a very different spirit, 
Ref.\refcite{RiemdS} finds de Sitter vacua
from flux compactifications on products of
Riemann surfaces, while  Ref.\refcite{noncrdS}
finds dS vacua in supercritical string theories.

\section{Distributions of flux vacua} \label{sec:stat}

It is somewhat clear from all the previous sections that the number of
possible string/M theory vacua is very large. Despite
the fact that as of today none of them fully
reproduces the Standard Model data 
(hierarchy of masses, CKM matrix, etc, which in string theory 
vacua depend on moduli vevs),
the hope is that adding sufficient number of ingredients
to the soup, many of them could. 
This raises the ``vacuum selection problem'': among the
very large number (not clear whether it is even finite)
of possible vacua, we have no idea which one is relevant,
and how to find it. If there is no vacuum selection
principle, i.e. no a priori condition that points toward
the right candidate vacuum, the only way to find
our vacuum seems to be by just enumerating all possible
vacua, and
testing each one against present observations. Since all
possible vacua are too many (may be even infinite),
Douglas
and collaborators \cite{Do,AsDo,DeDo,Dostr} have 
advocated for a statistical study of the ``landscape''
of vacua, which could give some guidance principle 
for the search of the right vacuum. Let us review
the basic ideas, and some of their results. \\

Given a set of effective $\N=1$ supergravity theories,
i.e. a set of K\"ahler potentials  $K_i$ and
superpotentials $W_i$, with the same
configuration space (the space in which the moduli
take values),
the first ingredient needed is the density $\rho$ of
(susy or non susy) vacua. Integrating this density
over a region $R$ in configuration space gives the number 
of vacua which stabilize moduli in that region, i.e.
\beq \label{N}
\N_{\rm{vac},R}=  \int_R d^{2n} z \, \rho(z)
\eeq 
where $z$ are the $n$ complex moduli fields. This density 
is given by
\beq \label{rhogen}
\rho(z)= \sum_i \delta^n(V'_i(z)) \, \, |{\rm det} \, V''_i (z)|
\eeq
where $V_i$ is the $\N=1$ potential (\ref{N=1pot})
for $K_i$, $W_i$. If the vacua are supersymmetric,
this can be written in terms of the K\"ahler covariant 
derivatives as
\beq \label{rhosusy}
\rho(z)_{\rm{susy}}= \sum_i \delta^n(DW_i(z)) \, 
\delta^n(\overline{DW}_i(\bar z)) \, \,
  |{\rm{det}} \, D^2 W_i (z)| \ ,
\eeq
where the Jacobian $D^2W$ is a $2n \times 2n$ matrix
\beq
D^2 W = \left( \begin{array}{cc}
D_{\bar I} D_{J} W(z) \  & \ D_{I} D_{J} W(z) \\ 
D_{\bar I} D_{\bar J} \bar W(\bar z) \ &  \ D_{I} D_{\bar J} \bar W(\bar z)
\end{array}
\right) \ .
\eeq
One can also find a density of supersymmetric vacua with a
given cosmological constant, by multiplying (\ref{rhosusy}) 
by $\delta( \Lambda - (-3 e^{K_i} |W_i(z)|^2))$. 

The ensemble of vacua that Douglas and collaborators
consider are flux compactifications of
F-theory on Calabi-Yau four-folds, or their
IIB orientifolds limit. In the case of IIB vacua 
with 3-form fluxes on a Calabi-Yau orientifold,
the K\"ahler potentials $K_i$ are all the same,
Eq.(\ref{KIIBO3}) (or an equivalent expression
with $\tau$ and $z$ together for the four-fold), 
while the superpotentials $W_i$,
given in (\ref{WO3})
 are labeled by 
the $4(h^{2,1}_-+1)$ fluxes $(e_K,m^K,e_{K \, RR},m^{K}_\RR)$.
The tadpole cancellation condition (\ref{tadpoleF})
gives an upper bound for the number of units of flux
(assuming we do not want to introduce $\overline{D3}$-branes),
given by
\beq \label{N*}
N_{\rm{flux}}= e_{K} m^K_{\RR}
- m^K e_{K\, \RR}=
N \eta N 
\le N_* \ , \qquad   N_*= \frac{\chi(X_4)}{24} \ .
\eeq

Let us illustrate this with the simplest example,
considered in Ref. \refcite{AsDo}, of a rigid Calabi-Yau 
3-fold ($h^{2,1}=0$). The only modulus of the theory appearing the
flux superpotential is
$\tau$, which is stabilized by the GVW superpotential (\ref{WO3})
\beq
W=(m^0 F_0 - e_0 Z^0) \tau + 
(-m^0_{\RR} F_0 + e_{0 \, \RR} Z^0 ) \equiv C \tau + D \ .
\eeq 
The constants $Z^0=\int_A \Ox;\, F^0 = \int_B \Omega$ are determined
by the geometry of the Calabi-Yau.

\begin{figure}[h]
\begin{center}
\includegraphics[height=5cm]{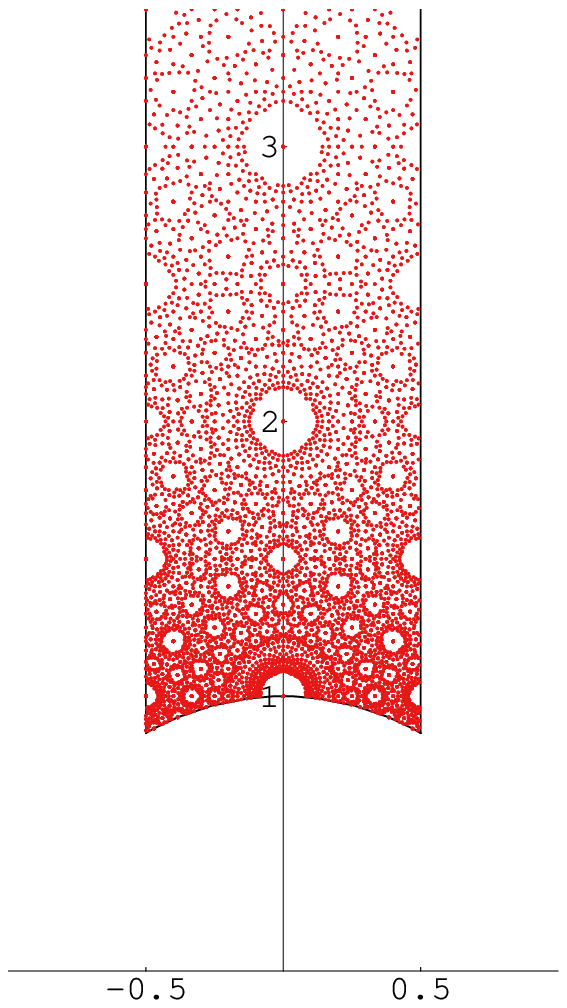}
\caption{\label{fig:taus} 
\text{Values of $\tau$ for rigid CY flux vacua with $N_* = 150$,
$(Z^0,\F_0)=(1,i)$ , taken from Ref.\refcite{DeDo}.}}
\end{center}
\end{figure}

Supersymmetric vacua obey
\beq
D_{\tau} W = \partial_{\tau} W - \frac{W}{\tau - \bar \tau}
= -\frac{C \bar \tau +D }{\tau - \bar \tau}=0 \  \ \ 
\Rightarrow \ \ \bar \tau=-\frac{D}{C} \ .
\eeq
One can now scan all the possible values of 
$(e_0,m^0,e_{0, \, \RR},m^{0}_\RR)$ satisfying the inequality
(\ref{N*}), and get the corresponding value of the stabilized axion-dilaton.
Taking $Z^0=1$, $\F_0=i$, $N_*=150$ and
doing if necessary an $SL(2,{\mathbb Z})$ transformation to 
each resulting $\tau$ such that it is in the fundamental domain
\footnote{The fundamental domain is 
${\mathcal F}= \{ \tau \in {\mathbb C}: \I \tau >0, 
|\tau| \ge 1, |\R \tau|<\tfrac{1}{2} \}$.}
, 
Ref. \refcite{DeDo} gets the distribution shown in Figure \ref{fig:taus}.

%\begin{figure}[pb]
%\centerline{\psfig{file=T2vacua.eps,width=3.5cm}}
%\vspace*{8pt}
%\caption{Values of $\tau$ for rigid CY flux vacua with $N_* = 150$,
%$(Z^0,\F_0)=(1,i)$ , taken from Ref.\refcite{DeDo}.
% \label{fig:taus}}
%\end{figure}

The simplest example of flux vacua already gives an intricate distribution,
from which it is hard to obtain any number
(although Douglas and collaborators succeed in doing so), like for example 
the total number of vacua. However, Refs. \refcite{AsDo,DeDo} show that
for a sufficiently large region, the density of vacua per unit volume in 
moduli space can be well
approximated
by a constant, equal to $2\pi N_*^2$. This is a good approximation
for disks of radii $R$ in moduli space if $R > \sqrt{1/N_*}$.
The total number of vacua,
Eq.(\ref{N}), is therefore
\bea \label{Ntau}
\N_{\rm{susy} \, {vac}} (N_{\rm{flux}}<N_*) &=& \int_{\M_{\tau}}
\rho(\tau) d^2\tau   \\
&\approx& \int_{\M_{\tau}}  2\pi N_*^2 \frac{d^2\tau}{(2 \I \tau)^2}= 
2\pi N_*^2 \int_{\M_{\tau}} 
g_{\tau \bar \tau}\,d^2\tau = 2\pi N_*^2 \, \frac{\pi}{12}\ \nn .
\eea
It is hard to believe looking at Figure \ref{fig:taus} 
that the density is constant. One sees on one hand an accumulation
of vacua close to the boundary, at $\tau$ =1, and with a sparser
distribution for larger $\tau$. Additionally, there are
voids around the points $\tau=ni$. The higher density
for lower values of $\I \tau$ is just due to the modular invariant 
metric in moduli space, $d^2\tau/(2 \I \tau)^2$. Regarding the holes,
these are interpreted in Ref. \refcite{DeDo} as consequences 
of the special conical shape of the region containing vacua.
In any case, there is a very large degeneracy of vacua {\it at}
the points $\tau=ni$ (there are for example 240 vacua at
$\tau=2i$) which offsets the empty spaces, making the constant
density approximation good for sufficiently large radii.

The rigid Calabi-Yau is a particularly simple case, we do not expect in 
general the density
of vacua to be well approximated by a constant.
But we do need a continuous approximation to the
density in order to extract numbers out of Eq.(\ref{N}).
Such continuous approximation will replace the sum over
the integer values of the fluxes by an integral, namely
\beq
\sum_{(e,m,e_{\RR},m_{\RR}) \ \in \mathbb{Z}} \rightarrow \int 
\prod_{K=0}^{h^{2,1}_-} d e_K \, d m^K\, d e_{K, \, RR} \, d m^{K}_{\RR} \ .
\eeq
(For a detailed discussion about the limitations
of this approximation, see Ref. \refcite{AsDo}.)
This integral should be cut off at a value 
given by the upper bound (\ref{N*}), i.e. it should be 
supplemented by a step function $\Theta(N_{\rm{flux}}-N_*)$.
Collecting all the pieces together, Ref \refcite{AsDo} arrives at
\bea \label{Ncont}
\N_{\rm{susy} \, {vac}} &=& 
\sum_{\rm{susy} \ {vac}} \Theta(N_{\rm{flux}}-N_*) \\
&\approx& \frac{1}{2\pi i } \int_C \frac{d \alpha}{\alpha}
e^{\alpha N_*} \int d^{2p} z \int d^{4p} N e^{-\frac{\alpha}{2} N \eta N}
 \delta^p(DW(z)) \, 
  |{\rm det} \, D^2 W (z)| \nn
\eea
where we have used the Laplace transform in $\alpha$ of the 
step function, and defined  $p=h^{2,1}_-+1$. 

Skipping the details of the calculation, which the reader is welcome 
to follow from Refs.~\refcite{AsDo,DeDo}, we quote the result. 
A good approximation for the number of type IIB 
O3/O7 supersymmetric
vacua inside a region $R$ in moduli space is given by
\beq \label{Ngen}
\N_{\rm{susy} \, {vac}}\,(N_{\rm{flux}} \le N_*) \approx
\frac{(2 \pi N_*)^{2p}}{\pi^{p} (2p)!} \int_R {\rm{det} }
({\mathbb R} + J \, {\mathbb I}) \ , 
\quad  p=h^{2,1}_-+1 \ .
\eeq
In this equation, ${\mathbb R}$ is the curvature two-form in the $2p$
dimensional moduli space expressed as a $p \times p$
matrix, 
namely 
$({\mathbb R})_{ab}=R_{k \bar l ab}\, dz^k \wedge d\bar z^l$ 
($a,b$ are orthonormal
frame indices, and $k,l$ tangent space indices in moduli space); 
$J$ is the K\"ahler two-form $J_{k \bar l}\, dz^k \wedge d\bar z^l $ and 
${\mathbb I}$ is the 
$p \times p$ identity matrix $\delta_{ab}$.

We see that the constant density for the case of a rigid Calabi-Yau
is replaced by a ``topological'' density
\beq \label{density}
\rho(z)_{\rm susy} = \frac{(2 \pi N_*)^{2p}}{\pi^{p} (2p)!} 
\rm{det} ({\mathbb R} + J \, {\mathbb I}) \ .
\eeq
This is actually an ``index'' density: it counts the number
of vacua with signs, i.e. dropping the absolute value
of the determinant in Eq.(\ref{rhosusy}). 
It therefore gives a lower
bound to the density of vacua.
Ref. \refcite{AsDo} argues that (\ref{density}) is a good
lower bound for the number of flux vacua for $N_* \gg 2p \gg 1$,
and that the total number of vacua is probably
(\ref{Ngen}) multiplied by 
$c^{2p}$, with some $c \sim 1$.

The index density (\ref{density}) agrees with the constant $2\pi  N_*^2$ 
per unit volume in $\M_{\tau}$ for the rigid
($p=1$) case. This means
that the vacuum expectation values of the dilaton, or the 
string coupling constant, is in good approximation
uniformly distributed. Integrating the density
over a fundamental domain in moduli space, gives the number
of supersymmetric flux vacua for a given Calabi-Yau. 

%As an illustration, for $p=2$, i.e. one complex
%structure modulus and the dilaton-axtion,  

Given the formula for the total number of vacua, the first 
question to ask is whether this number is finite.
It has been conjectured that the volumes of these
moduli spaces are finite \cite{HM}, which restricts
the question of finiteness to possible divergences
coming from the curvature. A typical case in which 
the curvature diverges is the neighborhood of
a conifold point. However, Ref. \refcite{AsDo}
shows that for the complex structure moduli space
 of the mirror quintic (expecting
conifold points on other CY's to have the same behavior),
in spite of $R$ being singular, the integral
is finite. With more moduli there might be more
complicated degenerations, but there is reasonable hope
that the number $\N_{\rm{susy} \, {vac}} (N_{\rm{flux}}<N_*)$
is finite \footnote{For a recent and very nice discussion 
about finiteness of string vacua, see M.Douglas' talk
at Strings 2005 \cite{Dougtalk}.}.

Let us now illustrate with the next to simplest example:
a moduli space of complex dimension $p=2$, i.e. consisting
of the dilaton-axion plus one complex structure modulus, 
$\M=\M_{\tau} \times \M_{Y}$.
The metric
${\mathbb R} + J \, {\mathbb I}$ is
\beq
{\mathbb R} + J \, {\mathbb I} = 
\left( 
\begin{array}{cc}
R_0 + J_0 +J_1 & 0 \\ 0 & 
R_1 + J_0 + J_1 
 \end{array} 
\right) 
\eeq
where we have defined
\beq \label{matr2}
\begin{array}{cc}
 R_0 \equiv R_{\tau \bar \tau}\,  d\tau \wedge d \bar\tau \ \ , \ \ \ &  
 J_0 \equiv J_{\tau \bar \tau}\,  d\tau \wedge d \bar\tau \\
 R_1 \equiv R_{z \bar z}\,  dz \wedge d \bar z \ \ ,  \ \ \ & 
 J_1 \equiv J_{z \bar z} \, dz \wedge d \bar z 
\end{array}
\eeq
It is easy to check that for the metric on $\M_{\tau}$,
given by the  K\"ahler potential 
$K=- \ln \left[ -i (\tau - \bar \tau) \right]$, $R_0=-2J_0$.
The determinant of (\ref{matr2}) is therefore given by
\beq \label{matr}
\rm{det} ({\mathbb R} + J \, {\mathbb I}) = 
(-J_0 + J_1) \wedge (R_1 + J_0 + J_1) =
-J_0 \wedge R_1  \ .
\eeq
Inserting this in (\ref{Ngen}), and 
using $\int_{\M_{\tau}} J_0=\tfrac{\pi}{12}$ (cf. Eq.(\ref{Ntau})), gives
\beq \label{N2}
\N_{\rm{susy} \, {vac}}\,(N_{\rm{flux}} \le N_*) \approx
- \frac{(2 \pi N_*)^{4}}{\pi^{2} 4!} \frac{\pi}{12} \int_{\M_Y}
R_{z \bar z}\,  dz \wedge d \bar z = - \frac{(2 \pi N_*)^4}{4!\, 6} \chi(\M_Y)
\eeq
In this next to simplest example, the number of vacua is therefore 
just proportional to the Euler 
characteristic of the complex structure modulus space (and
to the
usual power $N_*^{2p}=N_*^4$).

Let us apply this to a manifold $Y$ with a conifold degeneration 
\cite{DeDo}. 
As in the case of the deformed conifold of section \ref{sec:moduliconifold} 
(see below 
Eq.(\ref{WKS})), the periods are  $Z(z)=z$, 
$\F(z)= \frac{z}{2 \pi i} +$ analytic terms. Inserting
this in the K\"ahler potential (\ref{KIIBO3}), we get
$K \approx -\ln [\tfrac{1}{2\pi}|z|^2 \ln |z|^2]$. This gives
the following metric near $z=0$
\beq
g_{z \bar z} \approx c\, \, \ln \frac{1}{|z|^2}
\eeq
where $c=e^{K_0}/2\pi$, with $K_0$ the K\"ahler potential
at $z=0$. The curvature is therefore
\beq
R_{z \bar z} \approx -\frac{1}{|z|^2 \, \ln |z|^2}  \ .
\eeq 
This implies that the density (\ref{density}) diverges 
at the conifold point, $z=0$. The integral
over a finite domain $|z|<R$ is nevertheless finite,
namely
\beq
\N_{\rm{susy} \, {vac}}\,(N_{\rm{flux}} \le N_*) \approx
- \frac{(2 \pi N_*)^{4}}{\pi^{2} 4!} \frac{\pi}{12} \int_{|z|<R}
R_{z \bar z}\,  dz \wedge d \bar z \approx 
\frac{(\pi N_*)^{4}}{18 \, \ln \frac{1}{R^2}}
\eeq
The logarithmic dependence implies that a very large number of vacua
are extremely close to the conifold point. For
example, for $N_*=100$, there are about one million vacua with 
$|z| < 10^{-100}$. This is a good feature from 
the phenomenological point of view, since 
small vev's for the complex structure
modulus generate large hierarchies \cite{GKP},
as in Eq.(\ref{hierarchy}).

Very similarly to the example with $p=2$, we can compute the number
of vacua for tori with diagonal period matrix, i.e.
take $\kappa^{ij}$ in Eq. (\ref{torus}) to be $\kappa^{ij}=\kappa^i \, \delta^{ij}$.
There are 4 complex moduli: 3 complex structure $\kappa^i$ and the
dilaton-axion. Given the holomorphic 3-form in Eq. (\ref{oxsimple}), 
the K\"ahler potential, Eq.(\ref{KIIBO3}), is
\bea
K&=& -\ln [-i (\tau - \bar \tau)] 
-\ln \Big[ -i \int \Ox(\kappa) \wedge \overline \Ox(\bar \kappa)\Big] \nn \\
&=& -\ln [-i (\tau - \bar \tau)] -\ln [ \rm{det} \, \I \kappa] 
= - \sum_{p=0}^3 {\rm ln} [-i (w^p - \bar w^p)]
\eea
where ${\rm w^p}=(\tau,\kappa^i)$. There are therefore four copies 
of the moduli space $\M_{\tau}$. When building the 4 x 4 matrix
(\ref{matr}), we can use $R_i=-2 \, J_i$,
(see below Eq.(\ref{matr2})). We therefore have
\beq
\rm{det} ({\mathbb R} + J \, {\mathbb I}) = 8 \,
J_0 \wedge J_1 \wedge J_2 \wedge J_3 \ .
\eeq
This is a combinatoric factor times the volume form
of the total moduli space. 
The fundamental region has a permutation symmetry, 
i.e. is preserved by $SL(2,{\mathbb Z})^p \times S_p$, so
its volume is
\beq
V_p = \frac{1}{p!} \left(\frac{\pi}{12} \right)^p
\eeq 
The number of vacua is therefore 
\beq
\N_{\rm{susy} \, {vac}} (N_{\rm{flux}}<N_*) =
8 \frac{(2 \pi N_*)^8}{\pi^4 8!} \frac{1}{4!} \left(\frac{\pi}{12}\right)^4
=  \frac{1}{3} \frac{(2 \pi N_*)^8}{(12)^4 \,  8!} \ . 
\eeq
For  $N_*=16$, which is
the value for a $T^6$ with 64 O3-planes, 
we have about four million supersymmetric vacua. 

Typical Calabi-Yau's have many more complex structure moduli, 
leading to many more possible vacua. For example,  Ref. \refcite{GKTT} 
finds several supersymmetric flux vacua for 
the Calabi-Yau 3-fold 
hypersurface in $WP_{1,1,1,1,4}$. 
The 3-fold has $h^{2,1}=149$, and its orientifold 
descends from the F-theory Calabi-Yau 4-fold in $WP_{1,1,1,1,8,12}$,
which has $\chi/24=972$. We can roughly estimate the 
number of vacua to be of the order $10^{230}$.

After looking at these huge numbers of vacua, we can convince
ourselves that
searching for vacua one by one until we reach the right one
does not seem to be a clever idea. It is true nevertheless
that we are not imposing any observational constraint on these
vacua (like cosmological constant, spectrum...). Imposing 
such constraints the numbers should reduce. Clearly, none of these vacua
has a positive cosmological constant, but we could
nevertheless demand that $|W|$ is smaller than a given value
such that the KKLT (or its variations thereof) uplifting  
mechanism could work. Let us see then how
the constraint $|W|^2 \le \lambda_*$ reduces the
numbers.  

$|W|^2< \lambda$ is enforced by inserting
 a delta $\delta(|W|^2 - \lambda)$ in (\ref{Ncont})
and integrate over $\lambda$ up to $\lambda_*$.
This seems like an easy task to do. However, there is no
nice topological formula such as (\ref{Ngen})
for the index density when one includes this restriction.
Instead, Eq. (\ref{density}) becomes
\beq \label{densitylambda}
\rho(z,\lambda)_{\rm susy}= \Theta(N_* - \lambda) 
\frac{(2p)!\, (N_*-\lambda)^{2p-1}}{(\pi N_*)^p}
\sum_{k=0}^p \frac{c_k(z)}{(2p -1 -k)!} 
\left(\frac{\lambda}{N_* -\lambda}\right)^k \ ,
\eeq
where $c_k(z)$ are homogeneous polynomial functions.
We see that the cosmological constant is smaller
than $N_*$, and the density for $\lambda \sim N_*$ is 
suppressed by a factor $(N_* - \lambda)^{p-1}$. 
Because of this suppression, for large $p$ we expect the distribution 
to be peaked at small $\lambda$. Integrating
(\ref{densitylambda}) up to a small value 
$\lambda_*$, we see that the fraction of vacua
 with $|\lambda| \le \lambda_*$ behaves like
$\lambda_* / N_*$ (in units of $1/(2 \pi \sqrt{\alpha'})^4$).
This implies that the smallest cosmological constant
is of order $\alpha'^2/\N_{\rm{vac}}$. Vacuum multiplicity
would therefore help in solving the cosmological
constant problem.

Another important feature about the distribution of flux vacua, is that
it is suppressed in the large complex structure 
region. 
With one complex structure modulus, 
and in the large complex structure limit, the K\"ahler potential
$\int \Ox \wedge \bar \Ox$ has an expansion of the form
$-{\rm ln}[i (z-\bar z)^3]$, resembling the expression for
the K\"ahler moduli $t$, Eq.(\ref{KIIA}). We saw that for this
K\"ahler potential $R_{z \bar z} \propto g_{z \bar z}$, and therefore
the density of vacua per unit volume in moduli space is constant.     
The index density (\ref{density}) goes therefore like
$d \rho(z) \sim g_{z \bar z} d^2 z \sim \frac{d^2 z}{(z-\bar z)^2}$. 
This density behaves like that of Figure \ref{fig:taus}. 
We see therefore that the large complex structure limit region is  
strongly suppressed. In the mirror IIA picture, $\I z$ is
mapped to $\I t \sim V^{1/3}$ (where $V$ is the volume).
This gives a number of vacua that falls off with volume
as $V^{-1/3}$. For $n$ complex structure moduli
(or many K\"ahler moduli in the mirror picture), there
is a large volume falloff as $V^{-n/3}$.

Ref. \refcite{DeDosb} studies the density
of non supersymmetric (F-breaking) flux vacua, given by (\ref{rhogen}).
We will not review the details here, but just quote one of their
main results. The number of metastable vacua with F-term supersymmetry
breaking scale $M_{\rm{s\not usy}}<M_*$ inside a region
$R$ is related to that of supersymmetric vacua by
\beq
N_{\rm{s\not usy}, \rm{R}} (\Ms < M_*)= M^2_* N_{\rm{susy}, \rm{R}} \ .
\eeq
If one imposes additionally that the cosmological constant be
smaller than a given small value $\lambda_*$, this ratio is
\beq
N_{\rm{s\not usy}, \rm{R}} 
(\Ms < M_*,\lambda<\lambda_*)
= M^{12}_* \lambda_* N_{\rm{susy}, \rm{R}}  \ , \qquad 
\lambda_* < |F|,|W|
%\frac{1}{(2 \pi)^4 \alpha'^2} \ . 
\eeq
Low scale supersymmetry breaking is therefore disfavored

As mentioned, there is no constraint imposed regarding the matter
content of these flux vacua.
In that respect, 
Ref. \refcite{Do} gives a first estimate 
of the fraction of vacua that have Standard Model
spectrum (see Ref. \refcite{statSM} for some explicit 
D-brane models statistics). 
The case analyzed corresponds to SM matter arising from D6-branes
in IIA wrapping 3 cycles of Calabi-Yau's. 
The fraction of vacua that have a $U(3) \times
U(2) \times U(1)^2$ gauge group is roughly
$10^{-6}$.  While the
estimates of Ref \refcite{Do} are very
crude, the conjecture is that 
ignoring values for the couplings,
the fraction of models which 
realize the standard model spectrum
is closer to ${\mathcal O}(10^{-10})$ 
than to ${\mathcal O}(10^{-100})$.

Let us summarize the prominent features of the distribution
of supersymmetric vacua. \\

\noindent $\bullet$ The density of dilaton-axion vev's is  
constant per unit volume in moduli space. \\
$\bullet$ Vacua accumulate near conifold points. \\
$\bullet$ The distribution of values of cosmological
constant is uniform near zero.\\
$\bullet$ The distribution has a falloff at large complex 
structure, or in the mirror manifold, at large volume
as $\N \sim V^{-\frac{p-1}{3}}$.\\
$\bullet$ High scale supersymmetry breaking is favored.\\

Statistics of M-theory vacua have been 
put forward in the same spirit in Ref. \refcite{Mstat}, while some
``rudimentary statistics'' in a simple
IIA toy model are performed in Ref. \refcite{WGKT}.

Although still a bit premature to drive final
conclusions from the distributions of flux vacua,
Douglas and collaborators have made a big step
toward developing a statistical approach to
flux compactifications. The first conclusion seems
to be that even
after imposing metastability, acceptable
supersymmetry breaking and a Standard Model spectrum, the number 
of vacua seems to be very large, probably too large to 
be explored one by one, making it useful to
study their distribution statistically.
This distribution shows 
many simple properties,
such as large volume and low supersymmetry breaking suppression.

To finish this section, let us point out that the distribution of 
supersymmetric IIB flux vacua (\ref{density}), 
in particular its behavior
on the moduli space geometry as $\rm{det} ({\mathbb R} + J \, {\mathbb I})$; 
its growth with $N_*$ as $N_*^{2p}$ as well as its conifold attractor
point
have been tested using Monte Carlo experiments in Refs. \refcite{GKT,CoQ}
(see also Ref. \refcite{WGKTIIB}).
Ref. \refcite{GKT} studies the mirror of an orientifold of the 
CY hypersurface in $P_{1,1,1,1,4}$, which has
 $p=2$ (one complex structure modulus) and $N_*=972$.
Refs.~\refcite{CoQ,statWP} analyze the mirror of the 
CY hypersurface $P_{1,1,2,2,6}$, which
has two complex structure moduli.

\section{Summary and future prospects}

It is clear that
flux compactifications in string theory has been an extremely active and fruitful
area of research, particularly in the past five years.
The main reason for this flurry of activity 
is that fluxes are the only known perturbative mechanism  
that stabilizes moduli. 
In many setups, most
notably in all IIB compactifications on Calabi-Yau's or tori, closed string fluxes
are not enough to stabilize all moduli.  
Therefore, other perturbative and non perturbative mechanisms,
some of which we do not have enough theoretical control yet,
have to be invoked. However, 
fluxes were shown very recently to be enough
to stabilize all moduli in IIA compactifications on rigid tori and
on twisted tori,
and similarly all moduli can in principle be stabilized in IIB
by considering additionally open string fluxes.
While stabilizing moduli, fluxes break supersymmetry
partially or completely in a stable way, and generate 
warp factors, giving stringy mechanisms of realizing
large hierarchies.
Fluxes have also served as an ingredient in the 
construction of semi-realistic four-dimensional 
vacua with Standard Model spectra.

In all these models, the back-reaction of the geometry
is basically neglected. 
The back-reacted internal geometry 
involves, in all flux backgrounds, manifolds with torsion. In spite
of the recent progress
in understanding their mathematical properties,
and therefore the four-dimensional
resulting physics,
many of the very basic questions such as what
are their moduli spaces are still widely unanswered.  
However, 
conformally-Ricci-flat manifolds (like
conformal Calabi-Yau's or tori) 
allow for non trivial fluxes,
and give us a way of studying the
effective theories with the available
mathematical tools (in most cases
neglecting, however, the conformal factor).
In the absence of fluxes, these leave $\N=8,4,2$ unbroken
supersymmetries in four-dimensions, while 
turning on fluxes breaks these to $\N=4,3,2,1,0$.
Most of the analysis of moduli stabilization 
has been carried out in this class of compactifications.

There are explicit constructions of flux compactifications
with all moduli stabilized. In type IIB, these use
non-perturbative effects in sufficiently sophisticated 
internal manifolds that have the right cycles to support 
the corrections, and are non-supersymmetric. 
In type IIA, the explicit constructions involve 
rigid orientifolds of tori orbifolds,  where 
all moduli can be stabilized in  $\N=1$ supersymmetric AdS vacua,
or orientifolds of twisted tori,
where there are additional 
Minkowski vacua with all moduli fixed except for some axions.
Besides, metastable de Sitter vacua 
in IIB can also be attained using (supersymmetry breaking)
fluxes and non-perturbative 
effects. 

Flux compactifications have many additional interconnected 
aspects / applications
that we have not discussed in depth in this review, or have not discussed at all.
First of all, there is much more to report about
the gauged supergravity interpretation of certain
flux compactifications 
\cite{fluxandgauge,DKPZ} (for a short review, see
\cite{revfluxandgauge}). Besides,
we have not discussed the relation between flux compactifications on twisted
tori and  Scherk-Schwarz compactifications \cite{ScSc}
(see Refs.\refcite{DaHu,SSflux}).
Neither have we discussed non-geometric backgrounds \cite{nongeo},
which are sometimes dual to geometric flux vacua.
Another undiscussed area 
is that of topological strings \cite{revtop},
in particular the attractor mechanism \cite{attr} and the 
Ooguri-Strominger-Vafa conjecture \cite{OSV}.

On a more phenomenological level,
fluxes generate supersymmetric and soft supersymmetry breaking 
terms on D-branes \cite{GGJL,soft,soft7,softl}, which can stabilize
open string moduli \cite{BCMS,oms,LustMRS}.
(In the presence of fluxes, the supersymmetric cycles wrapped
by D-branes are also modified, in order to minimize
the action  \cite{London1,cali}.)
Soft-supersymmetry breaking effects and moduli stabilization
in KKLT type scenarios has been discussed extensively 
(see for example \cite{Quevedo,CFNO,AEB}).
Some other phenomenological applications
that we have not discussed much 
are, for example, the landscape of supersymmetry breaking vacua
and supersymmetry breaking scales \cite{Landsb}
(see Ref. \refcite{DeDo} for a statistical analysis);
and models of inflation embedded
in flux compactification scenarios, started most
notably by Ref. \refcite{KKLMMT}.

Some non-compact flux backgrounds play 
a very important role in AdS/CFT, as they
can realize string duals of confining gauge theories
(see \refcite{adscftreview} for a review).
Latest developments along these lines involve the 
recent explicit construction of conical Calabi-Yau
metrics over the Sasaki-Einstein 
manifolds $Y^{p,q}$ \cite{Ypq} and $L^{p,q,r}$ \cite{Lpqr}.
Their dual conformal field theories were found in
Ref. \cite{CFTYpq}. However, there are no complex
structure deformations on these manifolds, and
opposite to the case in Klebanov-Strassler's solution, 
adding 3-form fluxes does not seem to 
lead to regular supersymmetric backgrounds \cite{susybYpq}.

By turning on appropriate fluxes (which requires sometimes some
fine tuning), moduli are stabilized in the regime
where we can trust the effective field theory approximation.
This needs a (string frame) volume or radius  much 
bigger than the KK and string scales,
ans a small coupling constant (or more the more restrictive 
condition $g_s N_{\rm{flux}} \ll 1$).
Nevertheless, specific (small) numbers like the cosmological constant 
in non-supersymmetric solutions are subject to 
corrections which can be of the same order of magnitude as the leading
order value, and cannot therefore be fully trusted.
Corrections are however expected to be small if
the vacuum energy is itself much smaller than the string scale
and KK scale to the fourth power. 
In any case, leaving aside the phenomenology of scales,  
it is conceivable that in some of these backgrounds
a subsector of the theory   
develops an instability, and we are dropping in
the effective Lagrangian a KK or stringy mode which is
tachyonic. 
Backgrounds with NS fluxes can be studied
in the other extreme regime, where the internal space
is very small, but world-sheet techniques are 
powerful enough, like in orientifolds of Gepner models \cite{Gepner}
(see Ref \refcite{statgepner} for a search of vacua with
Standard Model spectra in Gepner models). 
The analysis of backgrounds with RR flux using world-sheet 
techniques is on the contrary still far from being available, 
although a lot of progress in this direction is expected 
(or hoped) to happen:
a new formalism for a covariant quantization
of the superstring which allows to study backgrounds
with RR flux was introduced a few years ago \cite{Berko}. As
as of today, however, loop amplitudes have only been obtained
in highly symmetric backgrounds like $AdS_5 \times S^5$
or the pp-wave.

A fair concluding remark would be
that we have made huge progress in the past years,
but there is still a lot of work ahead of us 
to see whether the available models of flux compactifications 
are as copious and rich as they appear today, and 
whether our four-dimensional world is one of them.

\section*{Acknowledgments}

We would like to thank Thomas Grimm, Jan Louis, Andrei Micu, Ruben Minasian,
Michela Petrini, Alessandro Tomasiello, Stefan Theisen,  Dan Waldram and Fabio Zwirner for 
useful discussions. This work has been supported by 
European Commission Marie Curie Postdoctoral
Fellowship under contract number MEIF-CT-2003-501485.
Partial support is provided additionally from
INTAS grant, 03-51-6346, CNRS PICS 2530,
RTN contracts MRTN-CT-2004-005104 and 
MRTN-CT-2004-503369 and by a European Union Excellence Grant,
MEXT-CT-2003-509661.

%... , mi papa, mi mama,
%mi abuela, mis gatas, por que no mi marido, pobre, que se banco un mes 
%de agosto del orto, y a todos los hombres del mundo que quieran habitar el suelo 
%argentino (y mujeres, consitucion machista!!, o es que ninguna mujer quiere
%habitar el suelo argentino??)...

\appendix

\section{Conventions} \label{ap:conventions}

\noindent $\bullet$ We use the following index notation

- $\mu, \nu, ...=0,..., 3$ label external coordinates.

- $m, n, ...=1,..., 6$ label real internal coordinates.

- $i,j=1,2,3$, $\bi, \bj=1,2,3$ label internal complex coordinates. 

- $M,N,...=0,...,9$ label all coordinates. 

\noindent $\bullet$ We switch back and forth between Einstein frame and string frame. 
Section \ref{sec:susy} is entirely in string frame.
Most of section \ref{sec:nogo} is also in string frame,
except Eqs. (\ref{branetension}), 
(\ref{nogosources}) and (\ref{Ftheory}) (all taken from Ref.\refcite{GKP}),
which are in Einstein frame. Sections \ref{sec:4D} and on are in
Einstein frame.
% except the superpotential subsection \ref{sec:fluxsup}
%which is in string frame.

\noindent $\bullet$ The different RR field strengths used 
(standard, modified, internal, external) are

-$\hat F_n=dC_{n-1}$ is the standard RR field strength, and 
$F^{(10)}$ is the modified one. Explicitly,
\beq
F^{(10)}=dC - H \wedge C + m \, e^B = \hat F- H \wedge C
\eeq

-The modified flux is split into internal and external components 
according to
\beq 
F^{(10)}_n = F_n + \rm{Vol}_4 \wedge \tilde F_{n-4} \ .
\eeq
Hodge dualities among the components given in Eqs (\ref{sd}, \ref{dualhats}).

\noindent $\bullet$ A $\star$ symbol means 10-dimensional Hodge duality, while we use
$*$ for internal (6D) Hodge duality.

\noindent $\bullet$ Whenever factors of $\alpha'$ are not written explicitly,
we are taking  $(2 \pi)^2 \ax'=1$.

\noindent $\bullet$ We use the standard 
decomposition 
of the
ten-dimensional gamma matrices $\Gamma^M = (\Gamma^\mu,\Gamma^m)$ as
\begin{equation}
   \Gamma^\mu=\gamma^\mu\otimes 1 ,\quad \mu=0,1,2,3\ ,\qquad
   \Gamma^m=\gamma_5\otimes\gamma^m\ ,\quad m=1,\ldots, 6\ , 
\end{equation}

and 
\beq
\gamma_5=\tfrac{i}{4!} \epsilon_{\mu\nu\lambda \rho} \gamma^{\mu\nu\lambda \rho}
\, , \qquad 
\gamma_7=-\tfrac{i}{6!} \epsilon_{mnpqrs} \gamma^{mnpqrs} \, , \qquad 
\gamma_{11}=\gamma_5 \, \gamma_7
\eeq

The $\gamma^m$ are Hermitean, as are the $\gamma^\mu$, except
$\gamma^0$ which is antihermitean.

\noindent  $\bullet$ A slash is defined in the following way
\beq
\sla \! {F_{n}} = \frac{1}{n!} F_{P_1...P_N} \Gamma^{P_1...P_N} \ , \qquad
\Gamma^{P_1...P_N}= \Gamma^{[P_1} ... \Gamma^{P_N]} \ .
\eeq

For the SU(3) structure, the norm of the normalized spinor is
\beq
\eta^{\dagger} \eta = \eta_+^{\dagger} \eta_+ +  \eta_-^{\dagger} \eta_-=1 \ . 
\eeq

\noindent  $\bullet$ The fundamental 2-form and holomorphic 3-form constructed from this 
spinor as in (\ref{bilinears}) obey
\beq
J \wedge J \wedge J = \frac{3i}{4} \, \Ox \wedge \bar \Ox  \ .
\eeq

\noindent  $\bullet$ The decomposition of a 2-,3-,4-,5- and 6-form in SU(3)
representations used in section \ref{sec:N=1} is 
\bea
F_1 &=& F_1^{(3)} + F_3^{(\bar 3)} \nn \\
F_2 &=& \frac{1}{3} F_{2}^{(1)}\, J +  \R (F_{2}^{(3)}\llcorner
\overline{\Ox}) + F_{2}^{(8)}  \, , \nn \\
F_3 &=& -\frac{3}{2}\, {\rm Im}(F_3^{(1)} \bar{\Omega}) + 
(F_3^{(3)}+F_3^{(\bar 3)}) \wedge J 
+ F_3^{(6)} + F_3^{(\bar 6)} \, , \nn \\
F_{4}&=& \frac16 F_{4}^{(1)} J \wedge J +   Re(F_{4}^{(3)}\wedge
\overline{\Ox}) + F_{4}^{(8)} \, , \nn\\
F_{5} &=&  (F_{5}^{(3)} +F_5^{(\bar 3)}) \wedge J \wedge J \, , \nn\\
F_{6} &=& \frac{1}{6}  F_{6}^{(1)} \,J \wedge J \wedge J \, ,
\eea

The inverse relations are
\bea \label{alvesre}
(F_1^{(3)})_i &=& F_i \nn \\
F_{2}^{(1)}&=& \frac{1}{2} F_{mn} J^{mn} =
F_{i \bj } J^{i \bj} \nn \\
F_3^{(1)}&=& -\frac{i}{36}   F^{ijk} \Ox_{ijk} \, ,\qquad \quad
(F_3^{(3)})_i = \frac14 F_{imn} J^{mn} \, , \qquad
(F_3^{(6)})_{ij} = F^{kl}\,_{(i} \Ox_{j)kl} \nn \\
F_{4}^{(1)}&=&\frac18 F^{mnpq} J_{mn} J_{pq} \, , \qquad 
(F_{4}^{(3)})_k= \frac{1}{24} F_{k}\,^{ijl} \Ox_{ijl} \,  \nn \\
(F_{5}^{(3)})_i&=& \frac{1}{16} F_i^{mnpq} J_{mn} J_{pq} \,  \nn\\
F_{6}^{(1)}&=& \frac{1}{48} F^{mnpqrs} J_{mn} J_{np} J_{qr} 
\eea

These can be used to obtain the torsion classes in (\ref{dJdOmega})
in terms of $dJ, d\Omega, J$ and $\Omega$. 
%Using the contracion
%on forms $\llcorner$, defined on the complex basis
%by $dz^i \llcorner d\bar z^{\bj}= g^{i \bj}$, the torsion classes are
%{\bf give these?}
%\bea
%W_1 &=& \frac{4i}{3} \Omega \llcorner dJ = \frac{4}{3} J^2 \llcorner d\Omega \nn \\
%W_2 &=& 4 J \llcorner \left(d\Omega - W_1 J^2 - \bar W_5 J\right) \nn \\
%W_3 &=& dJ - \frac{3}{2} \I (\bar W_1 \Omega) - W_4 \wedge J \nn \\
%W_4 &=& 2 J \llcorner dJ \nn \\
%W_5 &=& - \Omega \, \llcorner d \bar \Omega
%\eea

\section*{References}

%\begin{thebibliography}{000} %for 3 digits
%\begin{thebibliography}{00}  %for 2 digits

\end{document}